\documentstyle[12pt]{article}
\newcommand{\ret}{\nonumber \\}

\newcommand{\Section}[1]%
{\section{#1}\setcounter{equation}{0}%
\setcounter{theorem}{0}}
\newtheorem{theorem}{Theorem}
\newtheorem{lemma}[theorem]{Lemma}

\newtheorem{pro}[theorem]{Proposition}
\newtheorem{definition}[theorem]{Definition}

\newenvironment{proof}[1]%
{\par\noindent{\em #1:\ }}%
{~\rule{2mm}{2mm}\par\bigskip}
\begin{document}
\newpage\thispagestyle{empty}
{\topskip 2cm
\begin{center}
{\Large\bf Spectral Gaps of Quantum Hall Systems\\ 
\bigskip
with Interactions\\}
\bigskip\bigskip
{\large Tohru Koma$^\ast$\\}
\bigskip
\end{center}
\vfil
\noindent
A two-dimensional quantum Hall system without disorder for a wide class of 
interactions including any two-body interaction with finite range 
is studied by using the Lieb-Schultz-Mattis method 
[{\it Ann. Phys. (N.Y.)} {\bf 16}: 407 (1961)]. 
The model is defined on an infinitely long strip with a fixed 
large width, and the Hilbert space is restricted to the lowest 
$(n_{\rm max}+1)$ Landau levels with a large integer $n_{\rm max}$. 
We proved that, for a non-integer filling $\nu$ of the Landau 
levels, either (i) there is a symmetry breaking at zero temperature 
or (ii) there is only one infinite-volume ground state with a gapless 
excitation. We also proved the following two theorems: 
(a) If a pure infinite-volume ground state has a non-zero excitation 
gap for a non-integer filling $\nu$, then a translational symmetry 
breaking occurs at zero temperature. 
(b) Suppose that there is no non-translationally invariant infinite-volume 
ground state. Then, if a pure infinite-volume ground state has a non-zero 
excitation gap, the filling factor $\nu$ must be equal to a rational number. 
Here the ground state is allowed to have a periodic structure which is 
a consequence of the translational symmetry breaking. 
We also discuss the relation between our results and the quantized Hall 
conductance, and phenomenologically explain why odd denominators 
of filling fractions $\nu$ giving the quantized Hall conductance, 
are favored exclusively. 
\par\noindent
\bigskip
\hrule
\bigskip
\noindent
{\bf KEY WORDS:} Quantum Hall effect; fractional quantum Hall effect; 
Landau Hamiltonian; strong magnetic field; electron-electron interaction; 
spectral gap; translational symmetry breaking. 
\hfill
\bigskip 

\noindent
---$\!$---$\!$---$\!$---
\medskip

\noindent
$^\ast$ {\small Department of Physics, Gakushuin University, 
Mejiro, Toshima-ku, Tokyo 171, JAPAN}
\smallskip

\noindent
{\small\tt e-mail: koma@riron.gakushuin.ac.jp}
\vfil}\newpage
\tableofcontents
\newpage
\Section{Introduction}

Since the experimental discovery of the fractional quantum Hall effect 
\cite{TSG,SCTHG}, considerable theoretical efforts have been made 
to understand the nature of the ground state and of the low energy 
excitations above the ground state in the two-dimensional interacting 
electron gas in a strong magnetic field. 
Although there appeared many theories, mathematically rigorous 
or exact results are still fairly rare. 
Actually this is one of most difficult problems in solid state physics 
because the electron-electron interaction is essential to the fractional 
quantization of the Hall conductance. For the history of the quantum Hall 
effect, see refs.~\cite{PGBook,CPBook,SPBook} and references therein. 

In this paper we study the properties of infinite-volume ground states 
and of low energy excitations in a two-dimensional interacting electron gas 
in a uniform magnetic field without disorder for a wide class of 
electron-electron interactions. Although the class includes any two-body 
interaction with finite range, it does not include the standard Coulomb 
interaction proportional to $1/r$, where $r$ is the distance 
between two electrons. Owing to technical reasons, we define the model 
on an infinitely long strip with a fixed width, and restrict 
the Hilbert space to the lowest $(n_{\rm max}+1)$ Landau levels 
with a fixed integer $n_{\rm max}$. The precise form of the Hamiltonian is 
given in the next Section~\ref{mainresults}. 
In Section~\ref{cutoffs}, the reason why we must fix the width and 
the integer $n_{\rm max}$, will be explained, 
and with the present results, we will give a discussion about 
the two-dimensional infinite-volume system with no restriction 
on the width and the cutoff $n_{\rm max}$. 

We apply the Lieb-Schultz-Mattis method \cite{LSM} to the model. 
The method was developed to construct a low energy 
excitation above a finite-volume ground state for a lattice quantum 
spin system with a translational invariance. Later the method was applied 
to quantum spin chains in relation to the Haldane gap 
\cite{AffleckLieb,Affleck} and magnetization plateaus \cite{OYA}. 
Yamanaka, Oshikawa and Affleck  \cite{YOA} applied the method to 
a wide class of interacting fermions systems 
on a lattice.\footnote{The results of ref.~\cite{YOA} were revisited 
in specific cases by Gagliardini, Haas and Rice \cite{GHR}.} 
Among these works, Oshikawa, Yamanaka and Affleck \cite{OYA} pointed 
out the analogy between the magnetization plateaus in a quantum spin chain and 
the conductance plateaus in the quantum Hall system. In both systems, 
a non-zero excitation gap above a ground state indeed plays an important role. 

Using the Lieb-Schultz-Mattis method, an information about 
an infinite-volume ground state or a low energy excitation 
can be obtained for a translationally invariant system. 
In a quantum Hall system, it is believed that a non-zero excitation gap 
above a ground state leads to the quantization of the Hall conductance and 
the conductance plateaus. Therefore knowledge about a ground state 
and a low energy excitation is very important for the quantum Hall effect. 

\subsection{The main results of this paper}

Our results are as follows: 
\begin{itemize}
\item Let the filling $\nu$ of the Landau levels be a 
non-integer. Then, either (i) there is a symmetry breaking at zero temperature 
or (ii) there is only one infinite-volume ground state with a gapless 
excitation. 
\item If a pure infinite-volume ground state has a non-zero excitation 
gap for a non-integer filling $\nu$, then a translational symmetry 
breaking occurs at zero temperature. 
\item Suppose that there is no non-translationally invariant infinite-volume 
ground state. Then, if a pure infinite-volume ground state has a non-zero 
excitation gap, the filling factor $\nu$ must be equal to a rational number. 
Here the ground state is allowed to have a periodic structure which is 
a consequence of the translational symmetry breaking. 
\end{itemize}
Here we stress that these statements hold also for a fixed macroscopic width 
of the strip and a fixed integer $n_{\rm max}$ giving a macroscopic 
energy. But, in the proofs, the structure of the low energy excitation 
constructed by using the Lieb-Schultz-Mattis method strongly 
depends on the width of the strip and the energy cutoff $n_{\rm max}$. 
In particular, the size of the locally excited region must increase with 
increasing the cutoffs for keeping a small excitation energy. 
For this issue, we will give a discussion in 
Section~\ref{cutoffs}. 
In the next Section~\ref{mainresults}, the above results will be given again 
as our main theorems in a mathematically rigorous manner. 
The mathematically precise definitions of the filling factor $\nu$, 
an infinite-volume ground state and an excitation gap also will be given 
in the section. 

\subsection{Physical meaning of the results}
\label{Pmeaning}

Let us briefly discuss the physical meaning of the above our three results. 
To begin with, we remark the following: For an integer filling $\nu$, 
a ground state has a trivial non-zero excitation gap\footnote{It goes 
without saying that the integral quantization of the Hall conductance and 
the appearance of the conductance plateaus are non-trivial and suprising 
phenomena \cite{KDP,KawajiWakabayashi}.}
which comes from the Landau levels for the non-interacting system 
if the magnetic field is sufficiently strong compared to 
the electron-electron interaction. We also remark that, 
without an interaction, there is no non-trivial structure 
leading to the fractional quantization of the Hall conductance. 
Thus we are interested in the case with a non-integer filling $\nu$ and 
with an interaction.  

Since a non-trivial excitation gap above a ground state for 
a non-integer filling $\nu$ plays an important role for the fractional 
quantization of the Hall conductance, the first case (i) in the first result 
is of interest to us. In this situation, a translational symmetry breaking 
occurs at zero temperature. This is the second result. In addition, 
if the electron-electron interaction is repulsive, 
we can expect that there is no non-translationally 
invariant ground state.\footnote{As far as we know, there is no 
proof for this type of statement in the repulsive case.} 
But a pure infinite-volume ground state exhibits a periodic structure 
as a consequence of the translational symmetry breaking. 
Conversely, if the electron-electron interaction is attractive, 
we can expect that there is a phase separation which implies 
the existence of a non-translationally invariant ground state with no periodic 
structure. Thus, for the repulsive case, the assumption of the third 
result, i.e., the absence of non-translationally invariant ground states, 
is expected to be vaild. With this assumption, the third result states that 
the filling factor $\nu$ must be equal to a rational number in the case 
of interest that there is a non-zero excitation gap 
above the ground state. 
Physically this implies that there appears a commensurate phase at zero 
temperature with a rational filling $\nu$. 

\subsection{The relation between the results and the quantized Hall 
conductance}
\label{relationRQ}

Next we discuss the relation between the third result and the fractional 
quantization of the Hall conductance. The statements below in this subsection 
are not justified without additional assumptions to those of our present 
results. 

To begin with, we briefly state 
our result about the Hall conductance in a separate paper \cite{Koma}. 
We treated a two-dimensional electrons gas in a uniform magnetic field 
for a wide class of potentials including single-body potentials with 
disorder and repulsive electron-electron interactions. 
We stress that there is a wide class of common models 
which are included in both the class of ref.~\cite{Koma} 
and that of the present paper. 
We obtained the following result: If there is a non-zero excitation gap 
above the ground state(s), then the Hall conductance $\sigma_{xy}$ in 
the infinite-volume limit is given by\footnote{As is well known, 
an argument relying on a topological invariant of the Hall conductance 
\cite{topological} always yields an integral quantization 
of the Hall conductance without ad hoc assumptions \cite{adhoc}. 
Since we did not rely on such an argument in our derivation of 
the Hall conductance, 
our result includes both integral and fractional quantizations of the 
Hall conductance. For earlier theoretical works on the Hall 
conductance, see refs. \cite{earlyworks}.} 
\begin{equation}
\sigma_{xy}=-\frac{e^2}{h}\nu,
\label{Hallconductance}
\end{equation}
where $-e$ is the charge of electron, $h$ is the Planck constant, 
and we assumed a regularization about a uniform electric field in 
the derivation of the Hall conductance. See ref.~\cite{Koma} for 
the mathematically rigorous statement. Unfortunately, the condition of 
a gap is different from that of the third result in the present 
paper, and we do not know the relation between the two conditions in 
a mathematically rigorous sense. 
In the rest of this subsection, we will use the conductance formula 
(\ref{Hallconductance}) without carefully examining the condition of a gap. 

Let us consider a common model mentioned above, and make the assumptions for 
the third result in the present paper. Then we clearly have the fractional 
quantization of the Hall conductance by combining the rational filling $\nu$ 
of the third result with the conductance formula (\ref{Hallconductance}). 
Roughly speaking, a fractional filling factor $\nu$ 
with a non-zero excitation gap above a ground state gives the fractionally 
quantized Hall conductance. 
Next introduce weak disorder so that the non-zero excitation gap 
above the ground state in the clean system persists against disorder. 
Then we have the fractional quantization of the Hall conductance again 
because the conductance formula holds even for the presence of disorder. 

The appearance of a Hall conductance plateau due to disorder will 
be discussed with relation to localization of wavefunctions 
in another separate paper \cite{Koma2}. 

\subsection{A phenomenological explanation for the odd denominator rule}

Experimental results show the suprising fact that odd denominators of 
filling fractions $\nu$ for which the quantization of the Hall 
conductance occurs, are favored exclusively. 
Namely non-zero excitation gaps appear only for filling fractions $\nu$ with 
odd denominators except for a few filling fractions with even 
denominators \cite{WESTGE}. Having our results in mind, we shall discuss 
the reason. Consider first the problem of two electrons with a repulsive 
interaction in a uniform magnetic field. Clearly the two electrons exert 
opposing forces on each other. But they cannot separate in the large distance 
because of the magnetic field. From this naive observation, one can expect 
that two electrons favor a bound pair \cite{Halperin} in a quantum Hall 
system. 

Write $\nu=p/q$ with $p,q$ mutually prime integers. 
Then there are $p$ electrons and $q-p$ holes on $q$ lattice sites, 
where the lattice is defined by an identification with the set 
of wavenumbers for the eigenvectors of the single-electron Landau Hamiltonian 
with the Landau gauge.\footnote{See Sections~\ref{mainresults} and 
\ref{singleelectron} for the precise definition of the lattice.} 
Each wavenumber is identical to the center of a harmonic oscillator part 
of an eigenvector. The set of all the centers is identical to the 
one-dimensional lattice. Assume $q$ is an even integer. 
Then both $p$ and $q-p$ are odd. This implies that 
neither the electrons nor the holes are grouped into bound pairs 
on the $q$ lattice sites. To form a stable pairing state, we need $2q$ 
lattice sites which lead to a periodic structure with the period $2q$. 
Here the periodic structure is a consequence of a translational symmetry 
breaking. However, the filling $\nu=p/q$ is expected to lead to a structure 
with the period $q$, not $2q$. In consequence, we cannot expect a ground state 
with a non-zero excitation gap for an even denominator. 
Next assume $q$ is odd. Then there are two possibilities: 
(i) $p$ is odd and $q-p$ is even. (ii) $p$ is even and $q-p$ is odd. 
Namely, either the number of the holes or the number of the electrons is even. 
Therefore either the holes or the electrons are grouped into bound pairs 
on the $q$ lattice sites. In comparison to the case with an even denominator, 
we can expect a stable state, i.e., a ground state with a non-zero excitation 
gap. Unfortunately this is a phenomenological explanation which is still not 
justified. 

\subsection{Outline of this paper}

This paper is organized as follows: 
In Section~\ref{mainresults}, we give the precise definition of the model 
and some notions related to an infinite-volume ground state, 
and describe our main theorems in a mathematically rigorous manner. 
As preliminaries for the proofs of our theorems, we briefly review 
the eigenvalue problem of the single-electron Landau Hamiltonian 
and the degeneracy of finite-volume ground states in an interacting 
electrons system in Section~\ref{preliminary}. In Section~\ref{LSMmethod}, 
we construct a candidate for a low energy excitation above a ground state 
by using the Lieb-Schultz-Mattis method, and prove our main theorems. 
Section~\ref{Orthproof} is devoted to a proof of a proposition about 
the orthogonality between the excited and the ground states. 
The energy gaps are estimated in Section~\ref{estimateEgap}. 
For the convenience of readers, 
Appendices~\ref{MatsuiTheorem}-\ref{ProofintUphiphibound} are devoted 
to proofs of some technical theorem and lemmas. 

\Section{The model and the main theorems}
\label{mainresults}

The purpose of this section is to describe our main theorems in a 
mathematically rigorous manner after giving mathematically precise 
definitions of an infinite-volume ground state and of a excitation gap 
for the quantum Hall system we consider. 

\subsection{The Hamiltonian}
Consider a two-dimentional interacting electrons gas 
in a uniform magnetic field in a rectangular box 
$S:=[-L_x/2,L_x/2]\times[-L_y/2,L_y/2]$. 
Although we consider electrons without spin degrees of freedom 
in this paper, our method is applied also to a quantum Hall system 
with spin degrees of freedom or with multiple layers. 

The Hamiltonian of $N$ electrons without spin degrees of freedom is given by 
\begin{equation}
H^{(N)}=\sum_{j=1}^N
\frac{1}{2m_e}\left[(p_{x,j}-eBy_j)^2+p_{y,j}^2\right]
+\sum_{j=1}^N W(x_j)+U^{(N)}({\bf r}_1,{\bf r}_2,\ldots,{\bf r}_N),
\label{hamtot}
\end{equation}
where $m_e$ and $-e$ are, respectively, the mass of electron and 
the charge of electron, and $(0,0,B)$ is the uniform magnetic field 
perpendicular to the $x$-$y$ plane in which the electrons 
are confined; ${\bf r}_j=(x_j,y_j)$ is 
the j th Cartesian coordinate of the $N$ electrons, and 
\begin{equation}
p_{x,j}=-i\hbar \frac{\partial}{\partial x_j} \quad \mbox{and} \quad 
p_{y,j}=-i\hbar \frac{\partial}{\partial y_j}
\end{equation}
with the Planck constant $\hbar$. The single-body potential $W$ is 
a function of $x$ only such that $W$ is essentially bounded, i.e., 
$\Vert W\Vert_\infty<W_0<\infty$ with a positive constant $W_0$ 
which is independent of $L_x,L_y$, 
and that $W$ satisfies a periodic boundary 
condition as 
\begin{equation}
W(x+L_x)=W(x) \quad \mbox{for any}\ x\in{\bf R}.
\end{equation}
A simlpe example of $W$ is\footnote{The question of the applicability 
of our method to a quantum Hall system with a periodic potential was 
brought to the author by Mahito Kohmoto. Thus we have partially answered 
his question, although we still cannot treat a periodic potential 
modulating in both $x$ and $y$ directions.}
\begin{equation}
W(x)=W_0\cos\kappa x \quad \mbox{with}\ \kappa=\frac{2\pi}{L_x}n, 
\quad n\in{\bf Z},
\end{equation}
where $W_0$ is a real constant. The interaction $U^{(N)}$ is written in 
a sum of two-body interactions as 
\begin{equation}
U^{(N)}({\bf r}_1,\ldots,{\bf r}_N)=\sum_{i<j}U^{(2)}(x_i-x_j,y_i-y_j). 
\end{equation}
The two-body interaction $U^{(2)}$ is invariant under the exchange of two 
coordinates of the electrons, i.e., 
\begin{equation}
U^{(2)}(-x,-y)=U^{(2)}(x,y).
\end{equation}
We assume that $U^{(2)}$ satisfies the periodic boundary conditions 
\begin{equation}
U^{(2)}(x+L_x,y)=U^{(2)}(x,y+L_y)=U^{(2)}(x,y).
\label{periodicU2}
\end{equation}
Further we assume that $U^{(2)}$ is continuous on ${\bf R}^2$, and satisfies 
\begin{equation}
\left|U^{(2)}(x,y)\right|\le U_0\left\{1+[{\rm dist}(x,y)/r_0]^2
\right\}^{-\gamma/2}\quad \mbox{for}\ (x,y)\in{\bf R}^2 
\label{decayU2bound}
\end{equation}
with $\gamma>2$ and with positive constants $U_0,r_0$. 
Here the distance is given by 
\begin{equation}
{\rm dist}(x,y):=\sqrt{\min_{m\in {\bf Z}}\{\left|x-mL_x\right|^2\}
+\min_{n\in {\bf Z}}\{\left|y-nL_x\right|^2\}}.
\end{equation}
We take $L_xL_y=2\pi M\ell_B^2$ with a sufficienlty large positive integer 
$M$. Here $\ell_B$ is the magnetic length, i.e., $\ell_B:=\sqrt{\hbar/eB}$. 
For simplicity, we take $M$ even. This condition for $L_x,L_y$ 
is convenient for imposing periodic boundary conditions as follows: 
For an $N$-electron wavefunction $\Phi^{(N)}$, 
we impose periodic boundary conditions 
\begin{equation}
t_j^{(x)}(L_x)\Phi^{(N)}({\bf r}_1,{\bf r}_2,\ldots,{\bf r}_N)
=\Phi^{(N)}({\bf r}_1,{\bf r}_2,\ldots,{\bf r}_N),
\end{equation}
and
\begin{equation}
t_j^{(y)}(L_y)\Phi^{(N)}({\bf r}_1,{\bf r}_2,\ldots,{\bf r}_N)
=\Phi^{(N)}({\bf r}_1,{\bf r}_2,\ldots,{\bf r}_N) 
\end{equation}
for $j=1,2,\ldots,N$. Here $t^{(x)}(\cdots)$ and $t^{(y)}(\cdots)$ are 
magnetic translation operators \cite{Zak} defined as 
\begin{equation}
t^{(x)}(x')f(x,y)=f(x-x',y),\quad 
t^{(y)}(y')f(x,y)=\exp[iy'x/\ell_B^2]f(x,y-y')
\label{defmagnetictrans}
\end{equation}
for a function $f$ on ${\bf R}^2$, and a subscript $j$ of an operator 
indicates that the operator acts on the $j$-th coordinate of 
a function.\footnote{Throughout the present paper we use this convention.} 
The ranges of $x'$ and $y'$ are given 
by\footnote{See Section~\ref{singleelectron}.} 
\begin{equation}
x'=m\Delta x \quad \mbox{with} \ m\in {\bf Z}, \quad \mbox{and}
\quad 
y'=n\Delta y \quad \mbox{with} \ n\in {\bf Z},
\end{equation}
where the minimal units of the translations are given by 
\begin{equation}
\Delta x:=\frac{h}{eB}\frac{1}{L_y}, \quad \mbox{and}\quad 
\Delta y:=\frac{h}{eB}\frac{1}{L_x}.
\end{equation}

Owing to certain technical reasons,\footnote{See Section~\ref{cutoffs} 
for the detail.} we must restrict 
the whole Hilbert spapce to the lowest $(n_{\rm max}+1)$ 
Landau levels with a large positive integer $n_{\rm max}$. 
In order to give a more precise definition of the restriction, 
consider the Hamiltonian of a single electron given by 
\begin{equation}
{\cal H}=\frac{1}{2m_e}\left[(p_x-eBy)^2+p_y^2\right]
\label{singleham}
\end{equation}
with periodic boundary conditions 
\begin{equation}
\phi(x,y)=t^{(x)}(L_x)\phi(x,y), \quad \phi(x,y)=t^{(y)}(L_y)\phi(x,y) 
\end{equation}
for the wavefunction $\phi$, with $L_xL_y=2\pi M\ell_B^2$ with $M$ even. 
The explicit forms of the normalized eigenvectors $\phi_{n,k}^{\rm P}$ of 
the Hamiltonian ${\cal H}$ are given in Section~\ref{singleelectron}. 
Here $n\in\left\{0,1,2,\ldots\right\}$ is a Landau index, 
and $k$ is a wavenumber given by $k=2\pi m/L_x$ with 
$m\in\Lambda^{(M)}=\left\{-M/2+1,-M/2+2,\ldots,M/2\right\}$. 
The energy eigenvalue is given by 
\begin{equation}
{\cal E}_{n,k}:=\left(n+\frac{1}{2}\right)\hbar \omega_c
\end{equation}
with $\omega_c:=eB/m_e$. The system $\{\phi_{n,k}^{\rm P}\}_{n,k}$ is 
the orthonormal complete system. 

Now we define the restriction of the Hilbert space, i.e., the energy 
cutoff. For a non-negative integer $n_{\rm max}$, 
we define by $P(n_{\rm max})$ 
the spectral projection onto the subspace 
spanned by all the eigenvectors with the Landau indices 
$n\le n_{\rm max}$. Namely, by the projection $P(n_{\rm max})$, 
the whole Hilbert spapce is restricted to the lowest $(n_{\rm max}+1)$ 
Landau levels. The corresponding $N$ electrons Hamiltonian is given by 
\begin{equation}
H^{(N)}(n_{\rm max})=P^{(N)}(n_{\rm max})H^{(N)}P^{(N)}(n_{\rm max})
\label{hamHpro}
\end{equation}
with the projection 
\begin{equation}
P^{(N)}(n_{\rm max}):=\bigotimes_{j=1}^N P_j(n_{\rm max}). 
\end{equation}

\subsection{A $C^\ast$ algebraic approach}
Throughout the present paper, we consider the thermodynamic limit 
$L_y\rightarrow\infty$ for a fixed $L_x$ and a fixed $n_{\rm max}$. 
Namely we consider an infinitely long strip with a finite 
width\footnote{The reason why we must fix the width to finite will 
be explained in Section~\ref{cutoffs}.} $L_x$. 
In this limit, we also fix the filling factor $\nu$ which is 
given by $\nu=N/M$ for a finite volume with $L_xL_y=2\pi M\ell_B^2$. 
For treating the infinite-volume system, it is convenient to introduce 
the notion of local observables by following the idea of a $C^\ast$ algebra 
\cite{BraRob}. Although a $C^\ast$ algebra must be a fairly mathematical 
tool, it enables us to avoid confusion between the degeneracy 
of finite-volume ground states and that of infinite-volume ground 
states \cite{Haldane}. In addition it clarifies the notions of low energy 
excitations and of a gap above an infinite-volume ground state. 

In order to introduce the notion of local observables, we first consider 
the second quantized form of the Hamiltonian (\ref{hamHpro}). 
It is written as\footnote{A perturbative approach starting from 
the Hamiltonian (\ref{hamH2Qpro}) was treated in refs.~\cite{TaoThouless}. 
For other approximate methods, see refs.~\cite{Laughlin2}.}  
\begin{eqnarray}
H_{\Lambda^{(M)}}(n_{\rm max})&:=&\sum_{n=0}^{n_{\rm max}}
\sum_{m\in\Lambda^{(M)}}
\left(n+\frac{1}{2}\right)\hbar\omega_c c_{n,m}^\ast c_{n,m}
+\sum_{j,\alpha}\sum_{j',\alpha'}W(j,\alpha:j',\alpha')
c_{j,\alpha}^\ast c_{j',\alpha'}\ret
&+&\frac{1}{2}\sum_{j,\alpha}\sum_{\ell,\beta}
\sum_{j',\alpha'}\sum_{\ell',\beta'}
U^{(2)}(j,\alpha;\ell,\beta:j',\alpha';\ell',\beta')c_{j,\alpha}^\ast
c_{\ell,\beta}^\ast c_{j',\alpha'}c_{\ell',\beta'}
\label{hamH2Qpro}
\end{eqnarray}
with 
\begin{equation}
W(j,\alpha:j',\alpha'):=\int_S dxdy\left[\phi_{j,p}^{\rm P}(x,y)\right]^\ast
W(x)\phi_{j',p'}^{\rm P}(x,y)
\end{equation}
and
\begin{eqnarray}
& &U^{(2)}(j,\alpha;\ell,\beta:j',\alpha';\ell',\beta')\ret
&:=&\int_S dxdy\int_S dx'dy'\ \left[\phi_{j,p}^{\rm P}(x,y)\right]^\ast
\left[\phi_{\ell,q}^{\rm P}(x',y')\right]^\ast U^{(2)}(x-x',y-y')
\phi_{j',p'}^{\rm P}(x',y')\phi_{\ell',q'}^{\rm P}(x,y).\ret
\label{U2coeff}
\end{eqnarray}
Here we have written 
\begin{equation}
p=2\pi\alpha/L_x,\ q=2\pi\beta/L_x,\ p'=2\pi\alpha'/L_x,\ q'=2\pi \beta'/L_x,
\end{equation}
and $c_{n,m}$ and $c_{n,m}^\ast$ are, respectively, 
the electron annihilation and creation operators 
for the eigenstate $\phi_{n,k}^{\rm P}$ of the single electron Landau 
Hamiltonian ${\cal H}$ of (\ref{singleham}) 
with the wavenumber $k=2\pi m/L_x$. 
These annihilation and creation operators satisfy the canonical 
anti-commutation relations as 
\begin{equation}
\{c_{n,m},c_{n',m'}\}=0\quad, \quad \{c_{n,m},c_{n',m'}^\ast\}=
\delta_{n,n'}\delta_{m,m'}.
\end{equation}
We can identify the quantum number $m\in\Lambda^{(M)}$ with 
the lattice site $m$ in the one-dimentional lattice 
$\Lambda^{(M)}=\left\{-M/2+1,-M/2+2,\ldots,M/2\right\}$. 
In other words, the set of all the wavenumbers $k$ is identical to 
the one-dimentional lattice. A wavenumber $k$ is corresponding to 
the center of the harmonic oscillator part of 
the wavefunction\footnote{See Section~\ref{singleelectron}.} 
$\phi_{n,k}^{\rm P}$. From this identification, the present system 
of the Hamiltonian $H_{\Lambda^{(M)}}(n_{\rm max})$ of (\ref{hamH2Qpro}) 
is identical to a one-dimentional lattice fermions system with 
long-range interactions and without spin degrees of freedom. 
Then the original Landau levels with a wavenumber $k=2\pi m/L_x$ 
are interpreted as atomic levels at the corresponding lattice site $m$. 

We note that the electron-electron interaction $U^{(2)}$ of 
the present system satisfies the condition 
\begin{equation}
\lim_{\Lambda^{(M)}\uparrow{\bf Z}}\max_{j,\alpha}\sum_{\ell,\beta}
\sum_{j',\alpha'}\sum_{\ell',\beta'}
\left|U^{(2)}(j,\alpha;\ell,\beta:j',\alpha';\ell',\beta')\right|<\infty.
\end{equation}
One can easily prove this condition by using Lemmas~\ref{>intphiphiLemma} and 
\ref{intUphiphibound} below. By this condition, 
the total energy of a finte volume is of order of the volume. 
Further, the condition guarantees the existence of the time evolution 
of a local observable. Roughly speaking, the condition is equivalent to 
\begin{equation}
\int_{{\bf R}^2}dxdy \left|U^{(2)}(x,y)\right|<\infty.
\end{equation}
Clearly this condition is too strong. In fact, the standard Coulomb 
intercation does not satisfy the condition. 

Since the operator $c_{n,m}^\ast$ creates the single electron 
wavefunction $\phi_{n,k}^{\rm P}$ in the $L_x\times L_y$ 
rectangular box in the Fock space, the annihilation and creation 
operators $c_{n,m},c_{n,m}^\ast$ 
depend on the system size $L_y$. This fact is not convenient for 
introducing local observables in the following because it is very 
hard to treat the outside of the rectangular box with 
the operators $c_{n,m},c_{n,m}^\ast$. In order to avoid this difficulty, 
we introduce different abstract annihilation and creation operators 
${\tilde c}_{n,m},{\tilde c}_{n,m}^\ast$ with 
$m\in{\bf Z}:=\{\ldots,-2,-1,0,1,2,\ldots\}$. 
These operators also obey the canonical anti-commutation relations 
\begin{equation}
\{{\tilde c}_{n,m},{\tilde c}_{n',m'}\}=0\quad, 
\quad \{{\tilde c}_{n,m},{\tilde c}_{n',m'}^\ast\}=
\delta_{n,n'}\delta_{m,m'}.
\end{equation}
Namely ${\tilde c}_{n,m},{\tilde c}_{n,m}^\ast$ are defined on 
the infinite lattice ${\bf Z}$. We replace $c_{n,m},c_{n,m}^\ast$ 
with ${\tilde c}_{n,m},{\tilde c}_{n,m}^\ast$ in the Hamiltonian 
(\ref{hamH2Qpro}). As a result, we have the Hamiltonian 
\begin{eqnarray}
{\tilde H}_{\Lambda^{(M)}}(n_{\rm max})&:=&
\sum_{n=0}^{n_{\rm max}}\sum_{m\in\Lambda^{(M)}}
\left(n+\frac{1}{2}\right)\hbar\omega_c {\tilde c}_{n,m}^\ast
{\tilde c}_{n,m}
+\sum_{j,\alpha}\sum_{j',\alpha'}W(j,\alpha:j',\alpha')
{\tilde c}_{j,\alpha}^\ast{\tilde c}_{j',\alpha'}\ret
&+&\frac{1}{2}\sum_{j,\alpha}\sum_{\ell,\beta}
\sum_{j',\alpha'}\sum_{\ell',\beta'}
U^{(2)}(j,\alpha;\ell,\beta:j',\alpha';\ell',\beta'){\tilde c}_{j,\alpha}^\ast
{\tilde c}_{\ell,\beta}^\ast {\tilde c}_{j',\alpha'}{\tilde c}_{\ell',\beta'}
\label{hamH2Qproinf}
\end{eqnarray}
with the same periodic boundary conditions on the same finite lattice 
$\Lambda^{(M)}$ as in \hfill\break $H_{\Lambda^{(M)}}(n_{\rm max})$. 
Clearly ${\tilde H}_{\Lambda^{(M)}}(n_{\rm max})$ has the same 
spectrum as that of $H_{\Lambda^{(M)}}(n_{\rm max})$. 

Now we introduce local observables. 
Let $\Lambda$ be a subset of ${\bf Z}$. 
We denote by ${\cal A}_\Lambda$ the set of all the observables generated by 
all the annihilation ${\tilde c}_{n,m}$ and the creation 
${\tilde c}_{n',m'}^\ast$ operators with $m,m'\in\Lambda$ and 
with $n,n'\in\{0,1,\ldots,n_{\rm max}\}$. 
We define the set of the local observables ${\cal A}_{\rm loc}$ as 
\begin{equation}
{\cal A}_{\rm loc}:=\bigcup_{\Lambda\subset{\bf Z};|\Lambda|<\infty}
{\cal A}_\Lambda. 
\end{equation}
Let $\Lambda^c$ be the complement of the lattice $\Lambda$, i.e., 
$\Lambda^c={\bf Z}\backslash\Lambda$. Then ${\cal A}_{\Lambda^c}$ is 
the algebra for the outside of $\Lambda$. Roughly speaking, 
the algebra ${\cal A}_{\Lambda^c}$ is an algebra for the outside 
of the $L_x\times L_y$ rectangular box because an original wavenumber 
$k=2\pi m/L_x$ is identical to the center of the harmonic oscillator part of 
the wavefunction $\phi_{n,k}^{\rm P}$.  

Next we introduce a set of $U(1)$ global gauge transformations. 
A global gauge transformation $U_\theta$ in the set is defined as 
\begin{equation}
U_\theta({\tilde c}_{n,m})=e^{-i\theta}{\tilde c}_{n,m}\quad, \quad 
U_\theta({\tilde c}_{n,m}^\ast)=e^{i\theta}{\tilde c}_{n,m}^\ast
\end{equation}
with $\theta\in[0,2\pi)$. 
Namely a $U(1)$ gauge transformation $U_\theta$ is a global phase twist with 
a real angle $\theta$ for the quantum mechanical phase of wavefunctions. 
Following Matsui \cite{Matsui}, we define by ${\cal A}_{\rm loc}^{U(1)}$ 
the $U(1)$ gauge invariant part of ${\cal A}_{\rm loc}$, i.e., 
\begin{equation}
{\cal A}_{\rm loc}^{U(1)}:=
\left\{a\in{\cal A}_{\rm loc}\ \left|\ U_\theta(a)=a \quad \mbox{for all}\ \ 
\theta\in[0,2\pi)\right.\right\}. 
\end{equation}

\subsection{Infinite-volume ground states and excitation gaps} 

Let ${\tilde \Phi}_{\Lambda^{(M)}}^{(N)}$ be a normalized $N$ electrons ground 
state of the Hamiltonian ${\tilde H}_{\Lambda^{(M)}}^{(N)}(n_{\rm max})$ of 
(\ref{hamH2Qproinf}). 
Clearly ${\tilde \Phi}_{\Lambda^{(M)}}^{(N)}$ is identical to 
a ground state $\Phi_{L_y}^{(N)}$ of the Hamiltonian 
$H_{L_y}^{(N)}(n_{\rm max}):=H^{(N)}(n_{\rm max})$ of (\ref{hamHpro}) with 
the system size $L_y$ in the $y$ direction. 
Then an infinite-volume ground state $\omega$ can be constructed 
as\footnote{If necessary, we take a subsequence for the limit 
$\Lambda^{(M)}\uparrow{\bf Z}$.}
\begin{equation}
\omega(a)=\lim_{\Lambda^{(M)}\uparrow{\bf Z}}
\left\langle{\tilde \Phi}_{\Lambda^{(M)}}^{(N)},a
{\tilde \Phi}_{\Lambda^{(M)}}^{(N)}\right\rangle
\label{groundstateomegaPhi}
\end{equation}
for a local observable $a\in{\cal A}_{\rm loc}^{U(1)}$, and for 
fixed $L_x$, $n_{\rm max}$ and $\nu$. 
All the infinite-volume ground states thus obtained are not necessarily 
complete as physically natural ground states. See ref.~\cite{KomaTasaki} for 
example. We use a more general definition of infinite-volume ground states 
as follows: A state $\omega$, i.e., a positive normalized linear functional, 
on local observables ${\cal A}_{\rm loc}^{U(1)}$ is an infinite-volume 
ground state if and only if $\omega$ satisfies the local stability 
condition\footnote{For more details, see ref.~\cite{BraRob}.} 
\begin{equation}
\lim_{\Lambda\uparrow{\bf Z}}\omega\left(a^\ast
[{\tilde H}_{\Lambda}(n_{\rm max}),a]\right)\ge 0 
\label{defgroundstate}
\end{equation}
for any local observable $a\in{\cal A}_{\rm loc}^{U(1)}$. 
{From} the definition of the vector ${\tilde \Phi}_{\Lambda^{(M)}}^{(N)}$,
the infinite-volume ground state (\ref{groundstateomegaPhi}) satisfies 
the condition (\ref{defgroundstate}) as 
\begin{eqnarray}
\lim_{\Lambda\uparrow{\bf Z}}\omega\left(a^\ast
[{\tilde H}_{\Lambda}(n_{\rm max}),a]\right)&=&
\lim_{\Lambda^{(M)}\uparrow{\bf Z}}
\left\langle{\tilde \Phi}_{\Lambda^{(M)}}^{(N)},a^\ast
\left[{\tilde H}_{\Lambda^{(M)}}(n_{\rm max})-
E_{L_y}^{(N)}\right]a{\tilde \Phi}_{\Lambda^{(M)}}^{(N)}
\right\rangle\ret
&=& \lim_{L_y\uparrow\infty}
\left\langle\Phi_{L_y}^{(N)},{\hat a}^\ast
\left[H_{L_y}^{(N)}(n_{\rm max})-E_{L_y}^{(N)}\right]{\hat a}\Phi_{L_y}^{(N)}
\right\rangle\ge 0
\label{checkGSomega}
\end{eqnarray}
for $a\in{\cal A}_{\rm loc}^{U(1)}$. Here $E_{L_y}^{(N)}$ 
is the energy eigenvalue of $\Phi_{L_y}^{(N)}$ for 
the Hamiltonian $H_{L_y}^{(N)}(n_{\rm max})$ of (\ref{hamHpro}), and 
${\hat a}$ is the observable corresponding to the observable $a$. 

We denote by $\tau_j^{(y)}$ the shift operator by $j$ lattice sites 
in the $y$ direction. Namely the shift operator is defined as 
\begin{equation}
\tau_j^{(y)}({\tilde c}_{n,m})={\tilde c}_{n,m+j}\quad, \quad 
\tau_j^{(y)}({\tilde c}_{n,m}^\ast)={\tilde c}_{n,m+j}^\ast.
\end{equation}
Let $\omega$ be an infinite-volume ground state. 
We say that $\omega$ is translationally invariant with 
a period $q\in{\bf N}:=\{1,2,\ldots\}$ if and only if $\omega$ satisfies 
\begin{equation}
\omega\left(\tau_q^{(y)}\left(\cdots\right)\right)
=\omega(\cdots). 
\end{equation}
If a ground state $\omega$ has a non-trivial minimal period $q\ne 1$, 
then a translatinal symmetry beaking occurs at zero temperature. 
If a ground state $\omega$ has no period, then we say 
that $\omega$ is a non-translationally invariant ground state. 

Consider the Hamiltonian with a chemical potential $\mu$, 
\begin{equation}
{\tilde H}_{\Lambda,\mu}(n_{\rm max}):=
{\tilde H}_\Lambda(n_{\rm max})-\mu\sum_{n=0}^{n_{\rm max}}
\sum_{m\in\Lambda}{\tilde n}_{n,m}
\label{hamgrandcano}
\end{equation}
with the electron number operator 
\begin{equation}
{\tilde n}_{n,m}:={\tilde c}_{n,m}^\ast{\tilde c}_{n,m}. 
\label{electronnumber}
\end{equation}
For the grand-canonical emsemble, the definition of an 
infinite-volume ground state is given as follows: 
A state $\omega$ is an infinite-volume ground state if and only if 
\begin{equation}
\lim_{\Lambda\uparrow{\bf Z}}
\omega\left(a^\ast\left[{\tilde H}_{\Lambda,\mu}(n_{\rm max}),a\right]
\right)\ge 0
\label{defGSmu}
\end{equation}
for any $a\in{\cal A}_{\rm loc}$. 
Matsui \cite{Matsui} proved an equivalence between a canonical emsemble and a 
grand-canonical emsemble for a lattice fermion system with a certain 
interaction. The following theorem for the present quantum Hall system 
follows from the Matsui's result. 

\begin{theorem} 
\label{theoremMatsui}
Let $\omega$ be a translationally invariant infinite-volume ground state 
with a period for ${\cal A}_{\rm loc}^{U(1)}$. 
Then there exists a chemical potential $\mu$ 
such that the gauge invariant extension ${\tilde\omega}$ of $\omega$ 
to ${\cal A}_{\rm loc}$ is an infinite-volume ground state for 
${\cal A}_{\rm loc}$. 
\end{theorem}
A sketch of the proof is given in Appendix~\ref{MatsuiTheorem}. 

Next we shall introduce a definition of a gap 
above an infinite-volume ground state. For this purpose, we first 
define the time evolution of a local observable $a\in{\cal A}_{\rm loc}$ 
as 
\begin{equation}
\tau_{t,\Lambda}\left(a\right):=\exp[i{\tilde H}_{\Lambda,\mu}(n_{\rm max})
t/\hbar]a\exp[-i{\tilde H}_{\Lambda,\mu}(n_{\rm max})t/
\hbar]
\end{equation}
and its infinite-volume limit, 
\begin{equation}
\tau_t\left(a\right):=\lim_{\Lambda\uparrow{\bf Z}}\tau_{t,\Lambda}
\left(a\right). 
\end{equation}
In the sense of the norm, this limit exists uniformly for time $t$ in a 
compact set. Let ${\cal A}$ be the norm completion of ${\cal A}_{\rm loc}$. 
Then $\tau_t(a)$ is defined also for $a\in{\cal A}$. 
Further we define 
\begin{equation}
\tau_{\ast f}(a):=\int_{-\infty}^{+\infty}dt \ f(t)\tau_t(a)
\end{equation}
for a function $f$ on ${\bf R}$ and $a\in{\cal A}$ when the right-hand side 
exists. We denote by $C_0^\infty$ the set of infinitely differentiable 
functions with compact support. 

\begin{definition}
\label{definitiongap}
An infinite-volume ground state $\omega$ has a gap $\gamma$ 
if and only if the following condition is satisfied: 
Let $f$ be a function on ${\bf R}$ with Fourier transform 
${\hat f}\in C_0^{\infty}$ and ${\rm supp}{\hat f}\subseteq (0,\gamma)$, 
then 
\begin{equation}
\omega\left(\left[\tau_{\ast f}(a)\right]^\ast\tau_{\ast f}(a)\right)=0
\label{defgap}
\end{equation}
for all $a\in{\cal A}$. 
\end{definition}
This definition of a gap is slightly different from that 
in ref.~\cite{AffleckLieb}. For the gauge invariant 
extension ${\tilde \omega}$ of $\omega$ of (\ref{groundstateomegaPhi}), 
the left-hand side of (\ref{defgap}) becomes 
\begin{equation}
{\tilde \omega}
\left(\left[\tau_{\ast f}(a)\right]^\ast\tau_{\ast f}(a)\right)
=\lim_{\Lambda\uparrow{\bf Z}}\left\langle{\tilde \Phi}_\Lambda^{(N)},a^\ast
\left[{\hat f}\left(\left\{{\tilde H}_{\Lambda,\mu}(n_{\rm max})
-E_{L_y}^{(N)}+\mu N\right\}/\hbar\right)\right]^2a
{\tilde \Phi}_\Lambda^{(N)}\right\rangle. 
\end{equation}
Thus the above definition of a gap above an infinite-volume 
ground state is a physically natural definition for the states $\omega$ 
of (\ref{groundstateomegaPhi}). In particular, the gap condition 
(\ref{defgap}) becomes 
\begin{eqnarray}
& &\lim_{\Lambda\uparrow{\bf Z}}\left\langle{\tilde \Phi}_\Lambda^{(N)},a^\ast
\left[{\hat f}\left(\left\{{\tilde H}_\Lambda(n_{\rm max})-E_{L_y}^{(N)}
\right\}/\hbar\right)\right]^2
a{\tilde \Phi}_\Lambda^{(N)}\right\rangle\ret
&=&\lim_{L_y\rightarrow\infty}\left\langle\Phi_{L_y}^{(N)},{\hat a}^\ast
\left[{\hat f}\left(\left\{H_{L_y}(n_{\rm max})-E_{L_y}^{(N)}\right\}/\hbar
\right)\right]^2{\hat a}
\Phi_{L_y}^{(N)}\right\rangle=0
\end{eqnarray}
for $a\in{\cal A}_{\rm loc}^{U(1)}$. Here ${\hat a}$ is the observable 
corresponding to the observable $a$. 
We remark that $\omega$ is an infinite-volume ground state 
for ${\cal A}_{\rm loc}$ if and only if the following condition is satisfied: 
Let $f$ be a function on ${\bf R}$ with Fourier transform 
${\hat f}\in C_0^{\infty}$ and ${\rm supp}{\hat f}\subseteq(-\infty,0)$, 
then 
\begin{equation}
\omega\left(\left[\tau_{\ast f}(a)\right]^\ast\tau_{\ast f}(a)\right)=0
\end{equation}
for all $a\in{\cal A}$. See ref.~\cite{BraRob} for the detail. 

\subsection{Main theorems of this paper}

Now we describe our main theorems. In the following, we fix 
the width $L_x$ of the strip and the energy cutoff $n_{\rm max}$ 
to finite values. 

\begin{theorem}
\label{theorem1}
Suppose the filling factor $\nu$ is not an integer. 
Then, either (i) there is more than one infinite-volume ground state or 
(ii) there is only one infinite-volume ground state with a gapless excitation. 
\end{theorem}
In the case (i), there is a symmetry breaking at zero temperature. 
Since a non-zero excitation gap plays an important role for the quantization 
of the Hall conductance in a qunatum Hall system, we are not interested 
in the case (ii). 

\begin{theorem}
\label{theorem2}
Suppose that the filling factor $\nu$ is not an integer and that 
a pure infinite-volume ground state has a non-zero excitation gap. 
Then a translational symmetry breaking occurs at zero temperature. 
\end{theorem}
Thus a translational symmetry breaking inevitably occurs in the situation 
where there appears a fractional quantization of the Hall conductance 
which is observed with a non-zero excitation gap for 
a fractional filling. 
In a realistic situation where the electron-electron interaction is repulsive, 
we can expect that there is no non-translationally invariant ground state 
with no periodic structure as we mentioned in Section~\ref{Pmeaning}. 

\begin{theorem}
\label{theorem3}
Suppose that there is no non-translationally invariant infinite-volume 
\break ground state. Then, if a pure infinite-volume ground state $\omega$ has 
a non-zero excitation gap, the filling factor $\nu$ must be equal to 
a rational number. In particular, if the ground state has a periodic structure 
with a minimal period $q\in{\bf N}$ for the magnetic translation 
in the $y$ direction, the filling factor $\nu$ must satisfy $q\nu\in{\bf N}$. 
\end{theorem}
Here, if the period $q$ is equal to the denominator of 
the filling $\nu$ as in a usual commesurate phase, 
we have $\nu=p/q$ with $p,q$ mutually prime integers. 
The relation between Theorem~\ref{theorem3} and the fractional quantization 
of the Hall conductance was already discussed 
with the results of a separete paper \cite{Koma} in Section~\ref{relationRQ}. 
The appearance of the Hall conductance plateau will be discussed 
with relation to localization 
of wavefunctions in another separate paper \cite{Koma2}. 

\subsection{The finite width of the strip and the energy cutoff $n_{\rm max}$}
\label{cutoffs}

In the above we have fixed the width $L_x$ of the strip and 
the energy cutoff $n_{\rm max}$ to finite values. 
Although the statements of our three theorems hold even for a fixed 
macroscopic width and for a fixed $n_{\rm max}$ giving 
a macroscopic energy cutoff, the structure of the low energy excitation 
constructed by using the Lieb-Schultz-Mattis method strongly depends on 
these cutoffs. In particular, the size of the locally excited region must 
increase with increasing the cutoffs for keeping a small excitation energy. 
Before concluding this section, we shall give discussions about 
this cutoff dependence of the excitation and about 
the two-dimensional infinite-volume system with no such restrictions. 

Consider first the energy cutoff $n_{\rm max}$. We recall the model 
described by the Hamiltonian (\ref{hamH2Qproinf}). The model is 
identical to a one-dimentional lattice fermion system with 
long-range interactions. The range of the interactions strongly 
depends on the cutoff $n_{\rm max}$. Actually the effective range seems 
to increase with increasing the energy of a fermion state. 
As a result, the upper bound of the excitation energy of the state 
constructed by using the Lieb-Schultz-Mattis method depends 
on the cutoff $n_{\rm max}$ and is divergent as $n_{\rm max}$ tends to 
infinity. For the explicit cutoff dependence,\footnote{The cutoff 
$n_{\rm max}$ dependence of the energy bound is too complicated to be written 
explicitly here.}
see Section~\ref{estimateEgap}. 
Although we need an infinitesimally small upper bound of the excitation 
energy for a large volume, we can not get a desired one without the cutoff. 
This is nothing but the reason why we introduced the cutoff $n_{\rm max}$ 
into the Hilbert space. 
However, one can expect generally that the contribution of 
very high energy states to low energy quantities is negligibly small. 
In fact, the energy of the excitation constructed by the Lieb-Schultz-Mattis 
method can be written in the ground state expectation of an operator. 
(See Section~\ref{LSMmethod} for the detail.) 
Clearly the difference between the ground state expectation 
with the cutoff and that without the cutoff is determined 
by the high energy states which are cut off. 
If the contribution of the high energy states is negligibly small, 
then the upper bound of the energy of the excitation thus 
constructed is independent of the cutoff $n_{\rm max}$, and 
we can remove the cutoff. 
Unfortunately we could not get a useful estimate for the contribution of 
the high energy states. 

Next we give a discussion about the cutoff $L_x$ of the width 
of the strip. In order to prove our main three theorems, we 
construct a low energy excitation above a ground state 
by relying on the Lieb-Schultz-Mattis method. 
Here we stress that locality of the excitation is absolutely essential for 
the proofs. However, the constructed excitation is extended homogeneously 
from end to end in the $x$ direction. 
(See Section~\ref{LSMmethod} for the detail.) 
Moreover, in the $y$ direction it has a linear size $\delta y$ 
which strongly depends on the width $L_x$ as 
\begin{equation}
\delta y\propto L_x^{3+\epsilon}.
\end{equation}
Here $\epsilon$ is a positive small number. 
For the detail, see Section~\ref{EstDEI>}. 
In order to treat the two-dimensional infinite-volume system 
with no such a cutoff in this approach, 
we need to construct a low energy excitation state 
which is local in both $x$ and $y$ directions. 
Unfortunately we could not construct such a low energy state, and 
we fixed the width $L_x$ to a finite value. 
In order to overcome this difficulty, it seems to us that 
a new idea beyond the Lieb-Schultz-Mattis method is required. 

Although we failed to overcome the difficulty, we can give 
a physically plausible argument to show 
the existence of a low energy excitation which is local 
in both $x$ and $y$ directions. 
To begin with, we note the following folk statement which is 
not generally justified, but physically plausible: 
If a system with a volume has a low energy excitation, 
then the same system with a larger volume also has a similar 
excitation in the sense that the corresponding excitation in the larger system 
keeps the same orders of the spatial extent and the excitation energy 
as those of the small system. Having this folk statement in mind, 
let us consider the two quantum Hall systems of infinitely long strips with 
the widths $L_x$ and $L_x'\gg L_x$. Fix $L_x$. Then we have 
an excitation with a low energy $\Delta E$ and 
with the linear size $\delta y$ in the $y$ direction 
and $L_x$ in the $x$ direction, following the Lieb-Schultz-Mattis method. 
Here, if the above folk statement is true, we have 
a local excitation with a low energy of the same order $\Delta E$ 
and with the linear size of order $\delta y$ in the $y$ direction 
and of order $L_x$ in the $x$ direction for the system with 
the large width $L_x'$. Thus we can expect 
the existence of a low energy excitation which is local 
in both $x$ and $y$ directions. However, it is not so easy to 
construct such an excitation. In fact, we could not construct it. 

In conclusion, we believe that our three results hold also 
for the two-dimensional infinite-volume quantum Hall system 
without the energy and the spatial cutoffs $n_{\rm max}, L_x$, 
and that these conjectures will be justified in future studies.

\Section{Preliminaries}
\label{preliminary}

As preliminaries for the proofs of our main theorems, we briefly review 
the eigenvalue problem of the single-electron Landau Hamiltonian and the 
degeneracy of finite-volume ground states in a quantum Hall system of 
an interacting electrons gas. The degeneracy was found by Yoshioka, Halperin 
and Lee \cite{YHL}. For related works, see refs.~\cite{degeneracy}. 

\subsection{The single-electron Landau Hamiltonian in two dimensions}
\label{singleelectron}

Consider the eigenvalue problem of the single-electron Hamiltonian 
\begin{equation}
{\cal H}=\frac{1}{2m_e}\left[\left(p_x-eBy\right)^2+p_y^2\right] 
\label{resingleham}
\end{equation}
on the infinite plane ${\bf R}^2$. In order to obtain an eigenvector of 
the Hamiltonian ${\cal H}$, put its form as 
\begin{equation}
\phi(x,y)=e^{ikx}v(y)
\end{equation}
with a wavenumber $k\in{\bf R}$. Substituting this into the Schr\"odinger 
equation ${\cal H}\phi={\cal E}\phi$, one has 
\begin{equation}
\left[\frac{1}{2m_e}(\hbar k-eBy)^2+\frac{1}{2m_e}p_y^2\right]v(y)
={\cal E}v(y).
\end{equation}
Clearly this is identical to the eigenvalue equation of a quantum harmonic 
oscillator as 
\begin{equation}
\left[-\frac{\hbar^2}{2m_e}\frac{\partial^2}{\partial y^2}+
\frac{e^2B^2}{2m_e}\left(y-\frac{\hbar k}{eB}\right)^2\right]v(y)
={\cal E}v(y). 
\end{equation}
The eigenvectors are 
\begin{equation}
v_{n,k}(y):=v_n(y-y_k):=N_n \exp\left[-(y-y_k)^2/(2\ell_B^2)\right]
H_n\left[(y-y_k)/\ell_B\right],
\end{equation}
where $H_n$ is the Hermite polynomial, $y_k=\hbar k/eB$, and 
$N_n$ is the positive normalization constant so that 
\begin{equation}
\int_{-\infty}^{+\infty}dy|v_{n,k}(y)|^2=1.
\end{equation}
The eigenvalues are given by 
\begin{equation}
{\cal E}_{n,k}=\left(n+\frac{1}{2}\right)\hbar\omega_c\quad\mbox{for}\ 
n=0,1,2,\ldots
\label{eigenvalueEnk}
\end{equation}
with $\omega_c=eB/m_e$. Thus the eigenvectors of the Hamiltonian ${\cal H}$ 
of (\ref{resingleham}) are given by 
\begin{equation}
\phi_{n,k}(x,y)=e^{ikx}v_{n,k}(y). 
\end{equation}

Next we consider a single electron in $L_x\times L_y$ rectangular box 
$S=[-L_x/2,L_x/2]\times[-L_y/2,L_y/2]$ with $L_xL_y=2\pi M\ell_B^2$ 
with an even integer $M$. We impose periodic boundary conditions 
\begin{equation}
\phi(x,y)=t^{(x)}(L_x)\phi(x,y), \quad \phi(x,y)=t^{(y)}(L_y)\phi(x,y)
\label{PBC}
\end{equation}
for wavefunctions $\phi$ on ${\bf R}^2$. We claim that, 
if $f$ satisfies (\ref{PBC}), then the functions 
\begin{equation}
f_1(x,y)=t^{(x)}(x')f(x,y)
\end{equation}
and
\begin{equation}
f_2(x,y)=t^{(y)}(y')f(x,y)
\end{equation}
also satisfy the same periodic boundary conditions. As a result, 
$x'$ and $y'$ are restricted into the following values: 
\begin{equation}
x'=m\Delta x \quad \mbox{with} \ m\in {\bf Z}, \quad \mbox{and}
\quad 
y'=n\Delta y \quad \mbox{with} \ n\in {\bf Z},
\label{ranget}
\end{equation}
where 
\begin{equation}
\Delta x:=\frac{h}{eB}\frac{1}{L_y}, \quad \mbox{and}\quad 
\Delta y:=\frac{h}{eB}\frac{1}{L_x}.
\label{Deltaxy}
\end{equation}
One can easily show these statements. In fact one has 
\begin{eqnarray}
f_1(x,y)&=&f(x-x',y)\ret
&=&\exp[iL_y(x-x')/\ell_B^2]f(x-x',y-L_y)\ret
&=&\exp[-iL_yx'/\ell_B^2]\exp[iL_yx/\ell_B^2]f(x-x',y-L_y)\ret
&=&\exp[-iL_yx'/\ell_B^2]\exp[iL_yx/\ell_B^2]f_1(x,y-L_y)\ret
&=&\exp[-iL_yx'/\ell_B^2]t^{(y)}(L_y)f_1(x,y)\ret
&=&\exp[-iL_yx'/\ell_B^2]f_1(x,y).
\end{eqnarray}
by the definitions. This implies $L_yx'/\ell_B^2=2\pi m$ with an integer $m$. 
Similarly 
\begin{eqnarray}
f_2(x,y)&=&\exp[iy'x/\ell_B^2]f(x,y-y')\ret
&=&\exp[iy'x/\ell_B^2]f(x-L_x,y-y')\ret
&=&\exp[iy'L_x/\ell_B^2]\exp[iy'(x-L_x)/\ell_B^2]f(x-L_x,y-y')\ret
&=&\exp[iy'L_x/\ell_B^2]f_2(x-L_x,y)\ret
&=&\exp[iy'L_x/\ell_B^2]t^{(x)}(L_x)f_2(x,y)\ret
&=&\exp[iy'L_x/\ell_B^2]f_2(x,y).
\end{eqnarray}
Thus $y'L_x/\ell_B^2=2\pi n$ with an integer $n$. 
Throughout the present paper we restrict the ranges of the variables $x',y'$ 
in the magnetic translations to these values of (\ref{ranget}). 

Since 
\begin{equation}
t^{(y)}(y')(p_x-eBy)\left[t^{(y)}(y')\right]^{-1}=p_x-eBy 
\end{equation}
for any $y'$, the Hamiltonian ${\cal H}$ of (\ref{resingleham}) is invariant 
under all the magnetic translations $t^{(x)}(\cdots)$ and $t^{(y)}(\cdots)$. 
Consider wavefunctions 
\begin{equation}
\phi_{n,k}^{\rm P}(x,y)=L_x^{-1/2}\sum_{\ell=-\infty}^{+\infty}
e^{i(k+\ell K)x}v_{n,k}(y-\ell L_y)
\label{phiP}
\end{equation}
for $k=2\pi m/L_x$ with $m=-M/2+1,\ldots,M/2-1,M/2$, and with 
$K=L_y/\ell_B^2$. These wavefunctions are eigenvectors 
of the Hamiltonian ${\cal H}$ of (\ref{resingleham}) satisfying the periodic 
boundary conditions (\ref{PBC}), because $L_xL_y=2\pi M\ell_B^2$ with 
the even integer $M$. The eigenvalues of $\phi_{n,k}^{\rm P}$ are given by 
(\ref{eigenvalueEnk}). We identify the integer $m$ of a wavenumber $k$ 
with a lattice site $m$ in the one-dimensional lattice 
$\{-M/2+1,-M/2+2,\ldots,M/2-1,M/2\}$, with the periodic boundary conditions. 
Then there are $(n_{\rm max}+1)$ atomic levels at each site in 
the present quantum Hall system because we have restricted 
the Hilbert space to the lowest $(n_{\rm max}+1)$ Landau levels. 
An observable at a site in the system can be expressed by 
a $(n_{\rm max}+2)\times(n_{\rm max}+2)$ matrix. 
Therefore the present quantum Hall system is equivalent to 
a one-dimentional spinless fermion system with long-range interactions. 
Here we should remark that the lattice constant is given by 
$\Delta y=2\pi\hbar/(eBL_x)$ which tends to zero as $L_x\rightarrow\infty$. 
This causes us a technical problem for taking the limit 
$L_x\rightarrow\infty$ as we will show in Section~\ref{estimateEgap}. 
This is why we must fix $L_x$ to a finite value. 

In the rest of the present section, we review the properties of 
the eigenfunctions (\ref{phiP}) and check the completeness of 
the system of the eigenfunctions. 

One can easily get the following lemma: 

\begin{lemma} 
\label{Teigenvector}
The vector $\phi_{n,k}^{\rm P}$ of (\ref{phiP}) is an eigenvector of 
the magnetic translation \hfill\break $t^{(x)}(\Delta x)$, i.e., 
\begin{equation}
t^{(x)}(\Delta x)\phi_{n,k}^{\rm P}=e^{-ik\Delta x}\phi_{n,k}^{\rm P}
=e^{-i2\pi m/M}\phi_{n,k}^{\rm P}\quad \mbox{with} \ \ k=\frac{2\pi m}{L_x}, 
\end{equation}
and the magnetic translation $t^{(y)}(\Delta y)$ 
shifts the wavenumber $k$ of the vector $\phi_{n,k}^{\rm P}$ by 
one unit $2\pi/L_x$ as 
\begin{equation}
t^{(y)}(\Delta y)\phi_{n,k}^{\rm P}=\phi_{n,k'}^{\rm P}
\quad \mbox{with} \quad 
k'=k+\frac{\Delta y}{\ell_B^2}=k+\frac{2\pi}{L_x}. 
\end{equation}
\end{lemma}

As usual we denote by $L^2(S)$ the set of functions $f$ on $S$ such that 
\begin{equation}
\int_S dxdy \ |f(x,y)|^2=
\int_{-L_x/2}^{L_x/2}dx \int_{-L_y/2}^{L_y/2}dy \ |f(x,y)|^2 <\infty.
\end{equation}
Further we define the associate inner product $(f,g)$ as 
\begin{equation}
(f,g)=\int_S dxdy \ [f(x,y)]^\ast g(x,y)
\end{equation}
for $f,g\in L^2(S)$. 

\begin{lemma} Let $f,g$ be functions on ${\bf R}^2$ such that 
$f,g\in L^2(S)$, and that $f,g$ satisfy the boundary conditions (\ref{PBC}). 
Then 
\begin{equation}
(f,g)=\int_{-L_x/2}^{L_x/2}dx \int_{-L_y/2+y_0}^{L_y/2+y_0}dy \ 
[f(x,y)]^\ast g(x,y) 
\end{equation}
for any $y_0\in {\bf R}$. 
\label{inproind}
\end{lemma}

\begin{proof}{Proof}
By the periodic boundary condition $f(x,y)=t^{(x)}(L_x)f(x,y)$, 
the function $f$ can be expanded in Fourier series as 
\begin{equation}
f(x,y)=L_x^{-1/2}\sum_k e^{ikx}{\hat f}(k,y).
\label{Fourierf}
\end{equation}
Further, since 
\begin{eqnarray}
f(x,y)=t^{(y)}(L_y)f(x,y)&=&L_x^{-1/2}
\sum_k e^{i(k+K)x}{\hat f}(k,y-L_y) \ret
&=&L_x^{-1/2}\sum_k e^{ikx}{\hat f}(k-K,y-L_y),
\end{eqnarray}
the following relation holds: 
\begin{equation}
{\hat f}(k,y)={\hat f}(k-K,y-L_y).
\label{PBCk}
\end{equation}
Using this relation repeatedly, the function $f$ of (\ref{Fourierf}) 
can be rewritten as 
\begin{equation}
f(x,y)=\sum_{\{k=2\pi n/L_x\left|-M/2+1\le n\le M/2\right.\}}
L_x^{-1/2}\sum_{\ell=-\infty}^{+\infty}
e^{i(k+\ell K)x}{\hat f}(k,y-\ell L_y).
\end{equation}
This expression yields 
\begin{eqnarray}
(f,g)&=&\int_{-L_x/2}^{L_x/2}dx \int_{-L_y/2}^{L_y/2}dy \
\left[f(x,y)\right]^\ast g(x,y)\ret
&=&\sum_{\{k=2\pi n/L_x\left|-M/2+1\le n\le M/2\right.\}}\ 
\sum_{\ell=-\infty}^{+\infty} \int_{-L_y/2}^{L_y/2} dy \ 
\left[{\hat f}(k,y-\ell L_y)\right]^\ast
{\hat g}(k,y-\ell L_y)\ret
&=&\sum_{\{k=2\pi n/L_x\left|-M/2+1\le n\le M/2\right.\}}\ 
\int_{-\infty}^{+\infty} dy \ \left[{\hat f}(k,y)\right]^\ast
{\hat g}(k,y)\ret
&=&\sum_{\{k=2\pi n/L_x\left|-M/2+1\le n\le M/2\right.\}}\ 
\sum_{\ell=-\infty}^{+\infty} \int_{-L_y/2+y_0}^{L_y/2+y_0} dy \ 
\left[{\hat f}(k,y-\ell L_y)\right]^\ast
{\hat g}(k,y-\ell L_y)\ret
&=&\int_{-L_x/2}^{L_x/2}dx \int_{-L_y/2+y_0}^{L_y/2+y_0}dy \
\left[f(x,y)\right]^\ast g(x,y).
\label{exprin}
\end{eqnarray}
\end{proof}

Let us check that the set of the eigenvectors $\{\phi_{n,k}^{\rm P}\}$ 
of (\ref{phiP}) forms an orthonormal complete system. 
{From} the third equality in (\ref{exprin}) in the proof of 
Lemma~\ref{inproind}, the orthogonality is valid as 
\begin{eqnarray}
\left(\phi_{n',k'}^{\rm P},\phi_{n,k}^{\rm P}\right)
=\int_{-\infty}^{+\infty} dy \ 
v_{n',k}^\ast(y)v_{n,k}(y)\delta_{k,k'}=\delta_{n,n'}\delta_{k,k'}.
\label{innerprophiP}
\end{eqnarray}
Here $\delta_{k,k'}$ is the Kronecker delta. 
To show the completeness, consider a function $f$ satisfying 
the boundary conditions (\ref{PBC}). 
In the same way, 
\begin{equation}
\left(\phi_{n,k}^{\rm P},f\right)
=\int_{-\infty}^{+\infty} dy \ v_{n,k}^\ast(y){\hat f}(k,y).
\end{equation}
This implies that the function $f$ must be zero if the inner 
product $\left(\phi_{n,k}^{\rm P},f\right)$ is vanishing for all 
the vectors $\phi_{n,k}^{\rm P}$. 

\subsection{Degeneracy of finite-volume ground states}
\label{DegenarateFGS}

In this section, we review the degeneracy \cite{YHL} of the finite-volume 
ground states of a quantum Hall system. A wide class of quantum 
Hall systems without disorder has the property. 
As an example, we consider an interacting $N$ electrons gas 
in a uniform magnetic field, whose Hamiltoanian is given by 
\begin{equation}
H^{(N)}=\sum_{j=1}^N\frac{1}{2m_e}
\left[(p_{x,j}-eBy_j)^2+p_{y,j}^2\right]
+U^{(N)}({\bf r}_1,{\bf r}_2,\ldots,{\bf r}_N)
\label{Hamdegeneracy}
\end{equation}
which is the Hamiltonian $H^{(N)}$ of (\ref{hamtot})
with no single-body potential $W$. Clearly the system has 
the translational invariance. 

To begin with, we recall the properties of the magnetic translations. 
{From} the definitions (\ref{defmagnetictrans}) of the magnetic translations 
$t^{(x)}(\cdots)$ and $t^{(y)}(\cdots)$, one can easily get 
\begin{eqnarray}
t^{(x)}(x')t^{(y)}(y')f(x,y)&=&t^{(x)}(x')\exp[iy'x/\ell_B^2]f(x,y-y')\ret
&=&\exp[iy'(x-x')/\ell_B^2]f(x-x',y-y')\ret
&=&\exp[-ix'y'/\ell_B^2]t^{(y)}(y')t^{(x)}(x')f(x,y) 
\end{eqnarray}
for a function $f$. This implies 
\begin{equation}
t^{(x)}(x')t^{(y)}(y')=\exp[-ix'y'/\ell_B^2]t^{(y)}(y')t^{(x)}(x').
\label{tcom}
\end{equation}
We define the magnetic translations $T^{(N,x)}(x')$ and $T^{(N,y)}(y')$
for an $N$ electrons state as 
\begin{equation}
T^{(N,x)}(x')=\bigotimes_{j=1}^N t_j^{(x)}(x'),
\end{equation}
and
\begin{equation}
T^{(N,y)}(y')=\bigotimes_{j=1}^N t_j^{(y)}(y').
\end{equation}
{From} the commutation relation (\ref{tcom}), one has 
\begin{equation}
T^{(N,x)}(x')T^{(N,y)}(y')=\exp[-ix'y'N/\ell_B^2]
T^{(N,y)}(y')T^{(N,x)}(x').
\end{equation}
In particular, 
\begin{equation}
T^{(N,x)}(\Delta x)T^{(N,y)}(\Delta y)=\exp[-i2\pi\nu]
T^{(N,y)}(\Delta y)T^{(N,x)}(\Delta x),
\label{crTT}
\end{equation}
where $\nu=N/M$ with $M=L_xL_yeB/h$. The number $\nu$ is nothing but 
the filling factor for the Landau levels. 

Note that all the magnetic translations $T^{(N,y)}(\cdots)$ and 
$T^{(N)}(\cdots)$ commute with the Hamiltonian $H^{(N)}$ of 
(\ref{Hamdegeneracy}). 
Let $\Phi^{(N)}$ be a simultaneous eigenvector of the 
Hamiltonian $H^{(N)}$ and the magnetic translation 
operator $T^{(N,y)}(\Delta y)$, i.e., 
\begin{equation}
H^{(N)}\Phi^{(N)}=E^{(N)}\Phi^{(N)}\ \ , \quad 
T^{(N,y)}(\Delta y)\Phi^{(N)}=e^{i2\pi n/M}\Phi^{(N)}, \quad 
\mbox{with} \ n\in {\bf Z},
\end{equation}
where $E^{(N)}$ is the energy eigenvalue. 
Let $\Psi^{(N)}=T^{(N,x)}(\Delta x)\Phi^{(N)}$. Then 
the vector $\Psi^{(N)}$ is an eigenvector of $H^{(N)}$ 
with the same eigenvalue $E^{(N)}$. Further one can easily show 
\begin{eqnarray}
T^{(N,y)}(\Delta y)\Psi^{(N)}&=&T^{(N,y)}(\Delta y)T^{(N,x)}(\Delta x)
\Phi^{(N)}\ret
&=&e^{i2\pi\nu}T^{(N,x)}(\Delta x)T^{(N,y)}(\Delta y)\Phi^{(N)}\ret
&=&e^{i2\pi\nu}e^{i2\pi n/M}T^{(N,x)}(\Delta x)\Phi^{(N)}\ret
&=&e^{i2\pi\nu}e^{i2\pi n/M}\Psi^{(N)} 
\end{eqnarray}
by using the commutation relation (\ref{crTT}). 
Thus $\Psi^{(N)}$ is also an eigenvector of $T^{(N,y)}(\Delta y)$. 
{From} these observations, one can notice the fact that, 
{\em if $\nu=p/q$ with mutually prime positive integers $p$ and $q$, 
then any enegry level of finite volume is at least $q$-fold degenerate.} 

\Section{The Lieb-Schultz-Mattis method}
\label{LSMmethod}

In this section, we construct a candidate for a low energy 
excitation above a ground state by using the Lieb-Schultz-Mattis method 
\cite{LSM}. Our goal is to give the proofs of our main 
Theorems~\ref{theorem1}, \ref{theorem2} and \ref{theorem3}. 
For the convenience of readers, technical 
estimates in the proofs are given in Sections~\ref{Orthproof} and 
\ref{estimateEgap} and Appendices~\ref{ProofU2avlemma}, 
\ref{Proof>intphiphiLemma}, \ref{Proofpro:phiphiintbound} and 
\ref{ProofintUphiphibound}. 

We denote by ${\bf H}_{L_y}^{(N)}(n_{\rm max})$ the restricted 
$N$ electrons Hilbert space to the lowest $(n_{\rm max}+1)$ Landau 
levels with the system size $L_y$ in the $y$ direction. 
Throughout this section, we fix $n_{\rm max}$ and $L_x$ 
(the system size in the $x$ direction) to large numbers. 

Let $\Phi_{L_y}^{(N)}$ be a normalized $N$ electrons vector in 
${\bf H}_{L_y}^{(N)}(n_{\rm max})$. 
We expand $\Phi_{L_y}^{(N)}$ as 
\begin{equation}
\Phi_{L_y}^{(N)}=\sum_{\{\xi_j\}}a(\{\xi_j\})
{\rm Asym}\left[\phi_{\xi_1}^{\rm P}\otimes\phi_{\xi_2}^{\rm P}\otimes\cdots
\otimes\phi_{\xi_N}^{\rm P}\right]
\label{vectorPhi}
\end{equation}
in terms of the eigenvectors $\phi_{n,k}^{\rm P}$ of (\ref{phiP}) for 
the single-electron Hamiltonian ${\cal H}$ of (\ref{resingleham}). 
Here we have written 
\begin{equation}
\xi_j=(n_j,k_j)=(n_j,2\pi m_j/L_x) \quad\mbox{for}\ j=1,2,\ldots,N,
\end{equation}
i.e., $\phi_{\xi_j}^{\rm P}=\phi_{n_j,k_j}^{\rm P}$. Note that we have 
\begin{equation}
T^{(N,x)}(\Delta x)\Phi_{L_y}^{(N)}=\sum_{\{\xi_j\}}a(\{\xi_j\})
\left[\prod_{j=1}^Ne^{-i2\pi m_j/M}\right]
{\rm Asym}\left[\phi_{\xi_1}^{\rm P}\otimes\phi_{\xi_2}^{\rm P}\otimes\cdots
\otimes\phi_{\xi_N}^{\rm P}\right] 
\label{globaltwist}
\end{equation}
{from} Lemma~\ref{Teigenvector}. 
This vector $T^{(N,x)}(\Delta x)\Phi_{L_y}^{(N)}$ is a vector globally 
twisting the quantum mechanical phase for $\Phi_{L_y}^{(N)}$. 
As we saw in the preceding section, if $\Phi_{L_y}^{(N)}$ 
is a ground state of the Hamiltonian $H^{(N)}$ of 
(\ref{Hamdegeneracy}), the vector $T^{(N,x)}(\Delta x)\Phi_{L_y}^{(N)}$ is 
a ground state, too. As Haldane pointed out \cite{Haldane}, the degeneracy of 
the ground states does not directly lead to physical significance 
because the degeneracy is related to the degree of freedom for the center 
of the total mass. But we can construct a physically 
natural low energy excitation above a ground state for the present Hamiltonian 
$H^{(N)}(n_{\rm max})$ of (\ref{hamHpro}), 
by combining the translational invariance in the $y$ direction 
with the Lieb-Schultz-Mattis method. To do this, we replace the globally 
twisting phase change of (\ref{globaltwist}) with a local one. 
Namely we construct a locally perturbed state for a state $\Phi_{L_y}^{(N)}$ 
which is not necessarily a ground state. 

For this purpose, we introduce a unitary transformation $U^{(\ell)}_{\pm,q}$ 
with a compact support for $\ell,q\in{\bf N}$ as 
\begin{equation}
U^{(\ell)}_{\pm,q}\Phi_{L_y}^{(N)}
:=\sum_{\{\xi_j\}}a(\{\xi_j\})
\exp\left[\pm i2\pi\sum_{j=1}^N{\tilde m}(m_j)/\ell\right]
{\rm Asym}\left[\phi_{\xi_1}\otimes\phi_{\xi_2}\otimes\cdots
\otimes\phi_{\xi_N}\right], 
\label{defUnit}
\end{equation}
where 
\begin{equation}
{\tilde m}(m):=\cases{n & if $(n-1)q<m\le nq$ with $n=1,2,\ldots,\ell$ \cr
                              0, & otherwise\cr}
\label{deftildem}
\end{equation}
for $m\in{\bf Z}$. Consider two vectors 
\begin{equation}
\Psi_{\pm,L_y}^{(N)}:=U^{(\ell)}_{\pm,q}\Phi_{L_y}^{(N)} 
\label{Psipm}
\end{equation}
which are locally perturbed vectors for $\Phi_{L_y}^{(N)}$ of 
(\ref{vectorPhi}), and 
\begin{equation}
\Delta E_{L_y}^{(N)}=\eta_{L_y}^{(N)}\left(H_{L_y}^{(N)}(n_{\rm max})\right)
-\omega_{L_y}^{(N)}\left(H_{L_y}^{(N)}(n_{\rm max})\right),
\label{DE}
\end{equation}
where 
\begin{equation}
\eta_{L_y}^{(N)}(\cdots)=\frac{1}{2}
\left\langle\Psi_{+,L_y}^{(N)},(\cdots)\Psi_{+,L_y}^{(N)}\right\rangle
+\frac{1}{2}
\left\langle\Psi_{-,L_y}^{(N)},(\cdots)\Psi_{-,L_y}^{(N)}\right\rangle,
\end{equation}
and
\begin{equation}
\omega_{L_y}^{(N)}(\cdots)
=\left\langle\Phi_{L_y}^{(N)},(\cdots)\Phi_{L_y}^{(N)}\right\rangle. 
\end{equation}
Here $H_{L_y}^{(N)}(n_{\rm max})$ is the Hamiltonian $H^{(N)}(n_{\rm max})$ 
of (\ref{hamHpro}) with the system size $L_y$ in the $y$ direction. 
The vectors $\Psi_{\pm,L_y}^{(N)}$ are candidates for natural low energy 
excitations when $\Phi_{L_y}^{(N)}$ leads to an infinite-volume ground 
state. When $\Phi_{L_y}^{(N)}$ is a finite-volume ground state with 
the energy eigenvalue $E_{L_y}^{(N)}$, we have 
\begin{equation}
\Delta E_{L_y}^{(N)}=\frac{1}{2}
\left\langle\Psi_{+,L_y}^{(N)},(H_{L_y}^{(N)}(n_{\rm max})-E_{L_y}^{(N)})
\Psi_{+,L_y}^{(N)}\right\rangle
+\frac{1}{2}
\left\langle\Psi_{-,L_y}^{(N)},(H_{L_y}^{(N)}(n_{\rm max})-E_{L_y}^{(N)})
\Psi_{-,L_y}^{(N)}\right\rangle.
\end{equation}
Thus $\Delta E_{L_y}^{(N)}$ gives an upper bound for the energy gap. 

\begin{lemma}
\label{DeltaEestimate}
For any given small $\varepsilon>0$, there exist $\ell$ and $L$ 
such that 
\begin{equation}
\left|\Delta E_{L_y}^{(N)}\right|\le\varepsilon \quad
\mbox{for any}\ L_y\ge L. 
\end{equation}
\end{lemma}
This Lemma gives an estimate for an energy gap above a ground state. 
The proof is given in Section~\ref{estimateEgap}. 

Next we study a condition for which $\Psi_{\pm,L_y}^{(N)}$ is 
orthogonal to $\Phi_{L_y}^{(N)}$. We define a local charge operator 
${\hat n}_{m,n}$ as 
\begin{equation}
{\hat n}_{n,m}\phi_{n',k'}^{\rm P}=\delta_{n,n'}\delta_{m,m'}
\phi_{n',k'}^{\rm P} 
\end{equation}
with $k'=2\pi m'/L_x$. Further we define 
\begin{equation}
{\hat n}_m:=\sum_{n=0}^{n_{\rm max}}{\hat n}_{n,m}.
\end{equation}

\begin{pro} 
\label{orthopro}
Let $\Phi_{L_y}^{(N)}\in{\bf H}_{L_y}^{(N)}(n_{\rm max})$ be a normalized 
eigenvector of the magnetic translation $T^{(N,y)}(q\Delta y)$ with 
$q\in{\bf N}$. Write 
\begin{equation}
\omega(\cdots)=
w^\ast\mbox{-}\lim_{L_y\rightarrow\infty}\left\langle\Phi_{L_y}^{(N)},(\cdots)
\Phi_{L_y}^{(N)}\right\rangle,
\end{equation}
where the weak limit $L_y\rightarrow\infty$ is taken for a fixed 
$L_x$ and a fixed filling factor $\nu$. 
Suppose that the infinite-volume state $\omega$ satisfies 
\begin{equation}
\lim_{\ell\rightarrow\infty}\ 
\frac{1}{\ell^2}\sum_{i,j=1}^\ell
\omega\left({\hat n}_i{\hat n}_j\right)=\nu^2.
\label{noLRO}
\end{equation}
Then 
\begin{equation}
\lim_{\ell\rightarrow\infty}\omega\left(U_{\pm,q}^{(\ell)}\right)=0 
\end{equation}
for $q\nu\notin{\bf N}$. 
\end{pro}
The proof is given in Section~\ref{Orthproof}. The idea of the proof is 
due to Hal Tasaki \cite{Tasaki}. From the proof, one can see that 
the statement of Proposition~\ref{orthopro} holds for a wide class of systems 
with translational invariance. 

Before giving the proofs of our main Theorems~\ref{theorem1}, \ref{theorem2} 
and \ref{theorem3}, we recall the GNS 
representation of a $C^\ast$ algebra ${\cal A}$ on a Hilbert 
space.\footnote{For the GNS construction of a representation of a 
$C^\ast$ algebra, see ref.~\cite{BraRob}.}
Let $\omega$ be an infinite-volume state. Then there exist a Hilbert space 
${\bf H}_\omega$, a normalized vector $\Omega_\omega$ and a representation 
$\pi_\omega$ of ${\cal A}$ on ${\bf H}_\omega$ such that 
\begin{equation}
\omega(a)=\left(\Omega_\omega,\pi_\omega(a)\Omega_\omega\right)
\quad \mbox{for any}\ a\in{\cal A}. 
\end{equation}
Here, if $\omega$ is a ground state, there exist a self-adjoint operator 
$H_\omega\ge 0$ on ${\bf H}_\omega$ such that 
\begin{equation}
H_\omega\Omega_\omega=0\quad, \quad 
e^{itH_\omega/\hbar}\pi_\omega(a)e^{-itH_\omega/\hbar}
=\pi_\omega\left(\tau_t(a)\right)\quad \mbox{for any}\ a\in{\cal A}. 
\label{GSconditionGNS}
\end{equation}
Namely $H_\omega$ is the Hamiltonian in the infinite volume limit. 
Conversely, if the vector $\Omega_\omega$ satisfies the conditions 
(\ref{GSconditionGNS}) for a self-adjoint operator $H_\omega\ge 0$ on 
${\bf H}_\omega$, then the corresponding state 
$\omega(\cdots)=(\Omega_\omega,\pi_\omega(\cdots)\Omega_\omega)$ is 
a ground state. Using this representation, the gapful condition 
(\ref{defgap}) in Definition~\ref{definitiongap} can be written as 
\begin{equation}
\left(\Omega_\omega,\left[\pi_\omega(a)\right]^\ast
\left[{\hat f}(H_\omega/\hbar)\right]^2\pi_\omega(a)\Omega_\omega\right)
=0\quad\mbox{for any}\ a\in{\cal A}.
\end{equation}
\medskip

\begin{proof}{Proof of Theorem~\ref{theorem1}}
Let $\Phi_{L_y}^{(N)}$ be a normalized ground state of the Hamiltonian 
$H_{L_y}^{(N)}(n_{\rm max})$ of (\ref{hamHpro}) and eigenvector 
of $T^{(N,y)}(\Delta y)$, i.e., 
a translatinally invariant ground state for a finite volume. 
We fix the filling factor $\nu$ to a non-integer. 
Let ${\tilde \Phi}_\Lambda^{(N)}$ be the corresponding vector in the Fock 
space ${\bf H}_{L_y}(n_{\rm max}):=\bigoplus_{N\ge 0}{\bf H}_{L_y}^{(N)}
(n_{\rm max})$. We denote by $\omega$ the infinite-volume ground state, i.e., 
\begin{equation}
\omega(\cdots)=w^\ast\mbox{-}\lim_{\Lambda\uparrow{\bf Z}}
\left\langle{\tilde \Phi}_{\Lambda}^{(N)},(\cdots)
{\tilde \Phi}_{\Lambda}^{(N)}\right\rangle 
\end{equation}
for ${\cal A}_{\rm loc}^{U(1)}$. 
By Theorem~\ref{theoremMatsui}, there exists a chemical potential $\mu$ 
such that $\omega$ is an infinite-volume ground state for 
${\cal A}_{\rm loc}$. Assume that $\omega$ is the unique ground state 
with the chemical potential $\mu$. Since a unique pure ground state has 
the clustering property \cite{BraRob} 
\begin{equation}
\omega({\hat n}_i{\hat n}_j)-\omega({\hat n}_i)\omega({\hat n}_j)
\rightarrow 0 \quad \mbox{as}\ |i-j|\rightarrow\infty,
\end{equation}
we have 
\begin{equation}
0=\lim_{\ell\rightarrow\infty}
\lim_{L_y\rightarrow\infty}\left\langle\Phi_{L_y}^{(N)},U_{\pm,1}^{(\ell)}
\Phi_{L_y}^{(N)}\right\rangle
=\lim_{\ell\rightarrow\infty}\omega\left({\tilde U}_{\pm,1}^{(\ell)}\right)
=\lim_{\ell\rightarrow\infty}\left(\Omega_\omega,\pi_\omega
\left({\tilde U}_{\pm,1}^{(\ell)}\right)\Omega_\omega\right)
\end{equation}
{from} Proposition~\ref{orthopro}. Here ${\tilde U}_{\pm,1}^{(\ell)}$ is 
the extension of $U_{\pm,1}^{(\ell)}$ to that in the Fock space, 
$\pi_\omega$ is the GNS representation of ${\cal A}$ on the Hilbert space 
${\bf H}_\omega$, and $\Omega_\omega\in {\bf H}_\omega$ is the ground state 
corresponding to $\omega$. This implies that the vectors 
$\left[\pi_\omega\left({\tilde U}_{\pm,1}^{(\ell)}\right)
-\left(\Omega_\omega,\pi_\omega
\left({\tilde U}_{\pm,1}^{(\ell)}\right)\Omega_\omega\right)\right]
\Omega_\omega$ are excitations above the unique ground state $\Omega_\omega$ 
for a large $\ell$. Clearly the norms of these vectors go to one as 
$\ell\rightarrow\infty$. 

Next we show the existence of a gapless excitation. Note that 
\begin{eqnarray}
& &\left\langle\Phi_{L_y}^{(N)},\left(U_{\pm,1}^{(\ell)}\right)^\ast
H_{L_y}^{(N)}(n_{\rm max})U_{\pm,1}^{(\ell)}\Phi_{L_y}^{(N)}
\right\rangle-\left\langle\Phi_{L_y}^{(N)},H_{L_y}^{(N)}(n_{\rm max})
\Phi_{L_y}^{(N)}\right\rangle\ret
&=&\left\langle\Phi_{L_y}^{(N)},\left(U_{\pm,1}^{(\ell)}\right)^\ast
\left[H_{L_y}^{(N)}(n_{\rm max}),U_{\pm,1}^{(\ell)}\right]
\Phi_{L_y}^{(N)}\right\rangle\ret
&=&\left\langle{\tilde \Phi}_{\Lambda}^{(N)},
\left({\tilde U}_{\pm,1}^{(\ell)}\right)^\ast
\left[{\tilde H}_{\Lambda}(n_{\rm max}),{\tilde U}_{\pm,1}^{(\ell)}\right]
{\tilde \Phi}_{\Lambda}^{(N)}\right\rangle\ret
&=&\left\langle{\tilde \Phi}_{\Lambda}^{(N)},
\left({\tilde U}_{\pm,1}^{(\ell)}\right)^\ast
\left[{\tilde H}_{\Lambda,\mu}(n_{\rm max}),{\tilde U}_{\pm,1}^{(\ell)}\right]
{\tilde \Phi}_{\Lambda}^{(N)}\right\rangle.
\end{eqnarray}
Further we have 
\begin{equation}
\lim_{\Lambda\uparrow{\bf Z}}
\omega\left(\left({\tilde U}_{\pm,1}^{(\ell)}\right)^\ast
\left[{\tilde H}_{\Lambda,\mu}(n_{\rm max}),{\tilde U}_{\pm,1}^{(\ell)}\right]
\right)=\left(\Omega_\omega,\left[\pi_\omega
\left({\tilde U}_{\pm,1}^{(\ell)}\right)\right]^\ast
H_\omega\pi_\omega\left({\tilde U}_{\pm,1}^{(\ell)}\right)\Omega_\omega\right)
\end{equation}
in the thermodynamic limit because 
\begin{equation}
\lim_{\Lambda\uparrow{\bf Z}}
\omega\left(a^\ast\left[{\tilde H}_{\Lambda,\mu}(n_{\rm max}),a\right]\right)
=\left(\Omega_\omega,\left[\pi_\omega(a)\right]^\ast
\left[H_\omega,\pi_\omega(a)\right]\Omega_\omega\right)
\label{Homegadef}
\end{equation}
for any observable $a$ in a domain for the commutator.\footnote{Roughly 
speaking, the Hamiltonian $H_\omega$ in the infinite 
volume limit is defined by the relation (\ref{Homegadef}) 
because the left-hand side of (\ref{Homegadef}) is non-negative for 
any $a\in{\cal A}_{\rm loc}$. See ref.~\cite{BraRob} for the details.} 
Combining these observations with Lemma~\ref{DeltaEestimate}, 
we have the following: For any given small $\varepsilon>0$, 
there exists $\ell$ such that 
\begin{equation}
\left(\Omega_\omega,
\left[\pi_\omega\left({\tilde U}_{\pm,1}^{(\ell)}\right)\right]^\ast H_\omega
\pi_\omega\left({\tilde U}_{\pm,1}^{(\ell)}\right)\Omega_\omega\right)
\le\varepsilon.
\end{equation}
This implies that there exists a gapless excitation above the unique 
ground state. 
\end{proof}

\begin{proof}{Proof of Theorem~\ref{theorem2}}
Let the filling factor $\nu$ be a non-integer, and let 
$\omega$ be a pure ground state with  a non-zero excitation gap. 
Assume that all the infinite-volume ground states are translationally 
invariant with the period of one lattice unit, and we will find 
a contradiction. Without loss of generality, 
we can assume that there exists a sequence of vectors 
$\{{\tilde \Phi}_\Lambda\}$ such that 
\begin{equation}
\omega(\cdots)=w^\ast\mbox{-}\lim_{\Lambda\uparrow{\bf Z}}
\left\langle{\tilde \Phi}_\Lambda,
(\cdots){\tilde \Phi}_\Lambda\right\rangle.
\end{equation}
Each vector ${\tilde \Phi}_\Lambda$ for a finite lattice $\Lambda$ 
is expanded as 
\begin{equation}
{\tilde \Phi}_\Lambda=\sum_N \alpha_N{\tilde \Phi}_\Lambda^{(N)}
\end{equation}
in terms of the $N$ electrons vectors ${\tilde \Phi}_\Lambda^{(N)}$. 
Then we can assume, by the assumption about the translational invariance, 
that the expectation $\left\langle{\tilde \Phi}_\Lambda^{(N)},(\cdots)
{\tilde \Phi}_\Lambda^{(N)}\right\rangle$ is translationally invariant with 
the period $1$. Using the expansion, we have 
\begin{equation}
\sin^2(\pi\nu)\left|\omega\left({\tilde U}_{\pm,1}^{(\ell)}
\right)\right|^2\le\pi^2\left[\frac{1}{\ell^2}\sum_{s,t=1}^\ell
\omega\left({\tilde n}_s{\tilde n}_t\right)-\nu^2\right] 
\label{CLROboundomega}
\end{equation}
in the same way as in the proof of Proposition~\ref{orthopro}. 
Here ${\tilde n}_j$ is the number operator corresponding to ${\hat n}_j$. 
Since the ground state $\omega$ has the clustering property due to the purity, 
we obtain 
\begin{equation}
0=\lim_{\ell\rightarrow\infty}\omega\left({\tilde U}_{\pm,1}^{(\ell)}
\right)=\lim_{\ell\rightarrow\infty}
\left(\Omega_\omega,\pi_\omega\left({\tilde U}_{\pm,1}^{(\ell)}
\right)\Omega_\omega\right)
\label{orthogonalOmegaUOmega}
\end{equation}
from (\ref{CLROboundomega}) with the assumption $\nu\notin{\bf N}$. 
Here $\pi_\omega$ is the GNS representation of ${\cal A}$ on the 
Hilbert space ${\bf H}_\omega$, and $\Omega_\omega$ is the ground state 
corresponding to the state $\omega$. 

Consider a vector 
$\Xi=(1-G)\pi_\omega\left({\tilde U}_{\pm,1}^{(\ell)}\right)\Omega_\omega$, 
where $G$ is the orthogonal projection onto the sector of the ground states. 
We want to show that the norm of $\Xi$ is non-vanishing in the limit 
$\ell\rightarrow\infty$. Assume this is not true, and we find a contradiction. 
This assumption is rephrased as follows: For any given samll $\varepsilon>0$, 
there exist a positive integer $\ell_0$ such that 
\begin{equation}
\left|\left(\Omega_{\omega'},\pi_\omega\left({\tilde U}_{\pm,1}^{(\ell)}
\right)\Omega_\omega\right)-1\right|<\varepsilon\quad \mbox{for any}\ 
\ell>\ell_0,
\label{Omega'UOmegabound}
\end{equation}
where $\Omega_{\omega'}\in{\bf H}_\omega$ is a normalized ground state 
which may depend on the integer $\ell$. We decompose $\Omega_{\omega'}$ as 
\begin{equation}
\Omega_{\omega'}=c\pi_\omega\left({\tilde U}_{\pm,1}^{(\ell)}
\right)\Omega_\omega+\Omega'\quad \mbox{with}\ \ 
\left(\Omega',\pi_\omega\left({\tilde U}_{\pm,1}^{(\ell)}
\right)\Omega_\omega\right)=0, 
\end{equation}
where $c$ is a complex number. Immediately, we have 
\begin{equation}
|1-c|<\varepsilon\quad , \quad\left\Vert\Omega'\right\Vert\le 
\sqrt{2\varepsilon}.
\end{equation}
Using these inequalities, we get
\begin{eqnarray}
& &\left\Vert\omega'(\cdots)-\omega\left(\left[{\tilde U}_{\pm,1}^{(\ell)}
\right]^\ast(\cdots){\tilde U}_{\pm,1}^{(\ell)}\right)\right\Vert\ret
&=&\left\Vert\left(\Omega_{\omega'},(\cdots)\Omega_{\omega'}\right)
-\left(\Omega_\omega,\left[\pi_\omega\left({\tilde U}_{\pm,1}^{(\ell)}\right)
\right]^\ast(\cdots)\pi_\omega\left({\tilde U}_{\pm,1}^{(\ell)}\right)
\Omega_\omega\right)\right\Vert\le\varepsilon',
\label{omega'-omegaU}
\end{eqnarray}
where $\varepsilon'=2(2\varepsilon+\sqrt{2\varepsilon})$. 
Since $\omega'$ and $\omega$ are translationally invariant 
by the assumption, we have 
\begin{eqnarray}
\Vert a\Vert\varepsilon'&\ge&\left|\omega'(\tau_{-j}^{(y)}(a))-
\omega\left(\left[{\tilde U}_{\pm,1}^{(\ell)}
\right]^\ast\tau_{-j}^{(y)}(a){\tilde U}_{\pm,1}^{(\ell)}\right)\right|\ret
&=&\left|\omega'(a)
-\omega\left(\tau_j^{(y)}\left(\left[{\tilde U}_{\pm,1}^{(\ell)}
\right]^\ast\tau_{-j}^{(y)}(a){\tilde U}_{\pm,1}^{(\ell)}\right)\right)
\right|\ret
&=&\left|\omega'(a)
-\omega\left(\tau_j^{(y)}\left(\left[{\tilde U}_{\pm,1}^{(\ell)}
\right]^\ast\right)a\tau_j^{(y)}\left({\tilde U}_{\pm,1}^{(\ell)}\right)\right)
\right|
\end{eqnarray}
for any $a\in{\cal A}_{\rm loc}$. In the limit $j\rightarrow\infty$, we get
\begin{equation}
\left|\omega'(a)-\omega(a)\right|
=\left|\left(\Omega_{\omega'},\pi_\omega(a)\Omega_{\omega'}\right)-
\left(\Omega_\omega,\pi_\omega(a)\Omega_\omega\right)\right|\le\varepsilon'
\Vert a \Vert
\label{omega'-omega}
\end{equation}
for any $a\in{\cal A}_{\rm loc}$. We decompose $\Omega_{\omega'}$ as 
\begin{equation}
\Omega_{\omega'}=d\Omega_\omega+\Omega''\quad \mbox{with}\ \ 
\left(\Omega'',\Omega_\omega\right)=0, 
\label{Omega'decom2}
\end{equation}
where $d$ is a complex number. Taking the orthogonal projection onto 
$\Omega_\omega$ and the orthogonal projection onto $\Omega''$ 
as observables for\footnote{Since the state $\omega'$ is extended to 
that for the set of all bounded operators on ${\bf H}_\omega$, 
the inequality (\ref{omega'-omega}) is valid also for the set of 
all bounded operators by the Hahn-Banach theorem \cite{BraRob}.} 
(\ref{omega'-omega}), we obtain 
\begin{equation}
1-|d|^2\le\varepsilon'\quad ,\quad \left\Vert\Omega''
\right\Vert^2\le\varepsilon'.
\end{equation}
Substituting these inequalities and the decomposition (\ref{Omega'decom2}) 
into (\ref{Omega'UOmegabound}), we have 
\begin{equation}
\left|d^\ast\left(\Omega_\omega,\pi_\omega\left({\tilde U}_{\pm,1}^{(\ell)}
\right)\Omega_\omega\right)-1\right|\le\varepsilon+\sqrt{\varepsilon'}. 
\end{equation}
This inequality contradicts (\ref{orthogonalOmegaUOmega}). 
Thus the norm of $\Xi$ is non-vanishing in the limit $\ell\rightarrow\infty$. 

Next we show that the vector $\Xi$ gives a low energy excitation. 
Note that
\begin{eqnarray}
& &\left\langle{\tilde \Phi}_\Lambda,\left[{\tilde U}_{\pm,1}^{(\ell)}
\right]^\ast\left[{\tilde H}_{\Lambda,\mu}(n_{\rm max}),
{\tilde U}_{\pm,1}^{(\ell)}\right]{\tilde \Phi}_\Lambda\right\rangle\ret
&=&\sum_N\left|\alpha_N\right|^2
\left\langle{\tilde \Phi}_\Lambda^{(N)},\left[{\tilde U}_{\pm,1}^{(\ell)}
\right]^\ast\left[{\tilde H}_{\Lambda,\mu}(n_{\rm max}),
{\tilde U}_{\pm,1}^{(\ell)}\right]{\tilde \Phi}_\Lambda^{(N)}\right\rangle\ret
&=&\sum_N\left|\alpha_N\right|^2
\left\langle{\tilde \Phi}_\Lambda^{(N)},\left[{\tilde U}_{\pm,1}^{(\ell)}
\right]^\ast\left[{\tilde H}_\Lambda(n_{\rm max}),
{\tilde U}_{\pm,1}^{(\ell)}\right]{\tilde \Phi}_\Lambda^{(N)}\right\rangle\ret
&=&\sum_N\left|\alpha_N\right|^2\left[
\left\langle{\tilde \Phi}_\Lambda^{(N)},\left[{\tilde U}_{\pm,1}^{(\ell)}
\right]^\ast{\tilde H}_\Lambda(n_{\rm max})
{\tilde U}_{\pm,1}^{(\ell)}{\tilde \Phi}_\Lambda^{(N)}\right\rangle
-\left\langle{\tilde \Phi}_\Lambda^{(N)},{\tilde H}_\Lambda(n_{\rm max})
{\tilde \Phi}_\Lambda^{(N)}\right\rangle\right].
\end{eqnarray}
Combining this with the definition (\ref{DE}) of $\Delta E_{L_y}^{(N)}$, 
we have 
\begin{eqnarray}
& &\frac{1}{2}\left\langle{\tilde \Phi}_\Lambda,\left[{\tilde U}_{+,1}^{(\ell)}
\right]^\ast\left[{\tilde H}_{\Lambda,\mu}(n_{\rm max}),
{\tilde U}_{+,1}^{(\ell)}\right]{\tilde \Phi}_\Lambda\right\rangle
+\frac{1}{2}\left\langle{\tilde \Phi}_\Lambda,\left[{\tilde U}_{-,1}^{(\ell)}
\right]^\ast\left[{\tilde H}_{\Lambda,\mu}(n_{\rm max}),
{\tilde U}_{-,1}^{(\ell)}\right]{\tilde \Phi}_\Lambda\right\rangle\ret
&=&\sum_N\left|\alpha_N\right|^2\Delta E_{L_y}^{(N)}. 
\end{eqnarray}
Further, by using Lemma~\ref{DeltaEestimate} we obtain the following: 
For any given small $\varepsilon>0$, there exists $\ell$ such that 
\begin{equation}
\varepsilon\ge\lim_{\Lambda\uparrow{\bf Z}}\omega\left(
\left[{\tilde U}_{\pm,1}^{(\ell)}
\right]^\ast\left[{\tilde H}_{\Lambda,\mu}(n_{\rm max}),
{\tilde U}_{\pm,1}^{(\ell)}\right]\right)
=\left(\Omega_\omega,\left[\pi_\omega\left({\tilde U}_{\pm,1}^{(\ell)}
\right)\right]^\ast H_\omega\pi_\omega\left({\tilde U}_{\pm,1}^{(\ell)}
\right)\Omega_\omega\right). 
\end{equation}
This implies the existence of a gapless excitation above the ground state 
$\omega$, with the above result about the vector $\Xi$. 
Since there is no gapless excitation above $\omega$, the assumption 
that all the infinite-volume ground states are translationally 
invariant with the period $1$, is not valid. 
Namely a translational symmery breaking occurs. 
\end{proof}

\begin{proof}{Proof of Theorem~\ref{theorem3}}
Since the proof is very similar to that of Theorem~\ref{theorem2}, 
we roughly sketch it. 

Let $\omega$ be a translationally invariant pure ground state with 
a period $q\in{\bf N}$ and with a non-zero excitation gap. 
Assuming $q\nu\notin{\bf N}$, we find a contradiction. 
In the same way as in the proof of 
Theorem~\ref{theorem2}, we have  
\begin{equation}
0=\lim_{\ell\rightarrow\infty}\omega\left({\tilde U}_{\pm,q}^{(\ell)}\right)
=\lim_{\ell\rightarrow\infty}\left(\Omega_\omega,\pi\left(
{\tilde U}_{\pm,q}^{(\ell)}\right)\Omega_\omega\right). 
\end{equation}
Let $\Xi=(1-G)\pi_\omega\left({\tilde U}_{\pm,q}^{(\ell)}\right)\Omega_\omega$.
Then the norm of the vector $\Xi$ is non-vanishing in the limit 
$\ell\rightarrow\infty$ again. Further we have 
\begin{equation}
\left(\Omega_\omega,\left[\pi_\omega\left({\tilde U}_{\pm,q}^{(\ell)}\right)
\right]^\ast H_\omega\pi_\omega\left({\tilde U}_{\pm,q}^{(\ell)}\right)
\Omega_\omega\right)\le\varepsilon
\end{equation}
for large $\ell$. Thus there exists a gapless excitation above the 
ground state $\omega$. Since $\omega$ has a gap, the assumption 
$q\nu\notin{\bf N}$ is not valid. Namely $q\nu\in{\bf N}$. 
\end{proof}

\section[Orthogonality]{Orthogonality ---Proof of 
Proposition~\ref{orthopro}---}
\setcounter{equation}{0}
\setcounter{theorem}{0}
\label{Orthproof}

In order to prove Proposition~\ref{orthopro}, 
we first study the properties of the vectors $\Phi_{L_y}^{(N)}$ and 
$\Psi_{\pm,L_y}^{(N)}=U_{\pm,q}^{(\ell)}\Phi_{L_y}^{(N)}$ 
for the action of $T^{(N,y)}(q\Delta y)$. Note that 
\begin{eqnarray}
T^{(N,y)}(q\Delta y)\Phi_{L_y}^{(N)}&=&\sum_{\{\xi_j\}}
a(\{\xi_j\})T^{(N,y)}(q\Delta y){\rm Asym}\left[\phi_{\xi_1}^{\rm P}
\otimes\cdots\otimes\phi_{\xi_N}^{\rm P}\right]\ret
&=&\sum_{\{\xi_j\}}
a(\{\xi_j\}){\rm Asym}\left[\phi_{\xi_1'}^{\rm P}
\otimes\cdots\otimes\phi_{\xi_N'}^{\rm P}\right]\ret
&=&\sum_{\{\xi_j\}}
a(\{\xi_j''\}){\rm Asym}\left[\phi_{\xi_1}^{\rm P}
\otimes\cdots\otimes\phi_{\xi_N}^{\rm P}\right],
\end{eqnarray}
where $\xi_j'=(n_j,k_j+q\Delta k)$ and $\xi_j''=(n_j,k_j-q\Delta k)$
with $\xi_j=(n_j,k_j)$ and $\Delta k=2\pi/L_x$.
Since $\Phi_{L_y}^{(N)}$ is an eigenvector of $T^{(N,y)}(q\Delta y)$ 
with the eigenvalue $\exp[i2\pi n/M]$ with an integer $n$, we have 
\begin{equation}
a(\{\xi_j''\})=a(\{\xi_j\})\exp[i2\pi n/M].
\label{aarelation}
\end{equation}
Using the definition (\ref{defUnit}) of $U_{\pm,q}^{(\ell)}$, we have 
\begin{eqnarray}
& &T^{(N,y)}(q\Delta y)\Psi_{\pm,L_y}^{(N)}\ret
&=&\sum_{\{\xi_j\}}a(\{\xi_j\})
\exp[\pm i2\pi\sum_{j=1}^N{\tilde m}(m_j)/\ell]
{\rm Asym}\left[\phi_{\xi_1'}^{\rm P}
\otimes\cdots\otimes\phi_{\xi_N'}^{\rm P}\right]\ret
&=&
\sum_{\{\xi_j\}}a(\{\xi_j\})
\exp\left[\pm i2\pi\sum_{j=1}^N{\tilde m}(m_j')/\ell\right]
\exp\left[\mp i2\pi\sum_{s=1}^{q\ell}{\hat n}_s/\ell\right]
{\rm Asym}\left[\phi_{\xi_1'}^{\rm P}
\otimes\cdots\otimes\phi_{\xi_N'}^{\rm P}\right]\ret
&=&\sum_{\{\xi_j\}}a(\{\xi_j''\})
\exp\left[\pm i2\pi\sum_{j=1}^N{\tilde m}(m_j)/\ell\right]
\exp\left[\mp i2\pi\sum_{s=-q+1}^{q(\ell-1)}{\hat n}_s/\ell\right]
{\rm Asym}\left[\phi_{\xi_1}^{\rm P}
\otimes\cdots\otimes\phi_{\xi_N}^{\rm P}\right]\ret
&=&e^{2\pi in/M}\sum_{\{\xi_j\}}a(\{\xi_j\})
\exp\left[\pm i2\pi\sum_{j=1}^N\frac{{\tilde m}(m_j)}{\ell}\right]
\exp\left[\mp i2\pi\sum_{s=-q+1}^{q(\ell-1)}\frac{{\hat n}_s}{\ell}\right]
\ret& &\qquad\times
{\rm Asym}\left[\phi_{\xi_1}^{\rm P}
\otimes\cdots\otimes\phi_{\xi_N}^{\rm P}\right]\ret
&=&e^{2\pi in/M}\exp\left[\mp i\frac{2\pi}{\ell}
\sum_{s=-q+1}^{q(\ell-1)}{\hat n}_s\right]\Psi_\pm^{(N)},
\label{TyPsi}
\end{eqnarray}
where $k_j=2\pi m_j/L_x$, $k_j'=2\pi m_j'/L_x$, and we have used the 
relation (\ref{aarelation}).
\medskip

\begin{proof}{Proof of Proposition~\ref{orthopro}}
Following Tasaki \cite{Tasaki}, we prove the statement. 
{From} (\ref{TyPsi}), one has 
\begin{eqnarray}
& &\left\langle\Phi_{L_y}^{(N)},\Psi_{\pm,L_y}^{(N)}\right\rangle\ret
&=&
\left\langle T^{(N,y)}(q\Delta y)\Phi_{L_y}^{(N)},
T^{(N,y)}(q\Delta y)\Psi_{\pm,L_y}^{(N)}\right\rangle\ret
&=&\left\langle\Phi_{L_y}^{(N)},\exp\left[\mp i\frac{2\pi}{\ell}
\sum_{s=-q+1}^{q(\ell-1)}{\hat n}_s\right]
\Psi_{\pm,L_y}^{(N)}\right\rangle\ret
&=&e^{\mp i2\pi q\nu}\left\langle\Phi_{L_y}^{(N)},\Psi_{\pm,L_y}^{(N)}
\right\rangle
+\left\langle\Phi_{L_y}^{(N)},\left(\exp\left[\mp i\frac{2\pi}{\ell}
\sum_{s=-q+1}^{q(\ell-1)}{\hat n}_s\right]
-e^{\mp i2\pi q\nu}\right)\Psi_{\pm,L_y}^{(N)}\right\rangle.\ret
\label{innerPhiPsi}
\end{eqnarray}
Using the Schwarz inequality, the second term in the last line is evaluated as 
\begin{eqnarray}
& &
\left|\left\langle\Phi_{L_y}^{(N)},\left(\exp\left[\mp i\frac{2\pi}{\ell}
\sum_{s=-q+1}^{q(\ell-1)}{\hat n}_s\right]-e^{\mp i2\pi q\nu}\right)
\Psi_{\pm,L_y}^{(N)}\right\rangle\right|^2\ret
&\le&4\left\langle\Phi_{L_y}^{(N)},\sin^2\left[\pi\left(
\frac{1}{\ell}\sum_{s=-q+1}^{q(\ell-1)}{\hat n}_s-q\nu\right)\right]
\Phi_{L_y}^{(N)}\right\rangle\ret
&\le&4\pi^2\left\langle\Phi_{L_y}^{(N)},\left(
\frac{1}{\ell}\sum_{s=-q+1}^{q(\ell-1)}{\hat n}_s-q\nu\right)^2
\Phi_{L_y}^{(N)}\right\rangle\ret
&=&4\pi^2\left[\frac{1}{\ell^2}\sum_{s,t}\left\langle\Phi_{L_y}^{(N)},
{\hat n}_s{\hat n}_t\Phi_{L_y}^{(N)}\right\rangle-(q\nu)^2\right].
\label{CDLRObound}
\end{eqnarray}
Here, for getting the last equality we have used the identity 
\begin{equation}
\frac{1}{q}
\sum_{s=j+1}^{j+q}\left\langle\Phi_{L_y}^{(N)},{\hat n}_s
\Phi_{L_y}^{(N)}\right\rangle
=\nu \quad \mbox{for any lattice site}\ j. 
\end{equation}
This is a consequence of the translational invariance of the 
state $\left\langle\Phi_{L_y}^{(N)},(\cdots)\Phi_{L_y}^{(N)}\right\rangle$ 
for the action $T^{(N,y)}(q\Delta y)$. From (\ref{innerPhiPsi}) and 
(\ref{CDLRObound}), one can show 
\begin{equation}
\sin^2(\pi q\nu)\left|\omega\left(U_{\pm,1}^{(\ell)}\right)
\right|^2\le
\pi^2\left[\frac{1}{\ell^2}\sum_{s,t}\omega({\hat n}_s{\hat n}_t)
-(q\nu)^2\right].
\end{equation}
This right-hand side is nothing but the long range charge correlation which is 
vanishing in the limit $\ell\rightarrow\infty$ by the assumption 
(\ref{noLRO}). 
Therefore the statement of Proposition~\ref{orthopro} has been proved. 
\end{proof}

\Section{Estimate of the energy gap}
\label{estimateEgap}

In this section, we prove Lemma~\ref{DeltaEestimate}. 
For simplicity, we write $\Delta E^{(N)}$ by dropping the subscript 
$L_y$ of $\Delta E_{L_y}^{(N)}$ of (\ref{DE}). 

{From} the definition (\ref{DE}) of $\Delta E^{(N)}$, we have 
\begin{equation}
\Delta E^{(N)}
=\eta_{L_y}^{(N)}\left(H_{L_y}^{(N)}(n_{\rm max})\right)
-\omega_{L_y}^{(N)}\left(H_{L_y}^{(N)}(n_{\rm max})\right)
=\Delta E_W^{(N)}+\Delta E_U^{(N)}
\end{equation}
with 
\begin{equation}
\Delta E_W^{(N)}=\sum_{j=1}^N \left[\eta_{L_y}^{(N)}\left(W(x_j)\right)
-\omega_{L_y}^{(N)}\left(W(x_j)\right)\right]
\end{equation}
and
\begin{equation}
\Delta E_U^{(N)}=\eta_{L_y}^{(N)}\left(U^{(N)}\right)
-\omega_{L_y}^{(N)}\left(U^{(N)}\right). 
\end{equation}
In the following, we will estimate only $\Delta E_U^{(N)}$ 
because $\Delta E_W^{(N)}$ can be treated in a much easier way. 

To begin with, we note that 
$$
\Delta E_U^{(N)}=\sum_{ \{\xi_j\},\{\xi_j'\} }a^\ast\left(\{\xi_j\}\right)
a(\{\xi_j'\})\frac{1}{2}
\left[\prod_{j=1}^Ne^{i2\pi({\tilde m}(m_j')-{\tilde m}(m_j))/\ell}
+\prod_{j=1}^Ne^{-i2\pi({\tilde m}(m_j')-{\tilde m}(m_j))/\ell}-2\right]
$$
\begin{equation}
\times\left\langle{\rm Asym}\left[\phi_{\xi_1}^{\rm P}\otimes\cdots
\otimes\phi_{\xi_N}^{\rm P}\right],U^{(N)}
{\rm Asym}\left[\phi_{\xi_1'}^{\rm P}\otimes\cdots\otimes
\phi_{\xi_N'}^{\rm P}\right]\right\rangle. 
\end{equation}
Here we notice that the contribution from $\{\xi_j\}=\{\xi_j'\}$ is 
vanishing, and that the matrix element for $U^{(N)}$ is vanishing 
if $\{\xi_j\},\{\xi_j'\}$ differ by more than two pairs of 
single-body functions. Therefore $\Delta E_U^{(N)}$ can be written as 
\begin{equation}
\Delta E_U^{(N)}=\Delta E^{(N)}_{\rm I}+\Delta E^{(N)}_{\rm II}
\end{equation}
in terms of the two types of contributions, $\{\xi_j\},\{\xi_j'\}$ 
differing by one pair of functions, 
\begin{eqnarray}
\Delta E^{(N)}_{\rm I}&=&\sum_{\{\xi_j\}}\sum_{s=1}^N\sum_{\xi_s'}
a^\ast(\{\xi_j\})a(\{\xi_1,\ldots,\xi_s',\ldots,\xi_N\})
\left\{\cos\left[\frac{2\pi}{\ell}({\tilde m}(m_s')-{\tilde m}(m_s))\right]
-1\right\}
\ret
&\times&\left\langle{\rm Asym}\left[\phi_{\xi_1}^{\rm P}\otimes\cdots\otimes
\phi_{\xi_N}^{\rm P}\right],U^{(N)}{\rm Asym}
\left[\phi_{\xi_1}^{\rm P}\otimes\cdots\otimes
\phi_{\xi_s'}^{\rm P}\otimes\cdots\otimes\phi_{\xi_N}^{\rm P}\right]
\right\rangle,
\label{DEI}
\end{eqnarray}
and $\{\xi_j\},\{\xi_j'\}$ differing by two pairs of functions, 
\begin{eqnarray}
\Delta E^{(N)}_{\rm II}&=&\frac{1}{4}\sum_{\{\xi_j\}}\sum_{s=1}^N\sum_{t\ne s}
\sum_{\xi_s'}\sum_{\xi_t'}
a^\ast(\{\xi_j\})a(\{\xi_1,\ldots,\xi_s',\ldots,\xi_t',\ldots,\xi_N\})
\hspace{3.5cm}\ret
&\times&
\left\{\cos\left[\frac{2\pi}{\ell}
({\tilde m}(m_s')-{\tilde m}(m_s)+{\tilde m}(m_t')-{\tilde m}(m_t))\right]-1
\right\}
\nonumber
\end{eqnarray}
\begin{equation}
\qquad\quad\times\left\langle{\rm Asym}\left[\phi_{\xi_1}^{\rm P}\otimes\cdots
\otimes\phi_{\xi_N}^{\rm P}\right],U^{(N)}{\rm Asym}\left[\phi_{\xi_1}^{\rm P}
\otimes\cdots\otimes
\phi_{\xi_s'}^{\rm P}\otimes\cdots\otimes\phi_{\xi_t'}^{\rm P}\otimes\cdots
\otimes\phi_{\xi_N}^{\rm P}\right]\right\rangle.
\label{DEII}
\end{equation}

\subsection{Estimate of $\Delta E_{\rm I}^{(N)}$}

We first treat $\Delta E_{\rm I}^{(N)}$, and 
we will estimate $\Delta E_{\rm II}^{(N)}$ in Section~\ref{EstDEII}. 

To begin with, we decompose $\Delta E_{\rm I}^{(N)}$ into the following 
two parts:
\begin{eqnarray}
\Delta E_{{\rm I},<}^{(N)}&=&\sum_{\{\xi_j\}}\sum_{s=1}^N
\sum_{\xi_s'}a^\ast(\{\xi_j\})a(\{\xi_j'\})\chi
\left({\rm dist}^{(m)}(m_s,m_s')<\ell^\delta/2\right)\ret
&\times&\left\{
\cos\left[\frac{2\pi}{\ell}({\tilde m}(m_s')-{\tilde m}(m_s))\right]-1
\right\}\ret
&\times&
\left\langle{\rm Asym}\left[\phi_{\xi_1}^{\rm P}\otimes\cdots\otimes
\phi_{\xi_N}^{\rm P}\right],U^{(N)}{\rm Asym}\left[\phi_{\xi_1}^{\rm P}
\otimes\cdots\otimes
\phi_{\xi_s'}^{\rm P}\otimes\cdots\otimes\phi_{\xi_N}^{\rm P}\right]
\right\rangle,
\label{DEI<}
\end{eqnarray}
and 
\begin{eqnarray}
\Delta E_{{\rm I},\ge}^{(N)}&=&\sum_{\{\xi_j\}}\sum_{s=1}^N
\sum_{\xi_s'}a^\ast(\{\xi_j\})a(\{\xi_j'\})\chi
\left({\rm dist}^{(m)}(m_s,m_s')\ge\ell^\delta/2\right)\ret
&\times&\left\{
\cos\left[\frac{2\pi}{\ell}({\tilde m}(m_s')-{\tilde m}(m_s))\right]-1
\right\}\ret
&\times&
\left\langle{\rm Asym}\left[\phi_{\xi_1}^{\rm P}\otimes\cdots\otimes
\phi_{\xi_N}^{\rm P}\right],U^{(N)}{\rm Asym}\left[\phi_{\xi_1}^{\rm P}
\otimes\cdots\otimes
\phi_{\xi_s'}^{\rm P}\otimes\cdots\otimes\phi_{\xi_N}^{\rm P}\right]
\right\rangle,
\label{DE>}
\end{eqnarray}
where $\delta\in(0,1/4)$, and $\chi$ is the characteristic function given by 
\begin{equation}
\chi(Q)=\cases{1  & if $Q$ is true; \cr
               0, & otherwise,\cr}
\end{equation}
and
\begin{equation}
{\rm dist}^{(m)}(m_s,m_s'):=\min_{n\in {\bf Z}}\{|m_s-m_s'-nM|\}.
\end{equation}
Here we have written $\{\xi_j'\}=\{\xi_1,\xi_2,\ldots,\xi_{s-1},\xi_s',
\xi_{s+1},\ldots,\xi_N\}$. In the following, we fix $\delta$ 
to a number in the interval. 

\subsubsection{Estimate of $\Delta E_{{\rm I},<}^{(N)}$}
\label{EstDEII<}

As we will show in the following, we have a bound 
\begin{equation}
\left|\Delta E_{{\rm I},<}^{(N)}\right|
\le2\pi^2q{\cal C}^{(1)}(U^{(2)})(n_{\rm max}+1)^3
\frac{\ell^{3\delta}}{\ell},
\label{DEI<bound1}
\end{equation}
where ${\cal C}^{(1)}(U^{(2)})$ is a positive constant which depends 
only on\footnote{For simplicity, we do not write a magnetic length $\ell_B$ 
dependence which is not of interest here.} 
the interction $U^{(2)}$. 
Clearly $\Delta E_{{\rm I},<}^{(N)}$ is vanishing in the limit 
$\ell\rightarrow\infty$ because $\delta\in(0,1/4)$. 

In order to show the bound, we first evaluate the matrix element of 
$U^{(N)}$ in (\ref{DEI<}) as 
\begin{eqnarray}
& &\left|\left\langle{\rm Asym}\left[\phi_{\xi_1}^{\rm P}\otimes\cdots\otimes
\phi_{\xi_N}^{\rm P}\right],U^{(N)}{\rm Asym}\left[\phi_{\xi_1}^{\rm P}
\otimes\cdots\otimes
\phi_{\xi_s'}^{\rm P}\otimes\cdots\otimes\phi_{\xi_N}^{\rm P}\right]
\right\rangle\right|\ret
&\le&
\left|\sum_{\xi\in\{\xi_j\}}
\int_S dx_1dy_1\int_S dx_2dy_2\ {\phi_{\xi_s}^{\rm P}}^\ast({\bf r}_1)
{\phi_\xi^{\rm P}}^\ast({\bf r}_2)
U^{(2)}(x_1-x_2,y_1-y_2)\phi_{\xi_s'}^{\rm P}({\bf r}_1)
\phi_\xi^{\rm P}({\bf r}_2)\right|\ret
&+&
\left|\sum_{\xi\in\{\xi_j\}}
\int_S dx_1dy_1\int_S dx_2dy_2\ {\phi_{\xi_s}^{\rm P}}^\ast({\bf r}_1)
{\phi_\xi^{\rm P}}^\ast({\bf r}_2)
U^{(2)}(x_1-x_2,y_1-y_2)\phi_\xi^{\rm P}({\bf r}_1)\phi_{\xi_s'}^{\rm P}
({\bf r}_2)\right|.\ret
\label{mat1bound}
\end{eqnarray}

\begin{lemma} The following inequality is valid:
\label{U2avlemma}
\begin{equation}
\sum_{\{\xi=(n,k)| n\le n_{\rm max}\}}
\int_S dx dy 
\left|\phi_\xi({\bf r})\right|^2\left|U^{(2)}(x-x',y-y')\right|
\le{\cal C}^{(1)}(U^{(2)})(n_{\rm max}+1)\quad \mbox{for any}\ x',y',
\label{U2avbound}
\end{equation}
where ${\cal C}^{(1)}(U^{(2)})$ is the same constant as in 
(\ref{DEI<bound1}). 
\end{lemma}
The proof is given in Appendix~\ref{ProofU2avlemma}. Using 
Lemma~\ref{U2avlemma}, we have 
\begin{eqnarray}
& &
\left|\sum_{\xi\in\{\xi_j\}}
\int dx_1dy_1\int dx_2dy_2{\phi_{\xi_s}^{\rm P}}^\ast({\bf r}_1)
{\phi_\xi^{\rm P}}^\ast({\bf r}_2)
U^{(2)}(x_1-x_2,y_1-y_2)\phi_{\xi_s'}^{\rm P}({\bf r}_1)
\phi_\xi^{\rm P}({\bf r}_2)\right|\ret
&\le&{\cal C}^{(1)}(U^{(2)})(n_{\rm max}+1)
\int_S dxdy\left|{\phi_{\xi_s}^{\rm P}}({\bf r})\right|
\left|\phi_{\xi_s'}^{\rm P}({\bf r})\right|
\end{eqnarray}
for the first term in the right-hand side of (\ref{mat1bound}). 
Here we have replaced the sum about $\{\xi_j\}$ with the sum about 
the whole $\xi$ for getting the bound. From this bound, we obtain 
\begin{eqnarray}
& &
\left|a^\ast(\{\xi_j\})a(\{\xi_j'\})\right|\ret
&\times&
\left|\sum_{\xi\in\{\xi_j\}}
\int dx_1dy_1\int dx_2dy_2\ {\phi_{\xi_s}^{\rm P}}^\ast({\bf r}_1)
{\phi_\xi^{\rm P}}^\ast({\bf r}_2)
U^{(2)}(x_1-x_2,y_1-y_2)\phi_{\xi_s'}^{\rm P}({\bf r}_1)
\phi_\xi^{\rm P}({\bf r}_2)\right|\ret
&\le&\frac{1}{2}{\cal C}^{(1)}(U^{(2)})(n_{\rm max}+1)\left[
\left|a(\{\xi_j\})\right|^2
\int_S dxdy\left|{\phi_{\xi_s}^{\rm P}}({\bf r})\right|^2
+\left|a(\{\xi_j'\})\right|^2
\int_S dxdy\left|{\phi_{\xi_s'}^{\rm P}}({\bf r})\right|^2
\right]\ret
&=&\frac{1}{2}{\cal C}^{(1)}(U^{(2)})(n_{\rm max}+1)\left[
\left|a(\{\xi_j\})\right|^2
+\left|a(\{\xi_j'\})\right|^2\right]. 
\label{aaav}
\end{eqnarray}

On the other hand,  the second term in the right-hand side of 
(\ref{mat1bound}) is evalueted as 
\begin{eqnarray}
& &\left|\int dx_1dy_1\int dx_2dy_2\ {\phi_{\xi_s}^{\rm P}}^\ast({\bf r}_1)
{\phi_\xi^{\rm P}}^\ast({\bf r}_2)
U^{(2)}(x_1-x_2,y_1-y_2)\phi_\xi^{\rm P}({\bf r}_1)\phi_{\xi_s'}^{\rm P}
({\bf r}_2)\right|\ret
&\le&\sqrt{\int dx_1dy_1\int dx_2dy_2 \
\left|{\phi_{\xi_s}^{\rm P}}({\bf r}_1)\right|^2\left|U^{(2)}(x_1-x_2,y_1-y_2)
\right|
\left|{\phi_\xi^{\rm P}}({\bf r}_2)\right|^2}\ret
&\times&\sqrt{\int dx_1dy_1\int dx_2dy_2 \
\left|{\phi_{\xi_s'}^{\rm P}}({\bf r}_1)\right|^2\left|U^{(2)}(x_1-x_2,y_1-y_2)
\right|\left|{\phi_\xi^{\rm P}}({\bf r}_2)\right|^2}
\end{eqnarray}
by using the Schwarz inequality. In the same way as in (\ref{aaav}), we obtain 
\begin{eqnarray}
& &
\left|a^\ast(\{\xi_j\})a(\{\xi_j'\})\right|\ret
&\times&
\left|\int dx_1dy_1\int dx_2dy_2\ {\phi_{\xi_s}^{\rm P}}^\ast({\bf r}_1)
{\phi_\xi^{\rm P}}^\ast({\bf r}_2)
U^{(2)}(x_1-x_2,y_1-y_2)\phi_\xi^{\rm P}({\bf r}_1)\phi_{\xi_s'}^{\rm P}
({\bf r}_2)\right|\ret
&\le&\frac{1}{2}\left|a^\ast(\{\xi_j\})\right|^2
\int dx_1dy_1\int dx_2dy_2 \
\left|{\phi_{\xi_s}^{\rm P}}({\bf r}_1)\right|^2\left|U^{(2)}(x_1-x_2,y_1-y_2)
\right|
\left|{\phi_\xi^{\rm P}}({\bf r}_2)\right|^2\ret
&+&\frac{1}{2}\left|a^\ast(\{\xi_j'\})\right|^2
\int dx_1dy_1\int dx_2dy_2 \
\left|{\phi_{\xi_s'}^{\rm P}}({\bf r}_1)\right|^2\left|U^{(2)}(x_1-x_2,y_1-y_2)
\right|
\left|{\phi_\xi^{\rm P}}({\bf r}_2)\right|^2.\ret
\end{eqnarray}
Taking the sum over $\xi\in\{\xi_j\}$ and using Lemma~\ref{U2avlemma} in the 
same way, we have 
\begin{eqnarray}
& &\left|a^\ast(\{\xi_j\})a(\{\xi_j'\})\right|\ret
&\times&\left|\sum_{\xi}
\int dx_1dy_1\int dx_2dy_2\ {\phi_{\xi_s}^{\rm P}}^\ast({\bf r}_1)
{\phi_\xi^{\rm P}}^\ast({\bf r}_2)
U^{(2)}(x_1-x_2,y_1-y_2)\phi_\xi^{\rm P}({\bf r}_1)\phi_{\xi_s'}^{\rm P}
({\bf r}_2)\right|\ret
&\le&\frac{1}{2}{\cal C}^{(1)}(U^{(2)})(n_{\rm max}+1)\left[
\left|a^\ast(\{\xi_j\})\right|^2
+\left|a^\ast(\{\xi_j'\})\right|^2\right]. 
\label{aamat2}
\end{eqnarray}
Combining (\ref{mat1bound}), (\ref{aaav}) and (\ref{aamat2}), we obtain 
\begin{eqnarray}
& &\left|a^\ast(\{\xi_j\})a(\{\xi_j'\})
\left\langle{\rm Asym}\left[\phi_{\xi_1}^{\rm P}\otimes\cdots\otimes
\phi_{\xi_N}^{\rm P}\right],U^{(N)}{\rm Asym}\left[\phi_{\xi_1}^{\rm P}
\otimes\cdots\otimes
\phi_{\xi_s'}^{\rm P}\otimes\cdots\otimes\phi_{\xi_N}^{\rm P}\right]
\right\rangle\right|\ret
&\le&{\cal C}^{(1)}(U^{(2)})(n_{\rm max}+1)\left[\left|a(\{\xi_j\})\right|^2
+\left|a(\{\xi_j'\})\right|^2\right].
\label{aamatfbound}
\end{eqnarray}
By using this inequality, $\Delta E_{{\rm I},<}^{(N)}$ of 
(\ref{DEI<}) is evaluated as 
\begin{eqnarray}
\left|\Delta E_{{\rm I},<}^{(N)}\right|&\le&
2{\cal C}^{(1)}(U^{(2)})(n_{\rm max}+1)\sum_{\{\xi_j\}}\sum_{s=1}^N
\sum_{\xi_s'}\chi
({\rm dist}^{(m)}(m_s,m_s')<\ell^\delta/2)\ret
&\times&\left\{1-
\cos\left[\frac{2\pi}{\ell}({\tilde m}(m_s')-{\tilde m}(m_s))\right]\right\}
\left|a(\{\xi_j\})\right|^2.
\label{DEI<bound}
\end{eqnarray}
Note that 
\begin{eqnarray}
& &1-\cos\left[\frac{2\pi}{\ell}({\tilde m}(m_s')-{\tilde m}(m_s))\right]\ret
&=&\left\{1-\cos\left[\frac{2\pi}{\ell}({\tilde m}(m_s')-{\tilde m}(m_s))
\right]\right\}\ret
&\times&\left\{\chi(1\le m_s\le q\ell)+\left[1-\chi(1\le m_s\le q\ell)\right]
\chi(1\le m_s'\le q\ell)\right\}
\label{1-cosidentity}
\end{eqnarray}
from the definition (\ref{deftildem}) of ${\tilde m}(\cdots)$. 
Using this identity, we have 
\begin{eqnarray}
& &\left|\Delta E_{{\rm I},<}^{(N)}\right|\ret
&\le&2{\cal C}^{(1)}(U^{(2)})(n_{\rm max}+1)\ret&\times&
\sum_{\{\xi_j\}}\sum_{s=1}^N\sum_{\xi_s'}
\left\{\chi(1\le m_s\le q\ell)+\left[1-\chi(1\le m_s\le q\ell)\right]
\chi(1\le m_s'\le q\ell)\right\}\ret
&\times&\chi({\rm dist}^{(m)}(m_s,m_s')<\ell^\delta/2)
\left\{1-
\cos\left[\frac{2\pi}{\ell}({\tilde m}(m_s')-{\tilde m}(m_s))\right]\right\}
\left|a(\{\xi_j\})\right|^2\ret
&\le&{\cal C}^{(1)}(U^{(2)})(n_{\rm max}+1)\ret
&\times&\sum_{\{\xi_j\}}\sum_{s=1}^N
\sum_{\xi_s'}
\left\{\chi(1\le m_s\le q\ell)+\left[1-\chi(1\le m_s\le q\ell)\right]
\chi(1\le m_s'\le q\ell)\right\}\ret
&\times&\chi({\rm dist}^{(m)}(m_s,m_s')<\ell^\delta/2)
\left|a(\{\xi_j\})\right|^2\frac{\pi^2\ell^{2\delta}}{\ell^2}.
\label{DeltaEI>bound}
\end{eqnarray}
Note that
\begin{eqnarray}
& &\sum_{\{\xi_j\}}\sum_{s=1}^N
\sum_{\xi_s'}\chi(1\le m_s\le q\ell)
\chi({\rm dist}^{(m)}(m_s,m_s')<\ell^\delta/2)
\left|a(\{\xi_j\})\right|^2\ret
&\le&\sum_{\{\xi_j\}}\sum_{s=1}^N\chi(1\le m_s\le q\ell)
(n_{\rm max}+1)\ell^\delta
\left|a(\{\xi_j\})\right|^2\ret
&\le&\sum_{\{\xi_j\}}(n_{\rm max}+1)^2q\ell\ell^\delta
\left|a(\{\xi_j\})\right|^2=(n_{\rm max}+1)^2q\ell\ell^\delta,
\end{eqnarray}
and
\begin{eqnarray}
& &\sum_{\{\xi_j\}}\sum_{s=1}^N
\sum_{\xi_s'}\left[1-\chi(1\le m_s\le q\ell)\right]
\chi(1\le m_s'\le q\ell)
\chi({\rm dist}^{(m)}(m_s,m_s')<\ell^\delta/2)
\left|a(\{\xi_j\})\right|^2\ret
&\le&\sum_{\{\xi_j\}}\sum_{s=1}^N
\sum_{\xi_s'}\chi(-\ell^\delta/2+1\le m_s\le 0\ \mbox{or}\ 
q\ell+1\le m_s\le q\ell+\ell^\delta/2-1)\ret
&\times&\chi({\rm dist}^{(m)}(m_s,m_s')<\ell^\delta/2)
\left|a(\{\xi_j\})\right|^2\ret
&\le&\sum_{\{\xi_j\}}\sum_{s=1}^N
\chi(-\ell^\delta/2+1\le m_s\le 0\ \mbox{or}\ 
q\ell+1\le m_s\le q\ell+\ell^\delta/2-1)\ret
&\times&(n_{\rm max}+1)\ell^\delta
\left|a(\{\xi_j\})\right|^2\ret
&\le&\sum_{\{\xi_j\}}(n_{\rm max}+1)^2\ell^{2\delta}
\left|a(\{\xi_j\})\right|^2=(n_{\rm max}+1)^2\ell^{2\delta}.
\end{eqnarray}
Substituting these into (\ref{DeltaEI>bound}), we obtain 
\begin{eqnarray}
\left|\Delta E_{{\rm I},<}^{(N)}\right|
&\le&\pi^2{\cal C}^{(1)}(U^{(2)})(n_{\rm max}+1)^3
\left(\frac{q\ell^{3\delta}}{\ell}+\frac{\ell^{4\delta}}{\ell^2}\right)\ret
&\le&2\pi^2q{\cal C}^{(1)}(U^{(2)})(n_{\rm max}+1)^3
\frac{\ell^{3\delta}}{\ell}.
\end{eqnarray}

\subsubsection{Estimate of $\Delta E_{{\rm I},\ge}^{(N)}$}
\label{EstDEI>}

The main results in this subsection are summarized in Lemma~\ref{mainDEIg} 
below. The results include important information about 
the cutoff dependence of the size $\delta y$ of the local perturbation 
in the $y$ direction. 

For proceeding to this lemma, we make preparations. 
Using (\ref{mat1bound}), Lemma~\ref{U2avlemma} and (\ref{1-cosidentity})
in the same way as in the preceding Section~\ref{EstDEII<}, 
we can evaluate $\Delta E_{{\rm I},\ge}^{(N)}$ of (\ref{DE>}) as 
\begin{equation}
\left|\Delta E_{{\rm I},\ge}^{(N)}\right|
\le 2{\cal C}^{(1)}(U^{(2)})(n_{\rm max}+1)\Delta E_{{\rm I},\ge,1}^{(N)}
+2\Delta E_{{\rm I},\ge,1}^{(N)},
\label{DEI>Est1}
\end{equation}
where 
\begin{eqnarray}
\Delta E_{{\rm I},\ge,1}^{(N)}&=&\sum_{\{\xi_j\}}\sum_{s=1}^N
\sum_{\xi_s'}\left[\left|a(\{\xi_j\})\right|^2+\left|a(\{\xi_j'\})\right|^2
\right]\chi(1\le m_s\le q\ell)\ret
&\times&\chi({\rm dist}^{(m)}(m_s,m_s')
\ge \ell^\delta/2)\int_Sdxdy\ \left|\phi_{\xi_s'}^{\rm P}({\bf r})\right|
\left|\phi_{\xi_s}^{\rm P}({\bf r})\right|,
\end{eqnarray}
and
$$
\Delta E_{{\rm I},\ge,2}^{(N)}=\sum_{\{\xi_j\}}\sum_{s=1}^N
\sum_{\xi_s'}\left[\left|a(\{\xi_j\})\right|^2+\left|a(\{\xi_j'\})\right|^2
\right]\chi(1\le m_s\le q\ell)\chi({\rm dist}^{(m)}(m_s,m_s')
\ge \ell^\delta/2)
$$
\begin{equation}
\times\sum_{\xi\in\{\xi_j\}}\int dx_1dy_1\int dx_2dy_2
\left|\phi_{\xi_s}^{\rm P}({\bf r}_1)\right|
\left|\phi_{\xi}^{\rm P}({\bf r}_1)\right|
\left|U^{(2)}(x_1-x_2,y_1-y_2)\right|
\left|\phi_{\xi_s'}^{\rm P}({\bf r}_2)\right|
\left|\phi_{\xi}^{\rm P}({\bf r}_2)\right|.
\label{DEI>2}
\end{equation}

\begin{lemma}
\label{mainDEIg}
Suppose that $\pi(\ell^\delta/4-1)\ell_B/L_x>n_{\rm max}$ and 
$L_y>32n_{\rm max}\ell_B$. Then the following two bounds are valid:
\begin{equation}
\Delta E_{I,\ge,1}^{(N)}\le2\left(n_{\rm max}+1\right)q\ell
\epsilon^{(1)}(\ell^\delta/2-1,n_{\rm max},L_x,L_y),
\label{DEI>1bound}
\end{equation}
and
\begin{eqnarray}
\Delta E_{{\rm I},\ge,2}^{(N)}&\le&
4\left\Vert U^{(2)}\right\Vert_\infty(n_{\rm max}+1)q\ell
\epsilon^{(1)}(\ell^\delta/4-1,n_{\rm max},L_x,L_y)\ret
& &\times\left[2(n_{\rm max}+1)\ell^\delta+
\epsilon^{(1)}(\ell^\delta-1,n_{\rm max},L_x,L_y)\right],
\label{DEI>2bound}
\end{eqnarray}
where 
\begin{eqnarray}
\epsilon^{(1)}(\Delta\ell,n_{\rm max},L_x,L_y)
&:=&{\cal C}^{(2)}(n_{\rm max})\frac{L_x}{\ell_B}
\exp\left[-\left(\frac{\pi\ell_B}{L_x}\Delta\ell-n_{\rm max})\right)^2
\right]\ret
&+&{\cal C}^{(3)}(n_{\rm max})\frac{L_xL_y}{\ell_B^2}
\exp\left[-\frac{L_y^2}{32\ell_B^2}\left(1-\frac{32n_{\rm max}\ell_B}{L_y}
\right)^2\right].
\label{epsilon1}
\end{eqnarray}
Here the constants ${\cal C}^{(2)}(n_{\rm max})$ 
and ${\cal C}^{(3)}(n_{\rm max})$ depend on the energy cutoff 
$n_{\rm max}$ only.
\end{lemma}
Immediately, we have 
\begin{equation}
\lim_{\ell\rightarrow\infty}\lim_{L_y\rightarrow\infty}
\Delta E_{I,\ge,1}^{(N)}=0\quad, \quad
\lim_{\ell\rightarrow\infty}\lim_{L_y\rightarrow\infty}
\Delta E_{{\rm I},\ge,2}^{(N)}=0
\end{equation}
for a fixed $L_x$. Clearly these with (\ref{DEI>Est1}) yield 
\begin{equation}
\lim_{\ell\rightarrow\infty}\lim_{L_y\rightarrow\infty}
\Delta E_{I,\ge}^{(N)}=0
\end{equation}
for a fixed $L_x$. We remark that the size $\delta y$ 
of the local perturbation in the $y$ direction strongly depends 
on the width $L_x$ of the strip and the energy cutoff $n_{\rm max}$. 
To see this, we note that 
the size $\delta y$ is given by $\delta y=\ell \Delta y$, where 
$\Delta y$ is the lattice constant given by (\ref{Deltaxy}). 
From the lemma, the number $\ell$ must at least satisfy 
$\ell_B\ell^\delta \sim L_x n_{\rm max}$. Then we have 
\begin{equation}
\delta y \propto (L_x)^{1/\delta-1}(n_{\rm max})^{1/\delta}
=(L_x)^{3+\epsilon}(n_{\rm max})^{4+\epsilon},
\label{deltay}
\end{equation}
where we have taken $\epsilon$ to be a small positive number 
by using $\delta \in (0,1/4)$. This cutoff dependence is not 
a desired one. We believe that the size $\delta y$ does not 
depend on $L_x, n_{\rm max}$ in much better evaluation for 
the low energy excitations. 

First let us show the bound (\ref{DEI>1bound}). We can rewrite 
$\Delta E_{{\rm I},\ge,1}^{(N)}$ as 
\begin{eqnarray}
\Delta E_{{\rm I},\ge,1}^{(N)}&=&\sum_{\{\xi_j\}}\sum_{s=1}^N
\chi(1\le m_s\le q\ell)\left|a(\{\xi_j\})\right|^2\ret
&\times&\sum_{\xi_s'}\chi({\rm dist}^{(m)}(m_s,m_s')
\ge \ell^\delta/2)\int_Sdxdy\ \left|\phi_{\xi_s'}^{\rm P}({\bf r})\right|
\left|\phi_{\xi_s}^{\rm P}({\bf r})\right|\ret
&+&\sum_{\{\xi_j'\}}\left|a(\{\xi_j'\})\right|^2
\sum_{s=1}^N\sum_{\xi_s}\chi(1\le m_s\le q\ell)\ret
&\times&\chi({\rm dist}^{(m)}(m_s,m_s')
\ge \ell^\delta/2)\int_Sdxdy\ \left|\phi_{\xi_s'}^{\rm P}({\bf r})\right|
\left|\phi_{\xi_s}^{\rm P}({\bf r})\right|
\label{EI>1decom2}
\end{eqnarray}
Note that 
\begin{eqnarray}
& &\sum_{s=1}^N\sum_{\xi_s}\chi(1\le m_s\le q\ell)
\chi({\rm dist}^{(m)}(m_s,m_s')
\ge \ell^\delta/2)\int_Sdxdy\ \left|\phi_{\xi_s'}^{\rm P}({\bf r})\right|
\left|\phi_{\xi_s}^{\rm P}({\bf r})\right|\ret
&\le&\sum_\xi \chi(1\le m\le q\ell)\sum_{s=1}^N \chi({\rm dist}^{(m)}(m,m_s')
\ge \ell^\delta/2)\int_Sdxdy\ \left|\phi_{\xi_s'}^{\rm P}({\bf r})\right|
\left|\phi_{\xi}^{\rm P}({\bf r})\right|,
\label{EI>1decom21}
\end{eqnarray}
where the first sum in the right-hand side is taken over all the states 
$\xi=(n,k)$ with $k=2\pi m/L_x$. 

We note that the Hermite polynomial $H_n$ in the functions 
$\phi_{n,k}^{\rm P}$ satisfies 
\begin{equation}
\left|H_n(\zeta)\right|\le c_ne^{\beta_n|\zeta|}\quad 
\mbox{for}\ \zeta\in{\bf R},
\label{Hnbound}
\end{equation}
where the positive constants $c_n$ and $\beta_n$ depend only on 
the number $n$. The well-known values are given by 
\begin{equation}
c_n=\cases{(n-1)!! & for $n=$ even; \cr
            \quad n!!   & for $n=$ odd, \cr}
\end{equation}
and
\begin{equation}
\beta_n=\cases{\ \ \displaystyle{\sqrt{2n}} & for $n=$ even; \cr
                                            &                \cr
               \displaystyle{\sqrt{2(n-1)}} & for $n=$ odd. \cr}
\end{equation}
Here $(2n-1)!!=(2n-1)(2n-3)\cdots 3\cdot 1$ with $(-1)!!=1$. 
Using the bound (\ref{Hnbound}), we can get the following lemma: 

\begin{lemma}
\label{>intphiphiLemma}
Let $\phi_\xi^{\rm P}$ be an eigenvector (\ref{phiP}) of the Hamiltonian 
(\ref{resingleham}) with the periodic boundary conditions (\ref{PBC}) and 
with quantum numbers $\xi=(n,k)=(n,2\pi m/L_x)$, 
and let $\pi(\Delta \ell-1)\ell_B/L_x>n_{\rm max}$ 
and $L_y>32n_{\rm max}\ell_B$. Then the following bound is valid: 
\begin{equation}
\sum_{\xi'}\chi({\rm dist}^{(m)}(m,m')\ge\Delta\ell)
\int_Sdxdy\ \left|\phi_\xi^{\rm P}({\bf r})\right|
\left|\phi_{\xi'}^{\rm P}({\bf r})\right|
\le\epsilon^{(1)}(\Delta\ell-1,n_{\rm max},L_x,L_y),
\label{>intphiphibound}
\end{equation}
where the sum is over all the states $\xi'=(n',k')$ 
with $k'=2\pi m'/L_x$, and $\epsilon^{(1)}$ is given by (\ref{epsilon1}). 
\end{lemma}
The proof is given in Appendix~\ref{Proof>intphiphiLemma}. 
Combining (\ref{EI>1decom2}), (\ref{EI>1decom21}) and 
Lemma~\ref{>intphiphiLemma}, we obtain the desired bound 
(\ref{DEI>1bound}). 

Next consider $\Delta E_{{\rm I},\ge,2}^{(N)}$ of (\ref{DEI>2}), 
and we shall show the bound (\ref{DEI>2bound}). Using the inequality 
\begin{equation}
\chi\left({\rm dist}^{(m)}(m_s,m_s')\ge\frac{\ell^\delta}{2}\right)
\le\chi\left({\rm dist}^{(m)}(m_s,m)\ge\frac{\ell^\delta}{4}\right)
+\chi\left({\rm dist}^{(m)}(m_s',m)\ge\frac{\ell^\delta}{4}\right),
\end{equation}
we have 
\begin{equation}
\Delta E_{{\rm I},\ge,2}^{(N)}\le\left\Vert U^{(2)}\right\Vert_\infty
\sum_{i=1}^4\Delta E_{{\rm I},\ge,2,i}^{(N)},
\label{DEI>2bound2}
\end{equation}
where 
\begin{eqnarray}
\Delta E_{{\rm I},\ge,2,1}^{(N)}&=&\sum_{\{\xi_j\}}\sum_{s=1}^N
\left|a(\{\xi_j\})\right|^2\chi(1\le m_s\le q\ell)\ret
&\times&\sum_\xi \chi({\rm dist}^{(m)}(m_s,m)\ge \ell^\delta/4)
\int dx_1dy_1\left|\phi_{\xi_s}^{\rm P}({\bf r}_1)\right|
\left|\phi_{\xi}^{\rm P}({\bf r}_1)\right|\ret
&\times&\sum_{\xi_s'}\int dx_2dy_2\left|\phi_{\xi_s'}^{\rm P}({\bf r}_2)\right|
\left|\phi_{\xi}^{\rm P}({\bf r}_2)\right|,
\end{eqnarray}
\begin{eqnarray}
\Delta E_{{\rm I},\ge,2,2}^{(N)}&=&\sum_{\{\xi_j\}}\sum_{s=1}^N
\left|a(\{\xi_j\})\right|^2\chi(1\le m_s\le q\ell)\ret
&\times&\sum_\xi\int dx_1dy_1\left|\phi_{\xi_s}^{\rm P}({\bf r}_1)\right|
\left|\phi_{\xi}^{\rm P}({\bf r}_1)\right|\ret
&\times&\sum_{\xi_s'}\chi({\rm dist}^{(m)}(m_s',m)\ge \ell^\delta/4)
\int dx_2dy_2\left|\phi_{\xi_s'}^{\rm P}({\bf r}_2)\right|
\left|\phi_{\xi}^{\rm P}({\bf r}_2)\right|,
\end{eqnarray}
\begin{eqnarray}
\Delta E_{{\rm I},\ge,2,3}^{(N)}&=&\sum_{\{\xi_j'\}}
\left|a(\{\xi_j'\})\right|^2\sum_{\xi''}\chi(1\le m''\le q\ell)\ret
&\times&\sum_\xi\chi({\rm dist}^{(m)}(m'',m)\ge \ell^\delta/4)
\int dx_1dy_1\left|\phi_{\xi''}^{\rm P}({\bf r}_1)\right|
\left|\phi_{\xi}^{\rm P}({\bf r}_1)\right|\ret
&\times&\sum_{s=1}^N 
\int dx_2dy_2\left|\phi_{\xi_s'}^{\rm P}({\bf r}_2)\right|
\left|\phi_{\xi}^{\rm P}({\bf r}_2)\right|,
\end{eqnarray}
and
\begin{eqnarray}
\Delta E_{{\rm I},\ge,2,4}^{(N)}&=&\sum_{\{\xi_j'\}}
\left|a(\{\xi_j'\})\right|^2\sum_{\xi''}\chi(1\le m''\le q\ell)\ret
&\times&\sum_\xi\int dx_1dy_1\left|\phi_{\xi''}^{\rm P}({\bf r}_1)\right|
\left|\phi_{\xi}^{\rm P}({\bf r}_1)\right|\ret
&\times&\sum_{s=1}^N \chi({\rm dist}^{(m)}(m_s',m)\ge \ell^\delta/4)
\int dx_2dy_2\left|\phi_{\xi_s'}^{\rm P}({\bf r}_2)\right|
\left|\phi_{\xi}^{\rm P}({\bf r}_2)\right|.
\end{eqnarray}
We note that, from Lemma~\ref{>intphiphiLemma}, 
\begin{eqnarray}
& &\sum_{\xi'}\int_S dxdy \left|\phi_{\xi'}^{\rm P}({\bf r})\right|
\left|\phi_\xi^{\rm P}({\bf r})\right|\ret
&=&\sum_{\xi'}\left[\chi({\rm dist}^{(m)}(m,m')<\ell^\delta)+
\chi({\rm dist}^{(m)}(m,m')\ge\ell^\delta)\right]
\int_S dxdy \left|\phi_{\xi'}^{\rm P}({\bf r})\right|
\left|\phi_\xi^{\rm P}({\bf r})\right|\ret
&\le&2(n_{\rm max}+1)\ell^\delta+\epsilon^{(1)}
(\ell^\delta-1,n_{\rm max},L_x,L_y). 
\end{eqnarray}
Using this inequality and Lemma~\ref{>intphiphiLemma} again, we obtain 
\begin{eqnarray}
\Delta E_{{\rm I},\ge,2,i}^{(N)}&\le&(n_{\rm max}+1)q\ell
\epsilon^{(1)}(\ell^\delta/4-1,n_{\rm max},L_x,L_y)\ret
&\times&\left[2(n_{\rm max}+1)\ell^\delta+\epsilon^{(1)}
(\ell^\delta-1,n_{\rm max},L_x,L_y)\right]
\end{eqnarray}
for $i=1,2,3,4$. Substituting these into (\ref{DEI>2bound2}), we get 
the desired bound (\ref{DEI>2bound}). 

\subsection{Estimate of $\Delta E_{\rm II}^{(N)}$}
\label{EstDEII}

For $\Delta E_{\rm II}^{(N)}$ of (\ref{DEII}), one can easily obtain 
\begin{eqnarray}
\left|\Delta E_{\rm II}^{(N)}\right|&\le&
\frac{1}{2}\sum_{\{\xi_j\}}\sum_{s=1}^N\sum_{t\ne s}
\sum_{\xi_s'}\sum_{\xi_t'}
\left|a^\ast(\{\xi_j\})a(\{\xi_j''\})\right|\ret
&\times&
\left\{1-\cos\left[\frac{2\pi}{\ell}({\tilde m}(m_s')-{\tilde m}(m_s)
+{\tilde m}(m_t')-{\tilde m}(m_t))\right]\right\}\ret
&\times&\left|\int dx_1dy_1\int dx_2dy_2\ 
{\phi_{\xi_s}^{\rm P}}^\ast({\bf r}_1)
{\phi_{\xi_t}^{\rm P}}^\ast({\bf r}_2)
U^{(2)}(x_1-x_2,y_1-y_2)\phi_{\xi_s'}^{\rm P}({\bf r}_1)
\phi_{\xi_t'}^{\rm P}({\bf r}_2)\right|,\ret
\label{DEIIbound}
\end{eqnarray}
where we have written 
\begin{equation}
\{\xi_j''\}=\{\xi_1,\ldots,\xi_{s-1},\xi_s',\xi_{s+1},\ldots,\xi_{t-1},\xi_t',
\xi_{t+1},\ldots,\xi_N\}. 
\end{equation}
The right-hand side of (\ref{DEIIbound}) can be decomposed into the following 
two parts: 
\begin{eqnarray}
\Delta {\tilde E}_{{\rm II},<}^{(N)}&=&
\frac{1}{2}\sum_{\{\xi_j\}}\sum_{s=1}^N\sum_{t\ne s}
\sum_{\xi_s'}\sum_{\xi_t'}
\left|a^\ast(\{\xi_j\})a(\{\xi_j''\})\right|\ret
&\times&\left\{1-\cos\left[\frac{2\pi}{\ell}({\tilde m}(m_s')-{\tilde m}(m_s)
+{\tilde m}(m_t')-{\tilde m}(m_t))\right]\right\}\ret
&\times&\chi({\rm dist}^{(m)}(m_s,m_s')<\ell^\delta/2)
\chi({\rm dist}^{(m)}(m_t,m_t')<\ell^\delta/2)\ret
&\times&\left|\int dx_1dy_1\int dx_2dy_2\ 
{\phi_{\xi_s}^{\rm P}}^\ast({\bf r}_1)
{\phi_{\xi_t}^{\rm P}}^\ast({\bf r}_2)
U^{(2)}(x_1-x_2,y_1-y_2)\phi_{\xi_s'}^{\rm P}({\bf r}_1)
\phi_{\xi_t'}^{\rm P}({\bf r}_2)
\right|,\ret
\label{DtildeEII<}
\end{eqnarray}
and
\begin{eqnarray}
\Delta {\tilde E}_{{\rm II},>}^{(N)}&=&
\frac{1}{2}\sum_{\{\xi_j\}}\sum_{s=1}^N\sum_{t\ne s}
\sum_{\xi_s'}\sum_{\xi_t'}
\left|a^\ast(\{\xi_j\})a(\{\xi_j''\})\right|\ret
&\times&\left\{1-\cos\left[\frac{2\pi}{\ell}({\tilde m}(m_s')-{\tilde m}(m_s)
+{\tilde m}(m_t')-{\tilde m}(m_t))\right]\right\}\ret
&\times&[1-\chi({\rm dist}^{(m)}(m_s,m_s')<\ell^\delta/2)
\chi({\rm dist}^{(m)}(m_t,m_t')<\ell^\delta/2)]\ret
&\times&\left|\int dx_1dy_1\int dx_2dy_2\ 
{\phi_{\xi_s}^{\rm P}}^\ast({\bf r}_1)
{\phi_{\xi_t}^{\rm P}}^\ast({\bf r}_2)
U^{(2)}(x_1-x_2,y_1-y_2)\phi_{\xi_s'}^{\rm P}({\bf r}_1)
\phi_{\xi_t'}^{\rm P}({\bf r}_2)
\right|.\ret
\label{DtildeEII>}
\end{eqnarray}

\subsubsection{Estimate of $\Delta {\tilde E}_{{\rm II},<}^{(N)}$}
\label{SecEstDEII<}

As we will show in the following, we obtain 
\begin{equation}
\Delta {\tilde E}_{{\rm II},<}^{(N)}\le 4\pi^2q(n_{\rm max}+1)^4
{\cal C}^{(1)}(U^{(2)})\frac{\ell^{4\delta}}{\ell}.
\label{DEII<bound}
\end{equation}
for $\Delta {\tilde E}_{{\rm II},<}^{(N)}$ of (\ref{DtildeEII<}). 
Since $\delta\in(0,1/4)$, $\Delta {\tilde E}_{{\rm II},<}^{(N)}$ is 
vanishing in the limit $\ell\rightarrow\infty$. 

In order to show the bound (\ref{DEII<bound}), 
consider first the integral in the right-hand side of (\ref{DtildeEII<}). 
Using the Schwarz inequality, we have 
\begin{eqnarray}
& &\left|\int dx_1dy_1\int dx_2dy_2\ 
{\phi_{\xi_s}^{\rm P}}^\ast({\bf r}_1)
{\phi_{\xi_t}^{\rm P}}^\ast({\bf r}_2)
U^{(2)}(x_1-x_2,y_1-y_2)\phi_{\xi_s'}^{\rm P}({\bf r}_1)
\phi_{\xi_t'}^{\rm P}({\bf r}_2)
\right|\ret
&\le&\sqrt{\int dx_1dy_1\int dx_2dy_2\ 
\left|\phi_{\xi_s}^{\rm P}({\bf r}_1)\right|^2
\left|U^{(2)}(x_1-x_2,y_1-y_2)\right|
\left|\phi_{\xi_t}^{\rm P}({\bf r}_2)\right|^2}\ret
&\times&
\sqrt{\int dx_1dy_1\int dx_2dy_2\ 
\left|\phi_{\xi_s'}^{\rm P}({\bf r}_1)\right|^2
\left|U^{(2)}(x_1-x_2,y_1-y_2)\right|
\left|\phi_{\xi_t'}^{\rm P}({\bf r}_2)\right|^2}. 
\end{eqnarray}
Thereby 
$$
\quad\
\left|a^\ast(\{\xi_j\})a(\{\xi_j''\})
\int dx_1dy_1\int dx_2dy_2\ 
{\phi_{\xi_s}^{\rm P}}^\ast({\bf r}_1)
{\phi_{\xi_t}^{\rm P}}^\ast({\bf r}_2)
U^{(2)}(x_1-x_2,y_1-y_2)\phi_{\xi_s'}^{\rm P}({\bf r}_1)
\phi_{\xi_t'}^{\rm P}({\bf r}_2)
\right|
$$
\begin{eqnarray}
&\le&\frac{1}{2}\left[\left|a(\{\xi_j\})\right|^2\int dx_1dy_1\int dx_2dy_2\ 
\left|\phi_{\xi_s}^{\rm P}({\bf r}_1)\right|^2
\left|U^{(2)}(x_1-x_2,y_1-y_2)\right|
\left|\phi_{\xi_t}^{\rm P}({\bf r}_2)\right|^2
\right.\ret
& &+\left.\left|a(\{\xi_j''\})\right|^2\int dx_1dy_1\int dx_2dy_2\ 
\left|\phi_{\xi_s'}^{\rm P}({\bf r}_1)\right|^2
\left|U^{(2)}(x_1-x_2,y_1-y_2)\right|
\left|\phi_{\xi_t'}^{\rm P}({\bf r}_2)\right|^2\right].\qquad\quad
\end{eqnarray}
Substituting this into (\ref{DtildeEII<}), we get 
\begin{eqnarray}
\Delta {\tilde E}_{{\rm II},<}^{(N)}&\le&
\frac{1}{2}\sum_{\{\xi_j\}}\sum_{s=1}^N\sum_{t\ne s}
\sum_{\xi_s'}\sum_{\xi_t'}
\left|a(\{\xi_j\})\right|^2\ret
&\times&\left\{1-\cos\left[\frac{2\pi}{\ell}({\tilde m}(m_s')-{\tilde m}(m_s)
+{\tilde m}(m_t')-{\tilde m}(m_t))\right]\right\}\ret
&\times&\chi({\rm dist}^{(m)}(m_s,m_s')<\ell^\delta/2)
\chi({\rm dist}^{(m)}(m_t,m_t')<\ell^\delta/2)\ret
&\times&
\int dx_1dy_1\int dx_2dy_2\ 
\left|\phi_{\xi_s}^{\rm P}({\bf r}_1)\right|^2
\left|U^{(2)}(x_1-x_2,y_1-y_2)\right|
\left|\phi_{\xi_t}^{\rm P}({\bf r}_2)\right|^2
\end{eqnarray}
Here, if all of $m_s,m_s',m_t,m_t'$ are not in the interval $[1,q\ell]$, 
then the corresponding contributions are vanishing from 
the definitions (\ref{deftildem}) of ${\tilde m}(\cdots)$ and 
the cosine function. {From} this observation, we have 
\begin{eqnarray}
& &\Delta {\tilde E}_{{\rm II},<}^{(N)}\ret
&\le&\frac{1}{2}\sum_{\{\xi_j\}}\sum_{s=1}^N\sum_{t\ne s}
\sum_{\xi_s'}\sum_{\xi_t'}
\left|a(\{\xi_j\})\right|^2\ret
&\times&\left\{1-\cos\left[\frac{2\pi}{\ell}({\tilde m}(m_s')-{\tilde m}(m_s)
+{\tilde m}(m_t')-{\tilde m}(m_t))\right]\right\}\ret
&\times&\chi({\rm dist}^{(m)}(m_s,m_s')<\ell^\delta/2)
\chi({\rm dist}^{(m)}(m_t,m_t')<\ell^\delta/2)\ret
&\times&\left\{\chi(1\le m_s\le q\ell)+[1-\chi(1\le m_s\le q\ell)]
\chi(1\le m_t\le q\ell)\right.\ret
&+&[1-\chi(1\le m_s\le q\ell)][1-\chi(1\le m_t\le q\ell)]
\chi(1\le m_s'\le q\ell)[1-\chi(1\le m_t'\le q\ell)]\ret
&+&[1-\chi(1\le m_s\le q\ell)][1-\chi(1\le m_t\le q\ell)]
[1-\chi(1\le m_s'\le q\ell)]\left.\chi(1\le m_t'\le q\ell)\right\}
\ret&\times&\int dx_1dy_1\int dx_2dy_2\ 
\left|\phi_{\xi_s}^{\rm P}({\bf r}_1)\right|^2
\left|U^{(2)}(x_1-x_2,y_1-y_2)\right|
\left|\phi_{\xi_t}^{\rm P}({\bf r}_2)\right|^2. 
\label{DEIIbound2}
\end{eqnarray}
Note that
\begin{eqnarray}
& &\frac{1}{2}\sum_{\{\xi_j\}}\sum_{s=1}^N\sum_{t\ne s}
\sum_{\xi_s'}\sum_{\xi_t'}
\left|a(\{\xi_j\})\right|^2
\left\{1-\cos\left[\frac{2\pi}{\ell}({\tilde m}(m_s')-{\tilde m}(m_s)
+{\tilde m}(m_t')-{\tilde m}(m_t))\right]\right\}\ret
&\times&\chi({\rm dist}^{(m)}(m_s,m_s')<\ell^\delta/2)
\chi({\rm dist}^{(m)}(m_t,m_t')<\ell^\delta/2)\ret
&\times&\chi(1\le m_s\le q\ell)
\int dx_1dy_1\int dx_2dy_2\ 
\left|\phi_{\xi_s}^{\rm P}({\bf r}_1)\right|^2
\left|U^{(2)}(x_1-x_2,y_1-y_2)\right|
\left|\phi_{\xi_t}^{\rm P}({\bf r}_2)\right|^2\ret
&\le&
\pi^2\frac{\ell^{2\delta}}{\ell^2}
\sum_{\{\xi_j\}}\sum_{s=1}^N\sum_{t\ne s}
\sum_{\xi_s'}\sum_{\xi_t'}
\left|a(\{\xi_j\})\right|^2
\chi({\rm dist}^{(m)}(m_s,m_s')<\ell^\delta/2)\ret
&\times&\chi({\rm dist}^{(m)}(m_t,m_t')<\ell^\delta/2)
\chi(1\le m_s\le q\ell)\ret
&\times&\int dx_1dy_1\int dx_2dy_2\ 
\left|\phi_{\xi_s}^{\rm P}({\bf r}_1)\right|^2
\left|U^{(2)}(x_1-x_2,y_1-y_2)\right|
\left|\phi_{\xi_t}^{\rm P}({\bf r}_2)\right|^2\ret
&\le&
\pi^2(n_{\rm max}+1)^2\frac{\ell^{4\delta}}{\ell^2}
\sum_{\{\xi_j\}}\sum_{s=1}^N\sum_{t\ne s}
\left|a^\ast(\{\xi_j\})\right|^2\chi(1\le m_s\le q\ell)\ret
&\times&\int dx_1dy_1\int dx_2dy_2\ 
\left|\phi_{\xi_s}^{\rm P}({\bf r}_1)\right|^2
\left|U^{(2)}(x_1-x_2,y_1-y_2)\right|
\left|\phi_{\xi_t}^{\rm P}({\bf r}_2)\right|^2\ret
&\le&
\pi^2(n_{\rm max}+1)^3{\cal C}^{(1)}(U^{(2)})\frac{\ell^{4\delta}}{\ell^2}
\sum_{\{\xi_j\}}\sum_{s=1}^N
\left|a^\ast(\{\xi_j\})\right|^2\chi(1\le m_s\le q\ell)\ret
&\le&
\pi^2q(n_{\rm max}+1)^4{\cal C}^{(1)}(U^{(2)})\frac{\ell^{4\delta}}{\ell},
\end{eqnarray}
where we have used Lemma~\ref{U2avlemma} for getting the third inequality. 
Similarly we have 
\begin{eqnarray}
& &\frac{1}{2}\sum_{\{\xi_j\}}\sum_{s=1}^N\sum_{t\ne s}
\sum_{\xi_s'}\sum_{\xi_t'}
\left|a^\ast(\{\xi_j\})\right|^2
\left\{1-\cos\left[\frac{2\pi}{\ell}({\tilde m}(m_s')-{\tilde m}(m_s)
+{\tilde m}(m_t')-{\tilde m}(m_t))\right]\right\}\ret
&\times&\chi({\rm dist}^{(m)}(m_s,m_s')<\ell^\delta/2)
\chi({\rm dist}^{(m)}(m_t,m_t')<\ell^\delta/2)\ret
&\times&[1-\chi(1\le m_s\le q\ell)][1-\chi(1\le m_t\le q\ell)]
\chi(1\le m_s'\le q\ell)[1-\chi(1\le m_t'\le q\ell)]\ret
&\times&\int dx_1dy_1\int dx_2dy_2\ 
\left|\phi_{\xi_s}^{\rm P}({\bf r}_1)\right|^2
\left|U^{(2)}(x_1-x_2,y_1-y_2)\right|
\left|\phi_{\xi_t}^{\rm P}({\bf r}_2)\right|^2\ret
&\le&\pi^2\frac{\ell^{2\delta}}{\ell^2}
\sum_{\{\xi_j\}}\sum_{s=1}^N\sum_{t\ne s}
\sum_{\xi_s'}\sum_{\xi_t'}
\left|a^\ast(\{\xi_j\})\right|^2
\chi({\rm dist}^{(m)}(m_s,m_s')<\ell^\delta/2)\ret
&\times&\chi({\rm dist}^{(m)}(m_t,m_t')<\ell^\delta/2)\ret
&\times&[1-\chi(1\le m_s\le q\ell)][1-\chi(1\le m_t\le q\ell)]
\chi(1\le m_s'\le q\ell)[1-\chi(1\le m_t'\le q\ell)]\ret
&\times&\int dx_1dy_1\int dx_2dy_2\ 
\left|\phi_{\xi_s}^{\rm P}({\bf r}_1)\right|^2
\left|U^{(2)}(x_1-x_2,y_1-y_2)\right|
\left|\phi_{\xi_t}^{\rm P}({\bf r}_2)\right|^2\ret
&\le&\pi^2(n_{\rm max}+1)\frac{\ell^{3\delta}}{\ell^2}
\sum_{\{\xi_j\}}\sum_{s=1}^N\sum_{t\ne s}
\sum_{\xi_s'}\left|a^\ast(\{\xi_j\})\right|^2
\chi({\rm dist}^{(m)}(m_s,m_s')<\ell^\delta/2)\ret
&\times&[1-\chi(1\le m_s\le q\ell)]\chi(1\le m_s'\le q\ell)\ret
&\times&\int dx_1dy_1\int dx_2dy_2\ 
\left|\phi_{\xi_s}^{\rm P}({\bf r}_1)\right|^2
\left|U^{(2)}(x_1-x_2,y_1-y_2)\right|
\left|\phi_{\xi_t}^{\rm P}({\bf r}_2)\right|^2\ret
&\le&\pi^2(n_{\rm max}+1)^2{\cal C}^{(1)}(U^{(2)})
\frac{\ell^{3\delta}}{\ell^2}
\sum_{\{\xi_j\}}\sum_{s=1}^N\sum_{\xi_s'}\left|a^\ast(\{\xi_j\})\right|^2
\ret&\times&\chi({\rm dist}^{(m)}(m_s,m_s')<\ell^\delta/2)
[1-\chi(1\le m_s\le q\ell)]\chi(1\le m_s'\le q\ell).
\label{est3}
\end{eqnarray}
Note that 
\begin{eqnarray}
& &\sum_{s=1}^N\sum_{\xi_s'}\chi({\rm dist}^{(m)}(m_s,m_s')<\ell^\delta/2)
[1-\chi(1\le m_s\le q\ell)]\chi(1\le m_s'\le q\ell)\ret
&\le&\sum_{s=1}^N\sum_{\xi_s'}
\left[\chi(-\ell^\delta/2+1<m_s\le 0)+\chi(q\ell+1\le m_s<\ell^\delta/2+q\ell)
\right]\ret
&\times&\left[\chi(1\le m_s'<\ell^\delta/2)+\chi(q\ell+1-\ell^\delta/2
<m_s'\le q\ell)\right]\ret
&\le&(n_{\rm max}+1)^2\ell^{2\delta}. 
\end{eqnarray}
Substituting this into (\ref{est3}), we get 
\begin{eqnarray}
& &\frac{1}{2}\sum_{\{\xi_j\}}\sum_{s=1}^N\sum_{t\ne s}
\sum_{\xi_s'}\sum_{\xi_t'}
\left|a^\ast(\{\xi_j\})\right|^2
\left\{1-\cos\left[\frac{2\pi}{\ell}({\tilde m}(m_s')-{\tilde m}(m_s)
+{\tilde m}(m_t')-{\tilde m}(m_t))\right]\right\}\ret
&\times&\chi({\rm dist}^{(m)}(m_s,m_s')<\ell^\delta/2)
\chi({\rm dist}^{(m)}(m_t,m_t')<\ell^\delta/2)\ret
&\times&[1-\chi(1\le m_s\le q\ell)][1-\chi(1\le m_t\le q\ell)]
\chi(1\le m_s'\le q\ell)[1-\chi(1\le m_t'\le q\ell)]\ret
&\times&\int dx_1dy_1\int dx_2dy_2\ 
\left|\phi_{\xi_s}^{\rm P}({\bf r}_1)\right|^2
\left|U^{(2)}(x_1-x_2,y_1-y_2)\right|
\left|\phi_{\xi_t}^{\rm P}({\bf r}_2)\right|^2\ret
&\le&\pi^2(n_{\rm max}+1)^4{\cal C}^{(1)}(U^{(2)})
\frac{\ell^{5\delta}}{\ell^2}.
\end{eqnarray}
Since the rest of the contributions of the right-hand side of 
(\ref{DEIIbound2}) are treated in the same way, we obtain 
the bound (\ref{DEII<bound}).  

\subsubsection{Estimate of $\Delta {\tilde E}_{{\rm II},>}^{(N)}$}

Our goal in this subsection is to get the following bound: 
For $\pi(\ell^\delta/2-1)/L_x>n_{\rm max}$ and $L_y>32n_{\rm max}\ell_B$, 
\begin{equation}
\Delta{\tilde E}_{{\rm II},>}^{(N)}
\le 8(n_{\rm max}+1)q\ell\epsilon^{(1)}(\ell^\delta/2-1,n_{\rm max},L_x,L_y)
\kappa(n_{\rm max},L_x,L_y),
\label{boundDEII>}
\end{equation}
where 
\begin{eqnarray}
\kappa(n_{\rm max},L_x,L_y)&:=&\left\{
{\cal C}^{(7)}(n_{\rm max})+{\cal C}^{(8)}(n_{\rm max})\frac{L_x}{\ell_B}
\right.
\ret
&+&\left.{\cal C}^{(9)}(n_{\rm max})\frac{L_xL_y}{\ell_B^2}
\exp\left[-\frac{L_y^2}{32\ell_B^2}\left(1-\frac{32n_{\rm max}\ell_B}{L_y}
\right)^2\right]\right\}C^{(10)}(U^{(2)}).\ret
\label{defkappa}
\end{eqnarray}
Here the constants ${\cal C}^{(7)}(n_{\rm max})$, 
${\cal C}^{(8)}(n_{\rm max})$ and ${\cal C}^{(9)}(n_{\rm max})$ depend 
on the energy cutoff $n_{\rm max}$ only, and the constant $C^{(10)}(U^{(2)})$ 
depends on the potential $U^{(2)}$ only. 
{From} (\ref{epsilon1}) and (\ref{defkappa}), we have 
\begin{equation}
\lim_{\ell\rightarrow\infty}\lim_{L_y\rightarrow\infty}
\Delta{\tilde E}_{{\rm II},>}^{(N)}=0
\end{equation}
for a fixed $L_x$. 

Using the definition (\ref{deftildem}) of ${\tilde m}(\cdots)$, 
we can evaluate 
$\Delta{\tilde E}_{{\rm II},>}^{(N)}$ of (\ref{DtildeEII>}) as 
\begin{eqnarray}
\Delta {\tilde E}_{{\rm II},>}^{(N)}&\le&
2\sum_{\{\xi_j\}}\sum_{s=1}^N\sum_{t\ne s}
\sum_{\xi_s'}\sum_{\xi_t'}
\left[\left|a(\{\xi_j\})\right|^2+\left|a(\{\xi_j''\})\right|^2\right]
\chi(1\le m_s\le q\ell)\ret
&\times&[1-\chi({\rm dist}^{(m)}(m_s,m_s')<\ell^\delta/2)
\chi({\rm dist}^{(m)}(m_t,m_t')<\ell^\delta/2)]\ret
&\times&\left|\int dx_1dy_1\int dx_2dy_2\ 
{\phi_{\xi_s}^{\rm P}}^\ast({\bf r}_1)
{\phi_{\xi_t}^{\rm P}}^\ast({\bf r}_2)
U^{(2)}(x_1-x_2,y_1-y_2)\phi_{\xi_s'}^{\rm P}({\bf r}_1)
\phi_{\xi_t'}^{\rm P}({\bf r}_2)
\right|\ret
&\le&2\sum_{i=1}^4\Delta {\tilde E}_{{\rm II},>,i}^{(N)}
\label{DEII>decom}
\end{eqnarray} 
in a similar way as in (\ref{DEIIbound2}) in the preceding 
Section~\ref{SecEstDEII<}. Here 
$$
\Delta {\tilde E}_{{\rm II},>,1}^{(N)}:=
\sum_{\{\xi_j\}}\sum_{s=1}^N\sum_{t\ne s}
\sum_{\xi_s'}\sum_{\xi_t'}\left|a(\{\xi_j\})\right|^2
\chi(1\le m_s\le q\ell)\chi({\rm dist}^{(m)}(m_s,m_s')\ge\ell^\delta/2)
\qquad\quad
$$
\begin{equation}
\qquad\times\int dx_1dy_1\left|{\phi_{\xi_s}^{\rm P}}({\bf r}_1)\right|
\left|\phi_{\xi_s'}^{\rm P}({\bf r}_1)\right|
\int dx_2dy_2\ \left|U^{(2)}(x_1-x_2,y_1-y_2)\right|
\left|{\phi_{\xi_t}^{\rm P}}({\bf r}_2)\right|
\left|\phi_{\xi_t'}^{\rm P}({\bf r}_2)\right|, 
\end{equation}
$$
\Delta {\tilde E}_{{\rm II},>,2}^{(N)}:=
\sum_{\{\xi_j\}}\sum_{s=1}^N\sum_{t\ne s}
\sum_{\xi_s'}\sum_{\xi_t'}\left|a(\{\xi_j\})\right|^2
\chi(1\le m_s\le q\ell)\chi({\rm dist}^{(m)}(m_t,m_t')\ge\ell^\delta/2)
\qquad\quad
$$
\begin{equation}
\qquad\times\int dx_1dy_1\left|{\phi_{\xi_s}^{\rm P}}({\bf r}_1)\right|
\left|\phi_{\xi_s'}^{\rm P}({\bf r}_1)\right|
\int dx_2dy_2\ \left|U^{(2)}(x_1-x_2,y_1-y_2)\right|
\left|{\phi_{\xi_t}^{\rm P}}({\bf r}_2)\right|
\left|\phi_{\xi_t'}^{\rm P}({\bf r}_2)\right|, 
\end{equation}
$$
\Delta {\tilde E}_{{\rm II},>,3}^{(N)}:=
\sum_{\{\xi_j\}}\sum_{s=1}^N\sum_{t\ne s}
\sum_{\xi_s'}\sum_{\xi_t'}\left|a(\{\xi_j''\})\right|^2
\chi(1\le m_s\le q\ell)\chi({\rm dist}^{(m)}(m_s,m_s')\ge\ell^\delta/2)
\qquad\quad
$$
\begin{equation}
\qquad\times\int dx_1dy_1\left|{\phi_{\xi_s}^{\rm P}}({\bf r}_1)\right|
\left|\phi_{\xi_s'}^{\rm P}({\bf r}_1)\right|
\int dx_2dy_2\ \left|U^{(2)}(x_1-x_2,y_1-y_2)\right|
\left|{\phi_{\xi_t}^{\rm P}}({\bf r}_2)\right|
\left|\phi_{\xi_t'}^{\rm P}({\bf r}_2)\right|, 
\end{equation}
and
$$
\Delta {\tilde E}_{{\rm II},>,4}^{(N)}:=
\sum_{\{\xi_j\}}\sum_{s=1}^N\sum_{t\ne s}
\sum_{\xi_s'}\sum_{\xi_t'}\left|a(\{\xi_j''\})\right|^2
\chi(1\le m_s\le q\ell)\chi({\rm dist}^{(m)}(m_t,m_t')\ge\ell^\delta/2)
\qquad\quad
$$
\begin{equation}
\qquad\times\int dx_1dy_1\left|{\phi_{\xi_s}^{\rm P}}({\bf r}_1)\right|
\left|\phi_{\xi_s'}^{\rm P}({\bf r}_1)\right|
\int dx_2dy_2\ \left|U^{(2)}(x_1-x_2,y_1-y_2)\right|
\left|{\phi_{\xi_t}^{\rm P}}({\bf r}_2)\right|
\left|\phi_{\xi_t'}^{\rm P}({\bf r}_2)\right|, 
\end{equation}
and we have used 
\begin{eqnarray}
& &1-\chi({\rm dist}^{(m)}(m_s,m_s')<\ell^\delta/2)
\chi({\rm dist}^{(m)}(m_t,m_t')<\ell^\delta/2)\ret
&\le&
\chi({\rm dist}^{(m)}(m_s,m_s')\ge\ell^\delta/2)+
\chi({\rm dist}^{(m)}(m_t,m_t')\ge\ell^\delta/2).
\end{eqnarray}

Consider first $\Delta {\tilde E}_{{\rm II},>,1}^{(N)}$. It can be 
written as 
\begin{eqnarray}
\Delta {\tilde E}_{{\rm II},>,1}^{(N)}&=&
\sum_{\{\xi_j\}}\left|a(\{\xi_j\})\right|^2
\sum_{s=1}^N\chi(1\le m_s\le q\ell)\ret
&\times&\sum_{\xi_s'}\chi({\rm dist}^{(m)}(m_s,m_s')\ge\ell^\delta/2)
\int dx_1dy_1\left|{\phi_{\xi_s}^{\rm P}}({\bf r}_1)\right|
\left|\phi_{\xi_s'}^{\rm P}({\bf r}_1)\right|\ret
&\times&\sum_{t\ne s}\sum_{\xi_t'}
\int dx_2dy_2\ \left|U^{(2)}(x_2-x_1,y_2-y_1)\right|
\left|{\phi_{\xi_t}^{\rm P}}({\bf r}_2)\right|
\left|\phi_{\xi_t'}^{\rm P}({\bf r}_2)\right|. 
\label{expDtildeEII>1}
\end{eqnarray}
In order to evaluate the last two sums, we use the following lemma: 

\begin{lemma}
\label{intUphiphibound}
Let $L_y>32n_{\rm max}\ell_B$. Then 
\begin{equation}
\sum_\xi\sum_{\xi'}\int_Sdxdy\ \left|U^{(2)}(x-x',y-y')\right|
\left|\phi_\xi^{\rm P}(x,y)\right|\left|\phi_{\xi'}^{\rm P}(x,y)\right|
\le\kappa(n_{\rm max},L_x,L_y)
\label{intUphiphiinequality}
\end{equation}
for all $x',y'\in {\bf R}$. Here $\kappa$ is given by (\ref{defkappa}). 
\end{lemma}
The proof is given in Appendix~\ref{ProofintUphiphibound}. 
Using this Lemma~\ref{intUphiphibound}, (\ref{expDtildeEII>1}) and 
Lemma~\ref{>intphiphiLemma}, we get 
\begin{eqnarray}
\Delta {\tilde E}_{{\rm II},>,1}^{(N)}
&\le&\sum_{\{\xi_j\}}\left|a(\{\xi_j\})\right|^2
\sum_{s=1}^N\chi(1\le m_s\le q\ell)
\epsilon^{(1)}(\ell^\delta/2-1,n_{\rm max},L_x,L_y)
\kappa(n_{\rm max},L_x,L_y)\ret
&\le&(n_{\rm max}+1)q\ell\epsilon^{(1)}(\ell^\delta/2-1,n_{\rm max},L_x,L_y)
\kappa(n_{\rm max},L_x,L_y) 
\label{DEII>1bound}
\end{eqnarray}
for $\ell$ satisfying $\pi(\ell^\delta/2-1)\ell_B/L_x>n_{\rm max}$. 

Next consider $\Delta {\tilde E}_{{\rm II},>,3}^{(N)}$. 
One can easily get 
\begin{eqnarray}
\Delta {\tilde E}_{{\rm II},>,3}^{(N)}
&=&\sum_{\{\xi_j''\}}\sum_{s=1}^N\sum_{t\ne s}
\sum_{\xi_s}\sum_{\xi_t}\left|a(\{\xi_j''\})\right|^2
\chi(1\le m_s\le q\ell)\chi({\rm dist}^{(m)}(m_s,m_s')\ge\ell^\delta/2)\ret
&\times&\int dx_1dy_1\left|{\phi_{\xi_s}^{\rm P}}({\bf r}_1)\right|
\left|\phi_{\xi_s'}^{\rm P}({\bf r}_1)\right|\ret
&\times&\int dx_2dy_2\ \left|U^{(2)}(x_1-x_2,y_1-y_2)\right|
\left|{\phi_{\xi_t}^{\rm P}}({\bf r}_2)\right|
\left|\phi_{\xi_t'}^{\rm P}({\bf r}_2)\right|\ret 
&\le&\sum_{\{\xi_j''\}}\left|a(\{\xi_j''\})\right|^2
\sum_\xi\chi(1\le m\le q\ell)\ret
&\times&\sum_{s=1}^N
\chi({\rm dist}^{(m)}(m,m_s')\ge\ell^\delta/2)
\int dx_1dy_1\left|{\phi_\xi^{\rm P}}({\bf r}_1)\right|
\left|\phi_{\xi_s'}^{\rm P}({\bf r}_1)\right|\ret
&\times&
\sum_{t=1}^N\sum_{\xi_t}\int dx_2dy_2\ \left|U^{(2)}(x_2-x_1,y_2-y_1)\right|
\left|{\phi_{\xi_t}^{\rm P}}({\bf r}_2)\right|
\left|\phi_{\xi_t'}^{\rm P}({\bf r}_2)\right|.
\end{eqnarray}
Here the sum about $\xi$ is over all the states $\xi=(n,k)=(n,2\pi m/L_x)$ 
with the Landau index $n\le n_{\rm max}$. 
Therefore $\Delta{\tilde E}_{{\rm II},>,3}^{(N)}$ can be treated in the same 
way as $\Delta {\tilde E}_{{\rm II},>,1}^{(N)}$. Moreover the rest 
$\Delta {\tilde E}_{{\rm II},>,2}^{(N)}$ and 
$\Delta {\tilde E}_{{\rm II},>,4}^{(N)}$ also can be treated in the same way. 
In a consequence, we obtain the desired bound (\ref{boundDEII>}) from 
(\ref{DEII>decom}) and (\ref{DEII>1bound}).

\appendix

\section[Proof of Theorem 2.1]{Proof of Theorem~\ref{theoremMatsui}}
\setcounter{equation}{0}
\setcounter{theorem}{0}
\label{MatsuiTheorem}

Following Matsui, we sketch the proof of 
Theorem~\ref{theoremMatsui}. For the detail, see ref.~\cite{Matsui}. 

Let $\Lambda$ be a one-dimensional finite lattice, i.e., 
$\Lambda\subset{\bf Z}$. For simplicity, we assume $|\Lambda|=qL$ 
with positive integers $q,L$. Then there exists a self-adjoint operator 
${\tilde h}_0^{(q)}$, 
i.e., a local Hamiltonian, such that the Hamiltonian (\ref{hamH2Qproinf}) 
can be written as 
\begin{equation}
{\tilde H}_\Lambda(n_{\rm max})=\sum_{m=0}^{L-1}{\tilde h}_{qm}^{(q)}
\end{equation}
in terms of the translate ${\tilde h}_x^{(q)}:=
\tau_x^{(y)}\left({\tilde h}_0^{(q)}\right)$ of the local Hamiltonian, 
with the periodic boundary conditions. We introduce a number operator 
of electron on $q$ lattice sites as 
\begin{equation}
{\tilde n}_x^{(q)}:=\sum_{n=0}^{n_{\rm max}}\sum_{m=1}^q {\tilde n}_{n,x+m}. 
\end{equation}
Here ${\tilde n}_{n,m}$ is given by (\ref{electronnumber}). Thereby the 
Hamiltonian (\ref{hamgrandcano}) with a chemical potential $\mu$ is written as 
\begin{equation}
{\tilde H}_{\Lambda,\mu}(n_{\rm max})=\sum_{m=0}^{L-1}
\left[{\tilde h}_{qm}^{(q)}-\mu{\tilde n}_{qm}^{(q)}\right]
\end{equation}

To begin with, we recall the following two theorems:

\begin{theorem}
\label{GSgrandcano}
Let $\omega$ be a translationally invariant state with a period $q$. 
Then the following two conditions for the grand-canonical emsemble 
with a chemical potential $\mu$ are equivalent: 
\begin{itemize}
\item $\omega$ is a ground state for ${\cal A}_{\rm loc}$.
\item $\omega$ minimizes the local energy in the sense that 
\begin{equation}
\omega({\tilde h}_x^{(q)}-\mu{\tilde n}_x^{(q)})=\inf \psi({\tilde h}_x^{(q)}
-\mu{\tilde n}_x^{(q)}),
\end{equation}
where the $\inf$ is taken over all the translationally invariant 
states. 
\end{itemize}
\end{theorem}

\begin{theorem}
\label{GScano}
Let $\omega$ be a translationally invariant state with a period $q$ and 
with the local density $\omega({\tilde n}_x^{(q)})=\rho$. 
Then the following two conditions for the canonical emsemble are equivalent: 
\begin{itemize}
\item $\omega$ is a ground state for ${\cal A}_{\rm loc}^{U(1)}$.
\item $\omega$ minimizes the local energy in the sense that 
\begin{equation}
\omega({\tilde h}_x^{(q)})=\inf \psi({\tilde h}_x^{(q)}),
\end{equation}
where the $\inf$ is taken over all the translationally invariant 
states with the local density $\psi({\tilde n}_x^{(q)})=\rho$. 
\end{itemize}
\end{theorem}
In order to show these statements, one has only to estimate 
energy effects due to boundary conditions, by relying on 
the Bratteli-Kishimoto-Robinson theorem \cite{BKR}. 
See, for example, ref.~\cite{Matsui}. See also refs.~\cite{BraRob,KomaTasaki}. 

\begin{lemma}
Let $\omega$ be a translationally invariant ground state with an electron 
density $\omega({\tilde n}_x^{(q)})=\rho$ for ${\cal A}_{\rm loc}^{U(1)}$. 
Suppose that, for ${\cal A}_{\rm loc}$, there exists a translationally 
invariant ground state $\eta$ with the same electron density 
$\eta({\tilde n}_x^{(q)})=\rho$ and with a chemical potential $\mu$. 
Then the gauge invarinat extension ${\tilde \omega}$ of $\omega$ to 
${\cal A}_{\rm loc}$ is a ground state for ${\cal A}_{\rm loc}$, 
with the chemical potential $\mu$. 
\end{lemma}

\begin{proof}{Proof}
We note that, for $a\in{\cal A}_{\rm loc}^{U(1)}$, 
\begin{equation}
\lim_{\Lambda\uparrow{\bf Z}}
\eta\left(a^\ast\left[{\tilde H}_{\Lambda,\mu},a\right]\right)
=\lim_{\Lambda\uparrow{\bf Z}}
\eta\left(a^\ast\left[{\tilde H}_\Lambda,a\right]\right)\ge 0
\end{equation}
because the operator $a$ commutes with the total number operator of electron. 
This implies that $\eta$ is a translationally invariant ground state for 
${\cal A}_{\rm loc}^{U(1)}$. Therefore 
\begin{equation}
\eta\left({\tilde h}_x^{(q)}\right)=\omega\left({\tilde h}_x^{(q)}\right) 
\end{equation}
owing to Theorem~\ref{GScano}. Since $\eta$ and $\omega$ have the same 
electron density $\rho$, one has 
\begin{equation}
\eta\left({\tilde h}_x^{(q)}-\mu{\tilde n}_x^{(q)}\right)
=\omega\left({\tilde h}_x^{(q)}-\mu{\tilde n}_x^{(q)}\right). 
\end{equation}
This implies that ${\tilde \omega}$ is a translationally invariant ground 
state for ${\cal A}_{\rm loc}$, from Theorem~\ref{GSgrandcano}. 
\end{proof}

\begin{proof}{Proof of Theorem~\ref{theoremMatsui}}
By this Lemma, it is sufficient to show that, 
for any given electron density $\rho$, 
there exists a chemical potential $\mu$ such that 
a ground state $\eta$ for ${\cal A}_{\rm loc}$ with $\mu$ 
has the density $\rho$. 

Let ${\tilde \Phi}_{\Lambda,\mu}$ be a ground state of the Hamiltonian 
${\tilde H}_{\Lambda,\mu}$ with a chemical potential $\mu$ 
such that the corresponding expectation 
$\eta_{\Lambda,\mu}(\cdots)=\left\langle{\tilde \Phi}_{\Lambda,\mu},
(\cdots){\tilde \Phi}_{\Lambda,\mu}\right\rangle$ is translationally 
invariant. Then the corresponding infinite-volume state 
$\eta_\mu=w^\ast\mbox{-}\lim_{\Lambda\uparrow{\bf Z}}\eta_{\Lambda,\mu}$ 
is a translationally invariant ground state with the chemical 
potential $\mu$ for ${\cal A}_{\rm loc}$. From Theorem~\ref{GSgrandcano}, 
the following two inequalities are valid: 
\begin{equation}
\eta_\mu\left({\tilde h}_x^{(q)}\right)
-\eta_{\mu'}\left({\tilde h}_x^{(q)}\right)
\le\mu\left[\eta_\mu\left({\tilde n}_x^{(q)}\right)-
\eta_{\mu'}\left({\tilde n}_x^{(q)}\right)\right]
\end{equation}
and
\begin{equation}
\eta_{\mu'}\left({\tilde h}_x^{(q)}\right)
-\eta_\mu\left({\tilde h}_x^{(q)}\right)
\le\mu'\left[\eta_{\mu'}\left({\tilde n}_x^{(q)}\right)-
\eta_\mu\left({\tilde n}_x^{(q)}\right)\right]
\end{equation}
for the infinite-volume ground states $\eta_\mu$ and $\eta_{\mu'}$ with 
the chemical potentials $\mu$ and $\mu'$, respectively. 
By adding both sides, one has 
\begin{equation}
0\le(\mu-\mu')\left[\eta_\mu\left({\tilde n}_x^{(q)}\right)-
\eta_{\mu'}\left({\tilde n}_x^{(q)}\right)\right].
\end{equation}
This implies that the electron densty 
$\rho_\mu:=\eta_\mu\left({\tilde n}_x^{(q)}\right)$ is a non-decreasing 
function of the chemical potential $\mu$. As is well known, all 
the discontinuous points of a  non-decreasing function is at most countable. 
Assume that $\mu_0$ is such a discontinous point. Namely, 
\begin{equation}
\eta_{\mu_0}^-=\lim_{\mu\uparrow\mu_0}\eta_\mu\quad, \quad
\eta_{\mu_0}^+=\lim_{\mu\downarrow\mu_0}\eta_\mu
\end{equation}
with 
\begin{equation}
\rho_{\mu_0}^-:=\eta_{\mu_0}^-\left({\tilde n}_x^{(q)}\right)
\ne\eta_{\mu_0}^+\left({\tilde n}_x^{(q)}\right)=:\rho_{\mu_0}^+. 
\end{equation}
Consider the convex combination 
$\eta_{\mu_0}^\lambda:=\lambda\eta_{\mu_0}^-+(1-\lambda)\eta_{\mu_0}^+$ 
with $\lambda\in[0,1]$. Clearly the state $\eta_{\mu_0}^\lambda$ 
is a translationally invariant ground state for ${\cal A}_{\rm loc}$ 
and for any $\lambda\in[0,1]$. Further $\eta_{\mu_0}^\lambda$ 
has the electron density $\lambda\rho_{\mu_0}^-+(1-\lambda)
\rho_{\mu_0}^+$. This continuously interpolates between 
the two densities $\rho_{\mu_0}^-, \rho_{\mu_0}^+$. 
\end{proof}
\section[Proof of Lemma 6.1]{Proof of Lemma~\ref{U2avlemma}}
\setcounter{equation}{0}
\setcounter{theorem}{0}
\label{ProofU2avlemma}

Consider a density function given by 
\begin{equation}
\rho_n(x,y):=\sum_k\left|\phi_{n,k}^{\rm P}(x,y)\right|^2.
\end{equation}
{From} the expression (\ref{phiP}) of $\phi_{n,k}^{\rm P}$, 
this function $\rho_n$ is periodic in both $x$ and $y$ directions as 
\begin{equation}
\rho_n(x,y)=\rho_n(x+\Delta x,y)=\rho_n(x,y+\Delta y).
\label{periodicrhon}
\end{equation}
Here $\Delta x$ and $\Delta y$ are given by (\ref{Deltaxy}). 
Owing to this periodicity, the integral of $\rho_n$ on the unit cell 
$\Delta_{\ell,m}$ becomes 
\begin{equation}
\int_{\Delta_{\ell,m}}dxdy\ \rho_n(x,y)=\frac{1}{M},
\label{integralrhoncell}
\end{equation}
where 
\begin{equation}
\Delta_{\ell,m}:=[x_\ell,x_{\ell+1}]\times[y_m,y_{m+1}]
\end{equation}
with
\begin{equation}
x_\ell=-\frac{L_x}{2}+(\ell-1)\Delta x \quad \mbox{for} \ 
\ell=1,2,\ldots,M
\end{equation}
and
\begin{equation}
y_m=-\frac{L_y}{2}+(m-1)\Delta y \quad \mbox{for} \ 
m=1,2,\ldots,M.
\end{equation}
Clearly we have 
\begin{equation}
\sum_k\int_S dxdy\ \left|U^{(2)}(x-x',y-y')\right|
\left|\phi_{n,k}^{\rm P}(x,y)\right|^2
=\int_S dxdy\ \left|U^{(2)}(x-x',y-y')\right|\rho_n(x,y).
\end{equation}
Combining this, the periodiocity (\ref{periodicrhon}) of $\rho_n$ 
and the periodicity (\ref{periodicU2}) of $U^{(2)}$, 
we can assume 
\begin{equation}
|x'|\le\frac{\Delta x}{2}\quad,\quad
|y'|\le\frac{\Delta y}{2}.
\end{equation}
for showing the statement of Lemma~\ref{U2avlemma}. 

Since the function $U^{(2)}$ is continuous by 
the assumption, we have 
\begin{eqnarray}
\int_{\Delta_{\ell,m}}dxdy \ \left|U^{(2)}(x-x',y-y')\right|\rho_n(x,y)
&=&\left|U^{(2)}(\xi_{\ell,m}-x',\eta_{\ell,m}-y')\right|
\int_{\Delta_{\ell,m}}dxdy \ \rho_n(x,y)\ret
&=&\frac{1}{M}\left|U^{(2)}(\xi_{\ell,m}-x',\eta_{\ell,m}-y')\right|,
\label{localintUrho}
\end{eqnarray}
where $(\xi_{\ell,m},\eta_{\ell,m})$ is a point in the cell $\Delta_{\ell,m}$, 
and we have used (\ref{integralrhoncell}). Thereby 
\begin{eqnarray}
& &\int_S dxdy\ \left|U^{(2)}(x-x',y-y')\right|\rho_n(x,y)\ret
&=&\frac{eB}{h}\sum_{\ell,m}\left|U^{(2)}(\xi_{\ell,m}-x',\eta_{\ell,m}-y')
\right|\Delta x\Delta y \ret
&=&\frac{eB}{h}\sum_{r_{\ell,m}\le R}
\left|U^{(2)}(\xi_{\ell,m}-x',\eta_{\ell,m}-y')\right|\Delta x\Delta y\ret
&+&\frac{eB}{h}\sum_{r_{\ell,m}>R}
\left|U^{(2)}(\xi_{\ell,m}-x',\eta_{\ell,m}-y')\right|
\Delta x\Delta y, 
\end{eqnarray}
where $r_{\ell,m}=\sqrt{\xi_{\ell,m}^2+\eta_{\ell,m}^2}$. 
The first sum in the right-hand side of the second equality converges to 
\begin{equation}
\int_{x^2+y^2\le R^2}dxdy\ \left|U^{(2)}(x,y)\right|
\end{equation}
as $L_x,L_y\rightarrow +\infty$ for a fixed $R$ 
because $\left|U^{(2)}\right|$ is uniformly 
continuous. The second sum becomes small 
for a large $R$ from the assumption (\ref{decayU2bound}) of $U^{(2)}$ 
on the decay for a large distance. From these observations, we get 
\begin{equation}
\sum_n \int_S dxdy\ \left|U^{(2)}(x-x',y-y')\right|
\rho_n(x,y)\le {\cal C}^{(1)}(U^{(2)})(n_{\rm max}+1). 
\end{equation}
The finite constant ${\cal C}^{(1)}(U^{(2)})$ depends 
only on $U^{(2)}$. 
Thus the statement of Lemma~\ref{U2avlemma} has been proved. 

\section[Proof of Lemma 6.3]{Proof of Lemma~\ref{>intphiphiLemma}}
\setcounter{equation}{0}
\setcounter{theorem}{0}
\label{Proof>intphiphiLemma}

In order to prove Lemma~\ref{>intphiphiLemma}, 
we use the following estimate for the integral in the left-hand side 
of (\ref{>intphiphibound}): 

\begin{lemma}
\label{pro:phiphiintbound}
Let $L_y>32n_{\rm max}\ell_B$, and let $n,n'\le n_{\rm max}$. 
Then the following bound is valid:
\begin{equation}
\int_{-L_x/2}^{L_x/2}dx\int_{-L_y/2}^{L_y/2}dy\ 
\left|\phi_{n',k'}^{\rm P}({\bf r})\right|
\left|\phi_{n,k}^{\rm P}({\bf r})\right|\ret
\le\epsilon^{(2)}({\rm dist}^{(m)}(m,m'),n_{\rm max},L_y), 
\label{phiphiintbound}
\end{equation}
where $k=2\pi m/L_x$, $k'=2\pi m'/L_x$, and 
\begin{eqnarray}
\epsilon^{(2)}(\Delta\ell,n_{\rm max},L_y)
&:=&{\cal C}^{(4)}(n_{\rm max})
\exp\left[-\left(\frac{\pi\ell_B}{L_x}\Delta\ell
-n_{\rm max}\right)^2\right]\ret
&+&{\cal C}^{(5)}(n_{\rm max})
\exp\left[-\frac{L_y^2}{32\ell_B^2}\left(1-\frac{32n_{\rm max}\ell_B}{L_y}
\right)^2\right].
\label{epsilon2}
\end{eqnarray}
Here the constants ${\cal C}^{(4)}(n_{\rm max})$ and 
${\cal C}^{(5)}(n_{\rm max})$ depend on the enenrgy cutoff $n_{\rm max}$ only. 
\end{lemma}
The proof is given in the next Appendix~\ref{Proofpro:phiphiintbound}. 
By using the bound (\ref{phiphiintbound}), we have 
\begin{eqnarray}
& &\sum_{\xi'}\chi({\rm dist}^{(m)}(m,m')
\ge\Delta\ell)\int_Sdxdy\ \left|\phi_{\xi'}^{\rm P}({\bf r})\right|
\left|\phi_\xi^{\rm P}({\bf r})\right|\ret
&\le&2(n_{\rm max}+1){\cal C}^{(4)}(n_{\rm max})\sum_{\ell=\Delta\ell}^\infty 
\exp\left[-\left(\frac{\pi\ell_B}{L_x}\ell-n_{\rm max}\right)^2\right]\ret
&+&(n_{\rm max}+1)\frac{L_xL_y}{2\pi\ell_B^2}
{\cal C}^{(5)}(n_{\rm max})
\exp\left[-\frac{L_y^2}{32\ell_B^2}\left(1-\frac{32n_{\rm max}\ell_B}{L_y}
\right)^2\right]. 
\label{B3}
\end{eqnarray}
The sum in the right-hand side is evaluated as 
\begin{eqnarray}
& &\sum_{\ell=\Delta\ell}^\infty 
\exp\left[-\left(\frac{\pi\ell_B}{L_x}\ell-n_{\rm max}\right)^2\right]\ret
&\le&\int_{\Delta\ell-1}^\infty d\ell
\exp\left[-\left(\frac{\pi\ell_B}{L_x}\ell-n_{\rm max}\right)^2\right]\ret
&\le&\int_0^\infty d\ell
\exp\left[-\left\{\frac{\pi\ell_B}{L_x}(\ell+\Delta\ell-1)
-n_{\rm max}\right\}^2\right]\ret
&\le&\int_0^\infty d\ell
\exp\left[-\frac{\pi^2\ell_B^2}{L_x^2}\ell^2\right]
\exp\left[-\left\{\frac{\pi\ell_B}{L_x}(\Delta\ell-1)
-n_{\rm max}\right\}^2\right]\ret
&=&\frac{1}{2\sqrt{\pi}}\frac{L_x}{\ell_B}
\exp\left[-\left\{\frac{\pi\ell_B}{L_x}(\Delta\ell-1)
-n_{\rm max}\right\}^2\right].
\end{eqnarray}
Here we have used the assumption $\pi(\Delta \ell -1)\ell_B/L_x>n_{\rm max}$ 
of Lemma~\ref{>intphiphiLemma}. 
Substituting this into (\ref{B3}), we obtain the desired bound 
(\ref{>intphiphibound}) with (\ref{epsilon1}).

\section[Proof of Lemma C.1]{Proof of Lemma~\ref{pro:phiphiintbound}}
\setcounter{equation}{0}
\setcounter{theorem}{0}
\label{Proofpro:phiphiintbound}

Throughout the present Appendix we assume $L_y>32n_{\rm max}\ell_B$ which 
is the assumption of Lemma~\ref{pro:phiphiintbound}. 

Using the expression (\ref{phiP}) of $\phi_{n,k}^{\rm P}$, we evaluate 
the integral of the left-hand side of (\ref{phiphiintbound}) as 
\begin{eqnarray}
& &\int_{-L_x/2}^{L_x/2}dx\int_{-L_y/2}^{L_y/2}dy
\left|\phi_{n',k'}^{\rm P}({\bf r})\right|
\left|\phi_{n,k}^{\rm P}({\bf r})\right|\ret
&\le&\sum_{\ell,\ell'}\int_{-L_y/2}^{L_y/2}dy
\left|v_{n'}(y-y_{k'}-\ell'L_y)\right|
\left|v_n(y-y_k-\ell L_y)\right|\ret
&\le&\sum_{\ell,\ell'}\int_{-L_y/2}^{L_y/2}d{\tilde y}
\left|v_{n'}\left({\tilde y}+\frac{y_k-y_{k'}}{2}-\ell'L_y\right)\right|
\left|v_n\left({\tilde y}-\frac{y_k-y_{k'}}{2}-\ell L_y\right)\right|,
\label{basicvvbound}
\end{eqnarray}
where we have used the periodicity of the integrand, and 
have changed the variable as 
\begin{equation}
{\tilde y}=y-\frac{y_k+y_{k'}}{2} 
\end{equation}
for getting the second inequality. From the right-hand side of 
the first inequality, we can assume $|y_k-y_{k'}|\le L_y/2$ 
without loss of generality.

\begin{lemma}
\label{sumvest}
Let $|y|\le 3L_y/4$. Then 
\begin{equation}
\sum_{\ell=1}^\infty\left|v_n(y\pm\ell L_y)\right|
\le {\cal C}^{(6)}(n_{\rm max})
\exp\left[-\frac{L_y^2}{32\ell_B^2}\left(1-\frac{32n_{\rm max}\ell_B}{L_y}
\right)^2\right], 
\end{equation}
where the constant is given by 
\begin{equation}
{\cal C}^{(6)}(n_{\rm max}):=
\left(1+\frac{\sqrt{2\pi}}{16n_{\rm max}}\right)
\max_{n\le n_{\rm max}}\left\{c_nN_n\exp[{32\beta_n^2}]\right\}. 
\end{equation}
\end{lemma}

\begin{proof}{Proof}
Using the bound (\ref{Hnbound}) for the Hermite polynomial $H_n$ and 
the assumption $|y|\le 3L_y/4$, we have 
\begin{eqnarray}
\sum_{\ell=1}^\infty \left|v_n(y\pm\ell L_y)\right|&\le&
c_nN_n\sum_{\ell=1}^\infty\exp[\beta_n(\ell+3/4)L_y/\ell_B]
\exp\left[-\frac{\left(\ell-3/4\right)^2L_y^2}{2\ell_B^2}\right]\ret
&\le&
c_nN_n\sum_{\ell=1}^\infty\exp\left[\frac{2\beta_n\ell L_y}{\ell_B}\right]
\exp\left[-\frac{\ell^2L_y^2}{32\ell_B^2}\right]\ret
&=&c_nN_n\exp\left[32\beta_n^2\right]\sum_{\ell=1}^\infty
\exp\left[-\frac{L_y^2}{32\ell_B^2}\left(\ell-\frac{32\beta_n\ell_B}{L_y}
\right)^2\right].
\label{C5}
\end{eqnarray}
Here we have used 
\begin{equation}
\ell-\frac{3}{4}\ge \frac{\ell}{4},\quad \mbox{and}\quad 
\ell+\frac{3}{4}\le 2\ell
\end{equation}
for getting the second inequality. The sum in the last line of (\ref{C5}) 
is evaluated as 
\begin{eqnarray}
& &\sum_{\ell=1}^\infty
\exp\left[-\frac{L_y^2}{32\ell_B^2}\left(\ell-\frac{32\beta_n\ell_B}{L_y}
\right)^2\right]\ret&\le&
\exp\left[-\frac{L_y^2}{32\ell_B^2}\left(1-\frac{32\beta_n\ell_B}{L_y}
\right)^2\right]+\int_0^\infty dy
\exp\left[-\frac{L_y^2}{32\ell_B^2}\left(y+1-\frac{32\beta_n\ell_B}{L_y}
\right)^2\right]\ret
&\le&\exp\left[-\frac{L_y^2}{32\ell_B^2}\left(1-\frac{32\beta_n\ell_B}{L_y}
\right)^2\right]+\frac{1}{2}\sqrt{\frac{32\ell_B^2\pi}{L_y^2}}
\exp\left[-\frac{L_y^2}{32\ell_B^2}\left(1-\frac{32\beta_n\ell_B}{L_y}
\right)^2\right]\ret
&=&\left(1+2\sqrt{2\pi}\frac{\ell_B}{L_y}\right)
\exp\left[-\frac{L_y^2}{32\ell_B^2}\left(1-\frac{32\beta_n\ell_B}{L_y}
\right)^2\right]\ret
&\le&\left(1+\frac{\sqrt{2\pi}}{16n_{\rm max}}\right)
\exp\left[-\frac{L_y^2}{32\ell_B^2}\left(1-\frac{32\beta_n\ell_B}{L_y}
\right)^2\right]
\end{eqnarray}
by using the assumption $L_y>32n_{\rm max}\ell_B\ge 32\beta_n\ell_B$.
\end{proof}

Using the bound (\ref{Hnbound}) for the Hermite polynomial $H_n$, we have 
\begin{eqnarray}
\int_{-\infty}^{+\infty}dy\left|v_n(y)\right|
&\le&c_nN_n\int_{-\infty}^{+\infty}dy
\exp\left[\frac{\beta_n|y|}{\ell_B}\right]
\exp\left[-\frac{y^2}{2\ell_B^2}\right]\ret
&\le&2c_nN_n\int_0^{+\infty}dy
\exp\left[-\frac{1}{2\ell_B^2}(y-\beta_n\ell_B)^2\right]
\exp\left[\frac{1}{2}\beta_n^2\right]\ret
&\le&2c_nN_n\exp\left[\frac{1}{2}\beta_n^2\right]
\int_{-\infty}^{+\infty}dy
\exp\left[-\frac{y^2}{2\ell_B^2}\right]\ret
&=&2\sqrt{2\pi}c_nN_n\exp\left[\frac{1}{2}\beta_n^2\right]\ell_B.
\end{eqnarray}
{From} this inequality and Lemma~\ref{sumvest}, we have 
\begin{eqnarray}
& &\sum_{\ell,\ell'}\int_{-L_y/2}^{L_y/2}d{\tilde y}
\left|v_{n'}\left({\tilde y}+\frac{y_k-y_{k'}}{2}-\ell'L_y\right)\right|
\left|v_n\left({\tilde y}-\frac{y_k-y_{k'}}{2}-\ell L_y\right)\right|\ret
&\le&\int_{-L_y/2}^{L_y/2}d{\tilde y}
\left|v_{n'}\left({\tilde y}+\frac{y_k-y_{k'}}{2}\right)\right|
\left|v_n\left({\tilde y}-\frac{y_k-y_{k'}}{2}\right)\right|\ret
&+&\sum_{\ell}\sum_{\ell'\ne 0}\int_{-L_y/2}^{L_y/2}d{\tilde y}
\left|v_{n'}\left({\tilde y}+\frac{y_k-y_{k'}}{2}-\ell'L_y\right)\right|
\left|v_n\left({\tilde y}-\frac{y_k-y_{k'}}{2}-\ell L_y\right)\right|\ret
&+&\sum_{\ell'}\sum_{\ell\ne 0}\int_{-L_y/2}^{L_y/2}d{\tilde y}
\left|v_{n'}\left({\tilde y}+\frac{y_k-y_{k'}}{2}-\ell'L_y\right)\right|
\left|v_n\left({\tilde y}-\frac{y_k-y_{k'}}{2}-\ell L_y\right)\right|\ret
&\le&\int_{-L_y/2}^{L_y/2}d{\tilde y}
\left|v_{n'}\left({\tilde y}+\frac{y_k-y_{k'}}{2}\right)\right|
\left|v_n\left({\tilde y}-\frac{y_k-y_{k'}}{2}\right)\right|\ret
&+&{\cal C}^{(5)}(n_{\rm max})
\exp\left[-\frac{L_y^2}{32\ell_B^2}\left(1-\frac{32n_{\rm max}\ell_B}{L_y}
\right)^2\right],
\label{outvvest}
\end{eqnarray}
where 
\begin{equation}
{\cal C}^{(5)}(n_{\rm max}):=8\sqrt{2\pi}\ell_B
{\cal C}^{(6)}(n_{\rm max})\max_{n\le n_{\rm max}}
\left\{c_nN_n\exp[\beta_n^2/2]\right\}.
\end{equation}
Using the bound (\ref{Hnbound}) for the Hermite polynomial, 
the rest of the integral in (\ref{outvvest}) can be evaluated as 
\begin{eqnarray}
& &\int_{-L_y/2}^{L_y/2}d{\tilde y}\left|v_{n'}(y+\delta y)\right|
\left|v_n(y-\delta y)\right|\ret
&\le&c_nc_{n'}N_nN_{n'}\int_{-L_y/2}^{L_y/2}d{\tilde y}
\exp\left[n_{\rm max}\left(\frac{|{\tilde y}+\delta y|}{\ell_B}
+\frac{|{\tilde y}-\delta y|}{\ell_B}\right)\right]
\exp\left[-\frac{{\tilde y}^2}{\ell_B^2}-\frac{\delta y^2}{\ell_B^2}\right]\ret
&\le&c_nc_{n'}N_nN_{n'}\exp\left[-\frac{\delta y^2}{\ell_B^2}+
2n_{\rm max}\frac{|\delta y|}{\ell_B}\right]
\int_{-L_y/2}^{L_y/2}d{\tilde y}
\exp\left[-\frac{{\tilde y}^2}{\ell_B^2}
+2n_{\rm max}\frac{|{\tilde y}|}{\ell_B}\right]\ret
&\le&2\sqrt{\pi}\ell_B
c_nc_{n'}N_nN_{n'}\exp\left[2n_{\rm max}^2\right]
\exp\left[-\left(\frac{|\delta y|}{\ell_B}-n_{\rm max}\right)^2\right]
\label{mainvv}
\end{eqnarray}
for $\delta y\in {\bf R}$. Combining (\ref{basicvvbound}), (\ref{outvvest}) 
and (\ref{mainvv}), we obtain the desired bound (\ref{phiphiintbound}). 

\section[Proof of Lemma 6.4]{Proof of Lemma~\ref{intUphiphibound}}
\setcounter{equation}{0}
\setcounter{theorem}{0}
\label{ProofintUphiphibound}

Throughout the present Appendix, we assume $L_y>32n_{\rm max}\ell_B$ 
which is the assumption of Lemma~\ref{intUphiphibound}. 

Note that the right-hand side of (\ref{intUphiphiinequality}) 
is written as 
\begin{eqnarray}
& &\sum_\xi\sum_{\xi'}\int_Sdxdy\ \left|U^{(2)}(x-x',y-y')\right|
\left|\phi_\xi^{\rm P}(x,y)\right|\left|\phi_{\xi'}^{\rm P}(x,y)\right|\ret
&=&\sum_{n=0}^{n_{\rm max}}\sum_{n'=0}^{n_{\rm max}}\sum_{\Delta\ell}
\sum_k\int_Sdxdy\ \left|U^{(2)}(x-x',y-y')\right|
\left|\phi_\xi^{\rm P}(x,y)\right|\left|\phi_{\xi'}^{\rm P}(x,y)\right|\ret
&\le&(n_{\rm max}+1)^2\max_{n,n'}\left\{\sum_{\Delta\ell}
\sum_k\int_Sdxdy\ \left|U^{(2)}(x-x',y-y')\right|
\left|\phi_\xi^{\rm P}(x,y)\right|\left|\phi_{\xi'}^{\rm P}(x,y)\right|
\right\},\ret
\label{boundD1}
\end{eqnarray}
where $k'=k+2\pi\Delta\ell/L_x$. In order to estimate the right-hand 
side of the inequality, we use the following lemma which is 
an extension of Lemma~\ref{U2avlemma}: 

\begin{lemma}
Let $n,n'$ be indices of the Landau levels, and let $\Delta\ell$ 
be a positive integer. Then 
\begin{equation}
\sum_k\int_S dxdy\left|U^{(2)}(x-x',y-y')\right|
\left|\phi_{n,k}^{\rm P}({\bf r})\right|
\left|\phi_{n',k'}^{\rm P}({\bf r})\right|\le \epsilon^{(2)}
(\Delta\ell,n_{\rm max},L_x,L_y){\cal C}^{(10)}(U^{(2)})
\label{Urhonn}
\end{equation}
for $x',y'\in{\bf R}$, where $k'=k+2\pi\Delta\ell/L_x$, 
$\epsilon^{(2)}(\Delta\ell,n_{\rm max},L_x,L_y)$ is given by (\ref{epsilon2}), 
and ${\cal C}^{(10)}(U^{(2)})$ is a posiitve constant which depends on 
the potential $U^{(2)}$ only. 
\end{lemma}

\begin{proof}{Proof} 
Consider a density function
\begin{equation}
\rho_{n,n'}({\bf r};\Delta\ell)=\sum_k\left|\phi_{n,k}^{\rm P}({\bf r})\right|
\left|\phi_{n',k'}^{\rm P}({\bf r})\right|.
\end{equation}
Then we have 
\begin{eqnarray}
& &\sum_k\int_S dxdy\left|U^{(2)}(x-x',y-y')\right|
\left|\phi_{n,k}^{\rm P}({\bf r})\right|
\left|\phi_{n',k'}^{\rm P}({\bf r})\right|\ret
&=&\int_S dxdy\left|U^{(2)}(x-x',y-y')\right|\rho_{n,n'}
(x,y;\Delta\ell).
\label{sumkintU2phiphi}
\end{eqnarray}
{From} the expression (\ref{phiP}) of $\phi_{n,k}^{\rm P}$, 
one can notice that this density function is periodic in both 
$x$ and $y$ directions as 
\begin{equation}
\rho_{n,n'}(x+\Delta x,y;\Delta\ell)=\rho_{n,n'}(x,y+\Delta y;\Delta\ell)
=\rho_{n,n'}(x,y;\Delta\ell). 
\label{periodicrho}
\end{equation}
Owing to this property and the periodocity (\ref{periodicU2}) of $U^{(2)}$, 
we can assume 
\begin{equation}
|x'|\le\frac{1}{2}\Delta x\quad, \quad |y'|\le \frac{1}{2}\Delta y
\end{equation}
for evaluating (\ref{sumkintU2phiphi}). Further we have 
\begin{eqnarray}
\int_Sdxdy\rho_{n,n'}(x,y;\Delta\ell)&=&M^2\int_{\Delta_{\ell,m}} dxdy 
\rho_{n,n'}(x,y;\Delta\ell)\ret
&\le&M\max_k\int_Sdxdy
\left|\phi_{n,k}^{\rm P}({\bf r})\right|
\left|\phi_{n',k'}^{\rm P}({\bf r})\right|, 
\label{rhounit}
\end{eqnarray}
where $\Delta_{\ell,m}$ are the unit cells given by 
\begin{equation}
\Delta_{\ell,m}:=[x_\ell,x_{\ell+1}]\times[y_m,y_{m+1}]
\end{equation}
with 
\begin{equation}
x_\ell=-\frac{L_x}{2}+(\ell-1)\Delta x \quad \mbox{for}\ 
\ell=1,2,\ldots,M,
\end{equation}
\begin{equation}
y_m=-\frac{L_y}{2}+(m-1)\Delta y \quad \mbox{for}\
m=1,2,\ldots,M. 
\end{equation}
Since the right-hand side of the inequality in (\ref{rhounit}) 
is evaluated by using Lemma~\ref{pro:phiphiintbound}, we get 
\begin{equation}
\int_{\Delta_{\ell,m}} dxdy\rho_{n,n'}(x,y;\Delta\ell)\le 
\frac{1}{M}\epsilon^{(2)}(\Delta\ell,n_{\rm max},L_x,L_y). 
\end{equation}
{From} this inequality and the assumption that $U^{(2)}$ is continuous, 
we have 
\begin{eqnarray}
& &\int_{\Delta_{\ell,m}}dxdy \left|U^{(2)}(x-x',y-y')\right|
\rho_{n,n'}(x,y;\Delta\ell)\ret
&=&\left|U^{(2)}(\xi_{\ell,m}-x',\eta_{\ell,m}-y')\right|
\int_{\Delta_{\ell,m}}dxdy \rho_{n,n'}(x,y;\Delta\ell)\ret
&\le&\left|U^{(2)}(\xi_{\ell,m}-x',\eta_{\ell,m}-y')\right|
\frac{1}{M}\epsilon^{(2)}(\Delta\ell,n_{\rm max},L_x,L_y),
\end{eqnarray}
where $(\xi_{\ell,m},\eta_{\ell,m})\in\Delta_{\ell,m}$. 
Summing over all $\ell,m$, we obtain 
\begin{eqnarray}
& &\int_Sdxdy \left|U^{(2)}(x-x',y-y')\right|
\rho_{n,n'}(x,y;\Delta\ell)\ret&\le&\frac{eB}{h}
\epsilon^{(2)}(\Delta\ell,n_{\rm max},L_x,L_y)\sum_{\ell,m}
\left|U^{(2)}(\xi_{\ell,m}-x',\eta_{\ell,m}-y')\right|
\Delta x\Delta y. 
\label{U2epsilonCbound}
\end{eqnarray}
We write 
\begin{eqnarray}
& &\sum_{\ell,m}
\left|U^{(2)}(\xi_{\ell,m}-x',\eta_{\ell,m}-y')\right|
\Delta x\Delta y\ret
&=&\mathop{\sum_{\ell,m;}}_{r_{\ell,m}\le R}
\left|U^{(2)}(\xi_{\ell,m}-x',\eta_{\ell,m}-y')\right|
\Delta x\Delta y+
\mathop{\sum_{\ell,m;}}_{r_{\ell,m}>R}
\left|U^{(2)}(\xi_{\ell,m}-x',\eta_{\ell,m}-y')\right|
\Delta x\Delta y,\ret
\label{sumU2decompo}
\end{eqnarray}
with $r_{\ell,m}=\sqrt{\xi_{\ell,m}^2+\eta_{\ell,m}^2}$ and for a large 
positive number $R$. The first sum in the right-hand side is coverges to 
\begin{equation}
\int_{x^2+y^2\le R^2}dxdy\left|U^{(2)}(x,y)\right|
\end{equation}
as $L_x,L_y\rightarrow+\infty$ for a fixed large $R$ 
because $\left|U^{(2)}\right|$ is uniformly continuous from the assumtion 
on $U^{(2)}$. The second sum in the right-hand side of (\ref{sumU2decompo}) 
becomes small for a large $R$ from the assumtion 
(\ref{decayU2bound}) of $U^{(2)}$ about the decay for a large distance. 
Combining these observations with (\ref{sumkintU2phiphi}) and 
(\ref{U2epsilonCbound}), we obtain the desired bound (\ref{Urhonn}). 
\end{proof}

{From} (\ref{boundD1}), (\ref{Urhonn}) and (\ref{epsilon2}), 
we have 
\begin{eqnarray}
& &\sum_\xi\sum_{\xi'}\int_Sdxdy\ \left|U^{(2)}(x-x',y-y')\right|
\left|\phi_\xi^{\rm P}(x,y)\right|\left|\phi_{\xi'}^{\rm P}(x,y)\right|\ret
&\le&(n_{\rm max}+1)^2{\cal C}^{(10)}(U^{(2)})\left\{
{\cal C}^{(4)}(n_{\rm max})\sum_{\Delta\ell=-\infty}^{+\infty}
\exp\left[-\left(\frac{\pi\ell_B}{L_x}\Delta\ell-n_{\rm max}\right)^2\right]
\right.\ret
&+&\left.{\cal C}^{(5)}(n_{\rm max})\frac{L_xL_y}{2\pi\ell_B^2}
\exp\left[-\frac{L_y^2}{32\ell_B^2}\left(1-\frac{32n_{\rm max}\ell_B}{L_y}
\right)^2\right]\right\}. 
\end{eqnarray}
Since the sum in the right-hand side can be easily evaluated as 
\begin{equation}
\sum_{\Delta\ell=-\infty}^{+\infty}
\exp\left[-\left(\frac{\pi\ell_B}{L_x}\Delta\ell-n_{\rm max}\right)^2\right]
\le{\cal C}^{(11)}(n_{\rm max})+\frac{L_x}{\ell_B}{\cal C}^{(12)}(n_{\rm max}),
\end{equation}
we obtain the bound (\ref{intUphiphiinequality}) with (\ref{defkappa}). 
Here the constants 
${\cal C}^{(11)}(n_{\rm max})$ and ${\cal C}^{(12)}(n_{\rm max})$ depend 
on the energy cutoff $n_{\rm max}$ only. 


\bigskip
\bigskip

\noindent
{\Large\bf Acknowledgements} 
\medskip

\noindent
It is a pleasure to thank the following people for discussions 
and correspondence: 
Mahito Kohmoto, Hal Tasaki, and Masanori Yamanaka. 


\end{document}